\documentclass[preprint,12pt]{elsarticle}
\usepackage{amssymb}
\usepackage{graphicx}
\usepackage{multicol}
\usepackage{balance}
\usepackage{stfloats}
\usepackage[cmex10]{amsmath}
\usepackage{amsmath}
\usepackage{color}
\usepackage{CJK}
\usepackage{algorithm}
\usepackage{algorithmicx}
\usepackage{algpseudocode}

\begin{document}

\begin{frontmatter}



\title{Spatial-Frequency Domain Nonlocal Total Variation for Image Denoising}

\author[a]{Haijuan Hu}
\author[b]{Jacques Froment}
\author[c]{Baoyan Wang}
\author[d]{Xiequan Fan}
\address[a]{Northeastern University at Qinhuangdao,
              School of Mathematics and Statistics,  Hebei, 066004, China.
              ({huhaijuan61@126.com}).}

\address[b]{Univ Bretagne-Sud, CNRS UMR 6205 LMBA, Campus de Tohannic, F-56000 Vannes, France ({Jacques.Froment@univ-ubs.fr}).}
\address[c]{Northeastern University, School of Computer Science and Engineering, No.3-11, Wenhua Road, Heping District, Shenyang, 110819, China ({wangbaoyan2005@163.com}).}
\address[d]{Tianjin University, Center for Applied Mathematics, Tianjin, 300072,  China({fanxiequan@hotmail.com}).}
\begin{abstract}
Following the pioneering works of Rudin, Osher and Fatemi on total variation (TV) and of Buades,
Coll and Morel on non-local means (NL-means), the last decade has seen a large number of denoising
methods mixing these two approaches, starting with the nonlocal total variation (NLTV) model.
The present article proposes an analysis of the NLTV model for image denoising as well as a number of
improvements, the most important of which being to apply the denoising both in the space domain and
in the Fourier domain, in order to exploit the complementarity of the representation of image data
in both domains. A local version obtained by a regionwise implementation followed by an aggregation
process, called Local Spatial-Frequency NLTV (L-SFNLTV) model, is finally proposed as a new reference
algorithm for image denoising among the family of approaches mixing TV and NL operators. The
experiments show the great performance of L-SFNLTV, both in terms of image quality and of
computational speed, comparing with other recently proposed NLTV-related
methods.
\end{abstract}


\end{frontmatter}





 \section{Introduction}
 Gaussian noise removal is  a fundamental and simple inverse problem of image processing, and it  provides a good platform for theoretical research. Therefore it is still a hot topic in recent years, despite that there have been numerous works in the past 60 years or so. Among the numerous works,  two examples of well-known algorithms are total variational model and non-local means. The total variation model \cite{rudin_nonlinear_1992} is firstly proposed by Rudin, Osher, and Fatemi (ROF), which introduces the total variation (TV) to replace the more traditional Euclidean norm. Since TV allows for discontinuity, the  ROF model can recover image edges and piecewise constant regions very well. But at the same time the recovered images look piecewise constant. Another shortage is the loss of texture due to the fact that total variation disfavors oscillations.
  Non-local means (NL-means) method is proposed by  Buades, Coll, and Morel in \cite{buades2005review}, which exploits the fact that each patch in natural images generally has many similar patches.
 NL-means is a weighted average filter, where the weight between two pixels is determined by the Euclidean distance of two patches centered at the corresponding two pixels, which is more reliable than the weight determined by pixel distances.

In  \cite{gilboa2007nonlocal,gilboa_nonlocal_2008}, inspired by non-local means, a regularization functional called nonlocal total variation (NLTV) is proposed to generalize the total variation.
In \cite{lou2010image}, the nonlocal total variation term is combined with a fidelity term in $L^2$ norm for image deconvolution and tomographic reconstruction.  The applications of nonlocal operators  include
 image denoising \cite{kindermann_deblurring_2006,gilboa2006nonlocal,gilboa2007nonlocal},
image  deblurring \cite{kindermann_deblurring_2006,zhang2010bregmanized,lou2010image,tang2014non},
image inpainting \cite{gilboa_nonlocal_2008,zhang2010wavelet}, motion estimation \cite{werlberger2010motion}, etc.
 Some examples of  recently proposed image denoising methods  which improve the nonlocal total variation model  significantly are \cite{lefkimmiatis2015nonlocal,li2017regularized,liu2017block}.
 In \cite{lefkimmiatis2015nonlocal}, the authors propose a regularizer which employs  a novel non-local version of the
structure tensor.
 In \cite{li2017regularized}, the authors propose a regularization term on the weight function,  combined with the nonlocal total variation term to model the image prior.
In \cite{liu2017block}, a block-based nonlocal total variation is proposed to extend the original  point-based model, and the weighting function can also be adaptively determined by the cost function itself.

In the past years, many other hybrid methods appeared based on the ideas of ROF model, NL-means or  transform domain methods \cite{donoho1994ideal,chang2000adaptive}. For instance, the approaches \cite{elad2006image,dabov2007image,mairal2009non,gu2014weighted,hu2018note} are based on the fact that similar patches in images can be sparsely represented or the matrix composed of similar patches has low rank; the paper  \cite{louchet2011total} proposes a local total variation filter combining the idea of NL-means;
the researches \cite{knaus2013dual,pierazzo2014non} consider dual domain filters. Recent years have seen the emergence of a new paradigm for image denoising, where the
modeling of natural images is left to computers: the deep learning framework \cite{vincent2010stacked,xie2012image,Zhang2017BeyondAG,zhang2018ffdnet,lefkimmiatis2018universal}. This very promising domain will not be considered here and, on the contrary, we will make use of very old fashioned tools such as Fourier analysis.

To estimate the denoising performance without reference to original true image, an effective method is Stein's Unbiased Risk Estimation (SURE) \cite{stein1981estimation}, which provides  an unbiased estimation of the true MSE for the denoised image. When the analytical expression is difficult to obtain, SURE can be estimated using  Monte-Carlo method \cite{ramani2008monte}. In \cite{van2009sure}, the authors give the explicit analytical expression for NL-means method and use it for parameter selection. However, we do not find the computation of SURE for NLTV in the literature.

In this paper, we will study the NLTV model  in \cite{lou2010image}  for image denoising and improve it using Fourier transform. Note that the paper \cite{gilboa2007nonlocal} also considers image denoising with NLTV model, but the weight is different from \cite{lou2010image}.
We first study NLTV: we establish that NLTV is essentially a neighborhood filter as ROF model \cite{louchet2011total} in a simpler way than  \cite{louchet2011total}; we find that NLTV is better than ROF and comparable to non-local means in most of cases by experiments.  In addition, we derive SURE for NLTV, and show that it is better to choose the regularization parameter by it than randomly. Secondly, in order to benefit from the fact that the Fourier transform structures the image data in a totally different way from the standard basis, we consider frequency domain nonlocal total variation, called FNLTV. FNLTV is good at retaining image texture and details at the
cost of leaving some noise evident in homogeneous regions.  In addition, the regularization coefficient of FNLTV is not very sensitive. Thus we propose a spatial-frequency domain nonlocal total variation model (SFNLTV) to combine NLTV in spatial domain and frequency domain. Since the local application of FNLTV (L-FNLTV) greatly improves FNLTV, and local application of NLTV is similar to global NLTV, we finally take into account the local application of SFNLTV,
abbreviated as L-SFNLTV. 
 Experiments show that L-SFNLTV is  better than recently proposed NLTV related algorithms.

This paper is organized as follows: in Section 2, we introduce NLTV and SURE; in Section 3, we show the relations of NLTV with neighborhood filters, discuss parameter choices, and compute SURE for NLTV; in Section 4, the frequency domain nonlocal total variation (FNLTV) is studied; the spatial-frequency domain nonlocal total variation model (SFNLTV) and Local SFNLTV models are proposed in Section 5; simulations are provided in Section 6, and the conclusion is made in Section 7.

\section{Preliminaries}

As usual,  a digital image is denoted by a $M\times N$ matrix $u = \{u(i):
i\in I \}$, where $I=\{0, 1, ..., M-1\} \times \{0, 1, ..., N-1\}$,
and $0\leq u(i)\leq 255$. The additive Gaussian noise
model is:
\begin{equation}
v(i)=u_0(i)+\eta(i), \label{model}
\end{equation} 
 where $u_0$ and  $v$ are the
original and noisy images respectively, and $\eta$ is the
Gaussian noise: $\eta(i)$ are independent and identically distributed
Gaussian random variables with mean $0$ and positive standard deviation
 $\sigma$. A denoised image is denoted as $\bar{v}$. For simplicity, we assume symmetric boundary conditions in this paper.

\subsection{Nonlocal variation model}
\label{chanltv}



We consider the following variational model introduced in  \cite{gilboa_nonlocal_2008,lou2010image}:
 \begin{equation}
E(u):=\lambda \sum_{i\in I}  |\nabla_{{w}}u(i)|+\frac{1}2\sum_{i \in I}(u(i)-v(i))^2,\label{Eudis}
\end{equation}
where 
 \begin{equation}
|\nabla_{{w}}u(i)|=\sqrt{\sum_{j\in I}(u(i)-u(j))^2 w(i,j)} \label{ju}
\end{equation}
is called the nonlocal total variation of $u$.
The variational model (\ref{Eudis}) is denoted as NLTV. The denoised image $\bar{v}$ is the minimizer of the functional (\ref{Eudis}).
For $i \in I$ and $d$  an odd integer, let
$\mathcal{N}_{i}(d)= \{j\in I: \|j-i\|_{\infty} \leq (d-1)/2\}$  be the window with center
$i$ and size $d \times d$,
 simply written as
$\mathcal{N}_{i}$. Similarly, denote $U_i(D)$ the search window with center
$i$ and size $D \times D$, simply written as $U_i$.
We consider $w(i,j)$ used in \cite{lou2010image}:   $w(i,j)$
  is  taken as the one used in NL-means\cite{buades2005review} without normalization, which is different from the weight in \cite{gilboa2007nonlocal}.
 That is, the following weight is used
\begin{equation}
w(i,j) = \left\{\begin{array} {ll}
e^{- ||v(\mathcal{N}_i)-v(\mathcal{N}_j )||_a^{2}/ (2\sigma_r^2)} &  \mbox{if} \quad j\in U_i(D) \\
0 & \mbox{else} \end{array}\right., \label{nltvweight}
\end{equation}
where  $\sigma_r>0$ is a control parameter, and
\begin{equation}
     ||v(\mathcal{N}_{i})-v(\mathcal{N}_{j})||_a^2 =\frac{ \sum_{k\in \mathcal{N}_i(d)} a(i,k) |v(k)-v(\mathcal{T}(k))|^2}{\sum_{k\in \mathcal{N}_i(d)} a(i,k)}, \label{nlmdisnorm}
\end{equation}
with $\mathcal{T}=\mathcal{T}_{ij}$ being the translation mapping of $\mathcal{N}_i$ onto $\mathcal{N}_j$:  $\mathcal{T}(k)=k-i+j, k\in \mathcal{N}_i$, and $ a(i,k)=e^{-\|i-k\|^2/2\sigma_s^2}$ ($\sigma_s=(d-1)/4$ is a good choice).

The  weight $w(i,j)$ is used to estimate the similarity between two pixels $i$ and $j$.  When the two pixels $i$ and $j$ are similar, $w(i,j)$ is large, which makes the recovered values $\bar{v}(i)$ and
 $\bar{v}(j)$ close. Note that for $i=(i_1,i_2)$, if we use
\begin{equation}
w(i,j) = \left\{\begin{array} {ll}
1 &  \mbox{if} \quad j\in \{(i_1+1, i_2), (i_1, i_2+1)\} \\
0 & \mbox{else} \end{array}\right. \label{weighttv},
\end{equation}
then (\ref{Eudis}) reduces to the classical total variation model \cite{rudin_nonlinear_1992} denoted as ROF. Thus NLTV is more general and more adaptive to image content than ROF.

For any fixed $i$, $|\nabla_{{w}}u(i)|$ is a convex functional of $u$. Therefore $ \sum_{i\in I}  |\nabla_{{w}}u(i)|$ is convex. Since the fidelity term is strictly convex, so is the energy function $E(u)$.
Thus  the gradient descent method can be used to find the minimizer.
Write
\begin{equation}
W_u(i,j)=\frac{w(i,j)}{|\nabla_{{{w}}}u(i)|}+\frac{w(j,i)}{|\nabla_{{{w}}}u(j)|}, \label{W}
\end{equation}
then the gradient of $E(u)$ \cite{lou2010image} is $\nabla E(u)=\{\nabla E(u)(i)\}_{i\in I}$, with
\begin{equation}
\nabla E(u)(i)=\lambda\sum_{j\in I}\left(u(i)-u(j)\right)W_u(i,j)+u(i)-v(i). \label{eulernltv}
\end{equation}
Notice that the equation (\ref{W}) is not well defined at points $|\nabla_{{{w}}}u(i)|=0$. The common technique is to replace  $|\nabla_{{{w}}}u(i)|$ by $\sqrt{|\nabla_{{{w}}}u(i)|^2+\beta}$ with $\beta$ a small positive constant.
The  gradient descent algorithm can be expressed as follows,
\begin{equation}u^{k+1}(i)=u^k(i)-t_k\nabla E(u^k)(i), \quad k=0,1,2,\cdots \label{graddes} \end{equation}
where $u^0=v$ and $t_k$ is chosen such that $E(u^{k+1})<E(u^k)$.
 If $\|\nabla E(u^k)\|=0$, then $u^k$ is the desired solution, and the iterative process stops. Some stopping criterion is generally used, such as $\|u^{k+1}-u^k\|<\epsilon$, $|E(u^{k+1})-E(u^k)|<\epsilon$, and/or $k<n_{iter}$, where $\epsilon>0$ small enough, and $n_{iter}$ an integer large enough.

\subsection{Mean squared error and Stein's unbiased risk estimate}
The mean squared error (MSE) of the denoised image $\bar{v}$ with respect to the original image $u_0$ is
\begin{equation}
\mbox{MSE}(\bar{v})=\frac 1{MN}{\sum_{i\in I }(\bar{v}(i) -u_0 (i))^2}.
\label{mse}
\end{equation}

SURE provides an unbiased estimation of the true MSE for the denoised image of an algorithm. It is expressed by the following analytical equation \cite{stein1981estimation,van2009sure}:
\begin{equation}
\mbox{SURE}=\frac 1{MN}{\sum_{i\in I }(\bar{v}(i) -v (i))^2}-\sigma^2+2\sigma^2\frac{\mbox{div}_v\{\bar{v}\}}{MN}, \label{sure}
\end{equation}
where $\mbox{div}_v\{\bar{v}\} $ is the divergence of the denoised image $\bar{v}$ of the algorithm with respect to the noisy image $v$
\begin{equation}
\mbox{div}_v\{\bar{v}\} =\sum_{i\in I} \frac{\partial \bar{v}_i}{\partial v_i}, \label{surediv}
\end{equation}
and the variance of the noise $\sigma^2$  is supposed to be known. 
\section{Analysis of NLTV}
\subsection{Relations with neighborhood filters} \label{nltvneighbor}
We will establish that NLTV is essentially a neighborhood filter by analyzing the gradient descent algorithm.
Thus the image can be divided into local regions and NLTV is applied to each region, which can be implemented in parallel and speed up the algorithm.

\begin{figure}
\begin{center}
\includegraphics[width=0.50\linewidth]{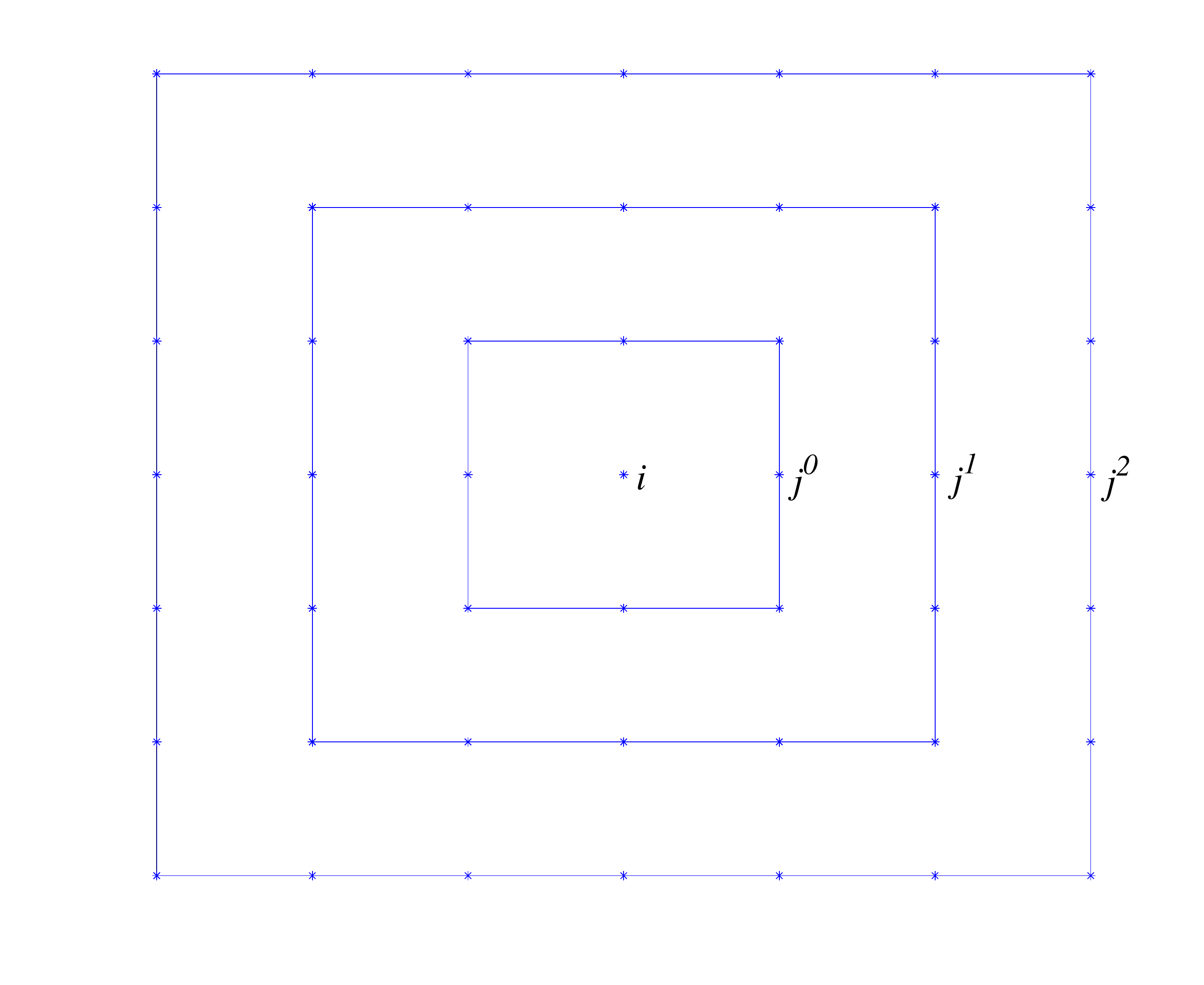}
\caption{ An illustration of the set $\mathcal{U}^2_i$ with $D=3$. }\label{huixingkuang}
\end{center}
\end{figure}

\begin{figure}
\begin{center}
\includegraphics[width=0.90\linewidth]{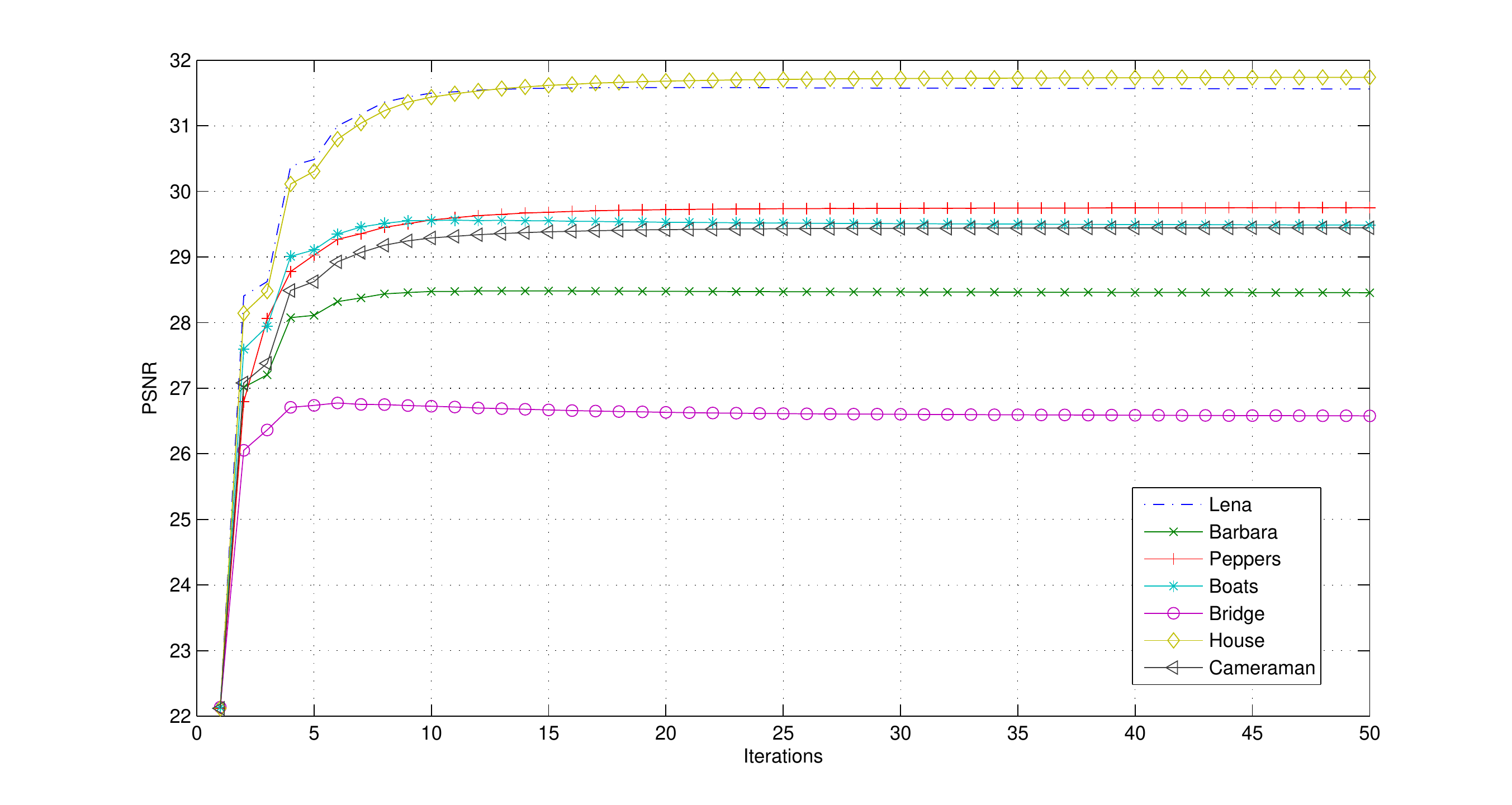}
\caption{ PSNR values versus iterations for different images for NLTV. }\label{nltvpsnriteration}
\end{center}
\end{figure}

Notice that by ($\ref{nltvweight}$), the summation of $j\in I$ in (\ref{eulernltv}) can be replaced by
$j\in U_i^0=U_i\backslash\{i\}$.
By (\ref{eulernltv}) and $(\ref{graddes})$, we have   
\begin{equation}
u^{k+1}(i)=\sum_{j\in U_i} u^k(j)W^k(i,j)+t_k v(i), \label{nltvneidisiter}
\end{equation}
where
\begin{equation}
W^k(i,j)=\left\{\begin{array}{ll}
t_k\lambda W_{u^k}(i,j) & \mbox{if} \quad i\neq j \\
1-t_k-t_k\lambda\sum_{j\in U_i^0} W_{u^k}(i,j) & \mbox{if} \quad i=j.
\end{array}\right. \label{nltvweighted}
\end{equation}
Note that
$$\sum_{j\in U_i}W^k(i,j)+t_k=1.
$$
By (\ref{nltvneidisiter}), we can obtain
\begin{eqnarray}  
u^{k+1}(i)&=&\sum_{(j^0,j^1,\cdots,j^k)\in \mathcal{U}^k_i}
W^k(i,j^0) W^{k-1}(j^0,j^1)\cdots  W^0(j^{k-1},j^k)u^0(j^k) \nonumber \\
&&+t_0\sum_{(j^0,j^1,\cdots,j^{k-1})\in \mathcal{U}^{k-1}_i} 
W^k(i,j^0)\cdots W^1(j^{k-2},j^{k-1})v(j^{k-1}) \nonumber \\
&&+\cdots \nonumber\\
&&+t_{k-1}\sum_{j^0\in \mathcal{U}^0_i=U_i}W^k(i,j^0)v(j^0)\nonumber \\
&&+t_k v(i). \nonumber
\end{eqnarray}
with $$\mathcal{U}^k_i=\{(j^0,j^1,\cdots,j^k) : j^0\in {U}_i, j^1\in {U}_{j^0},\cdots, j^{k-1}\in {U}_{j^{k-2}}, j^k\in {U}_{j^{k-1}}\}.$$
Recalling that $U_i=U_i(D)$,  in our experiments, $D=3$ is used. An illustration of the set $\mathcal{U}^2_i$ with $D=3$ is shown in Figure \ref{huixingkuang}.
Thus we can see that  $u^{k+1}(i)$ is the weighted average of its  neighbors $v(j), j\in {U}_i(Dk+D-k)$ if we do not consider that $W^l(s,t)$ depends on the neighbors of $s$ and $t$. In Figure $\ref{nltvpsnriteration}$, we plot the PSNR values versus iterations for different images. After 15 iterations, the iterative process for all the image approximately converges. Thus  NLTV model is also a local operator as ROF model stated in \cite{louchet2011total}. 

In addition,
by (\ref{nltvneidisiter}) and (\ref{nltvweighted}), 
when $i$ is an isolated point, there are no or few similar points in its neighbors, so
$\sum_{j\in U_i^0(D)} W(i,j)$ is small. Thus the recovered value is close to the noisy one. This explains that NLTV is good at preserving  isolated values unlike NL-means (visual results can be seen in Sec. \ref{sec6_nltv}).

\subsection{Parameter Choices}
We have some remarks about the choices of the parameters. 

\begin{figure}
\begin{center}
\center 
\renewcommand{\arraystretch}{0.5} \addtolength{\tabcolsep}{0pt} \vskip3mm %
\fontsize{8pt}{\baselineskip}\selectfont
\begin{tabular}{ccc}
\includegraphics[width=0.30\linewidth]{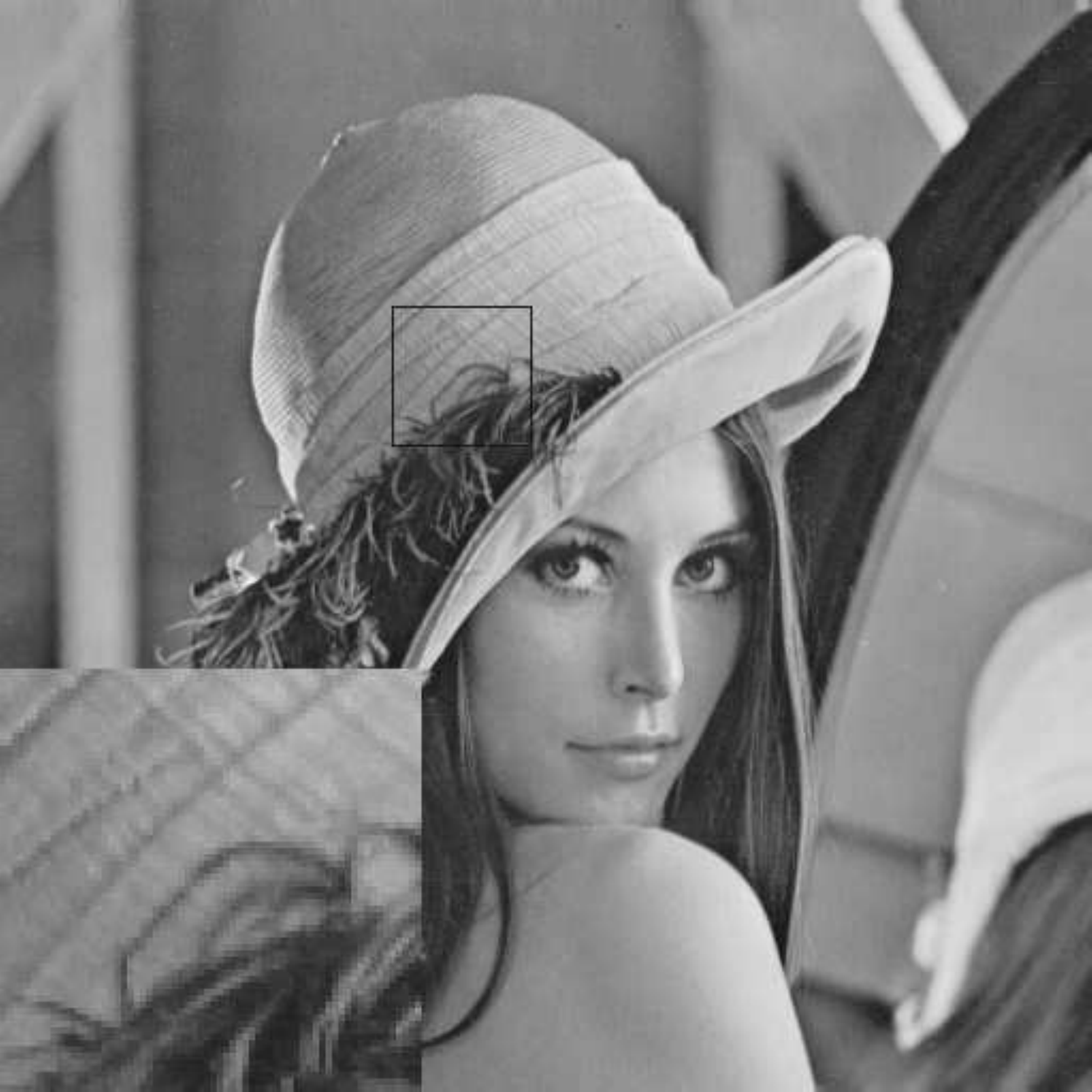}
& \includegraphics[width=0.30\linewidth]{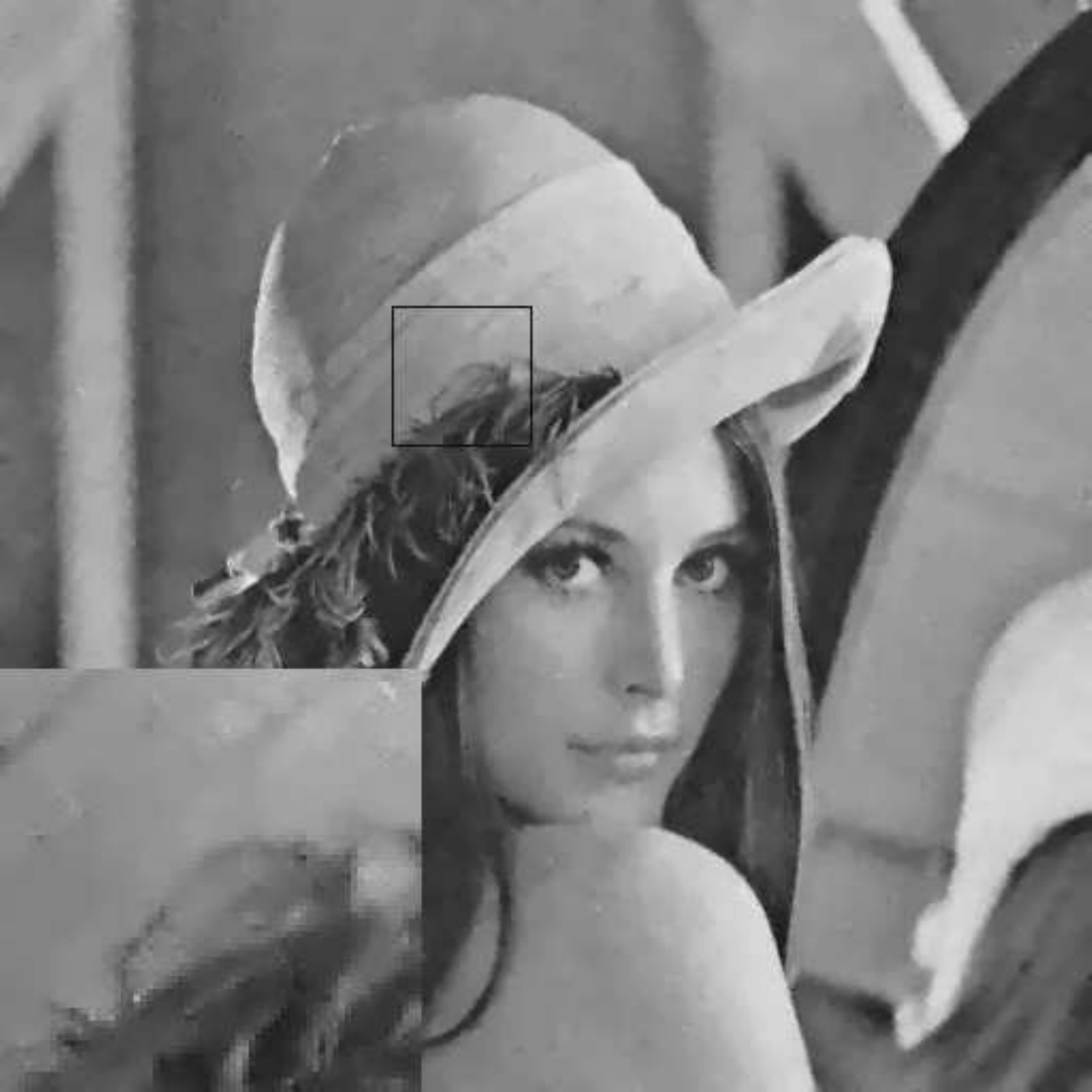}
&\includegraphics[width=0.30\linewidth]{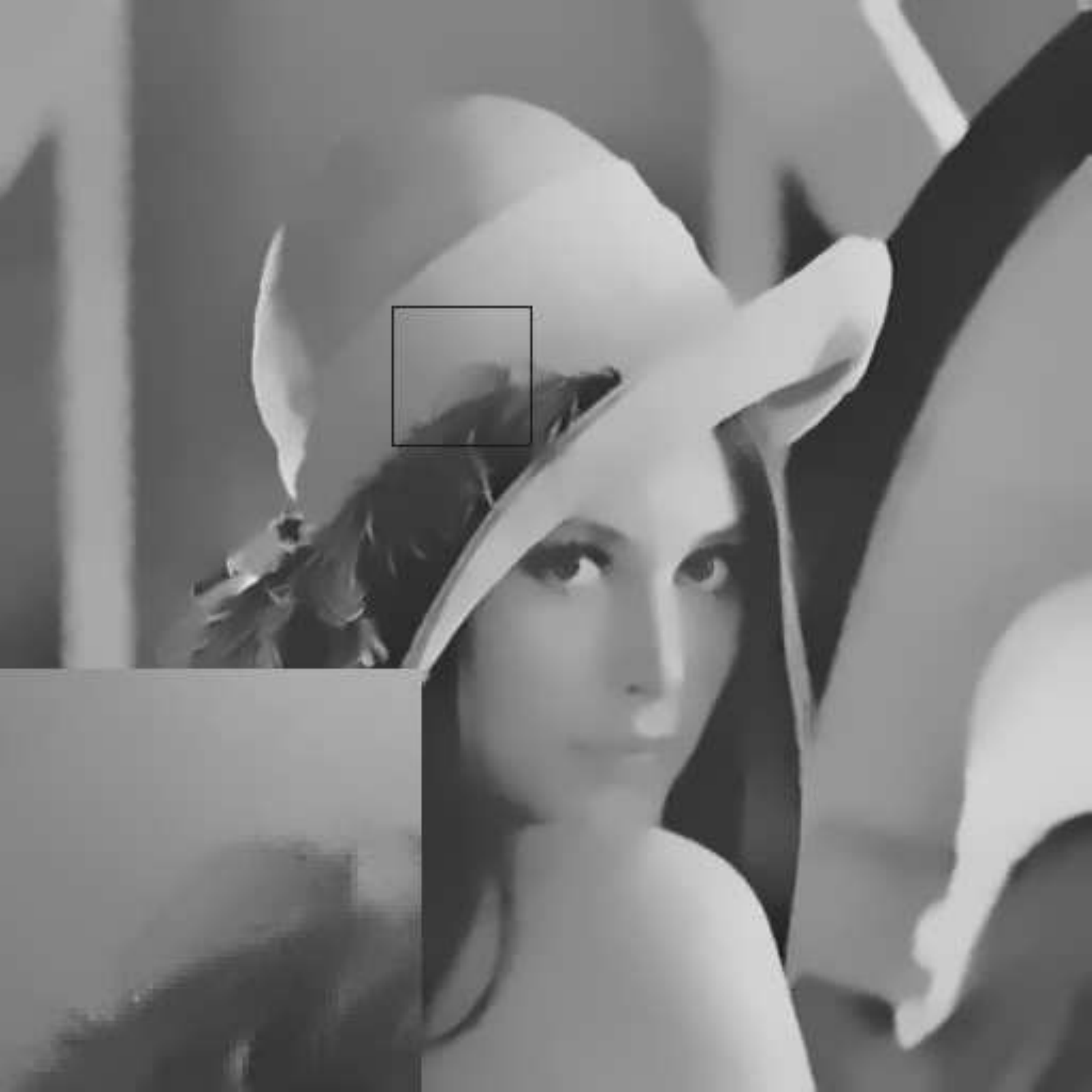}\\
\small{Original} & \small{$D=3$} & \small{$D=7$}\\
\includegraphics[width=0.30\linewidth]{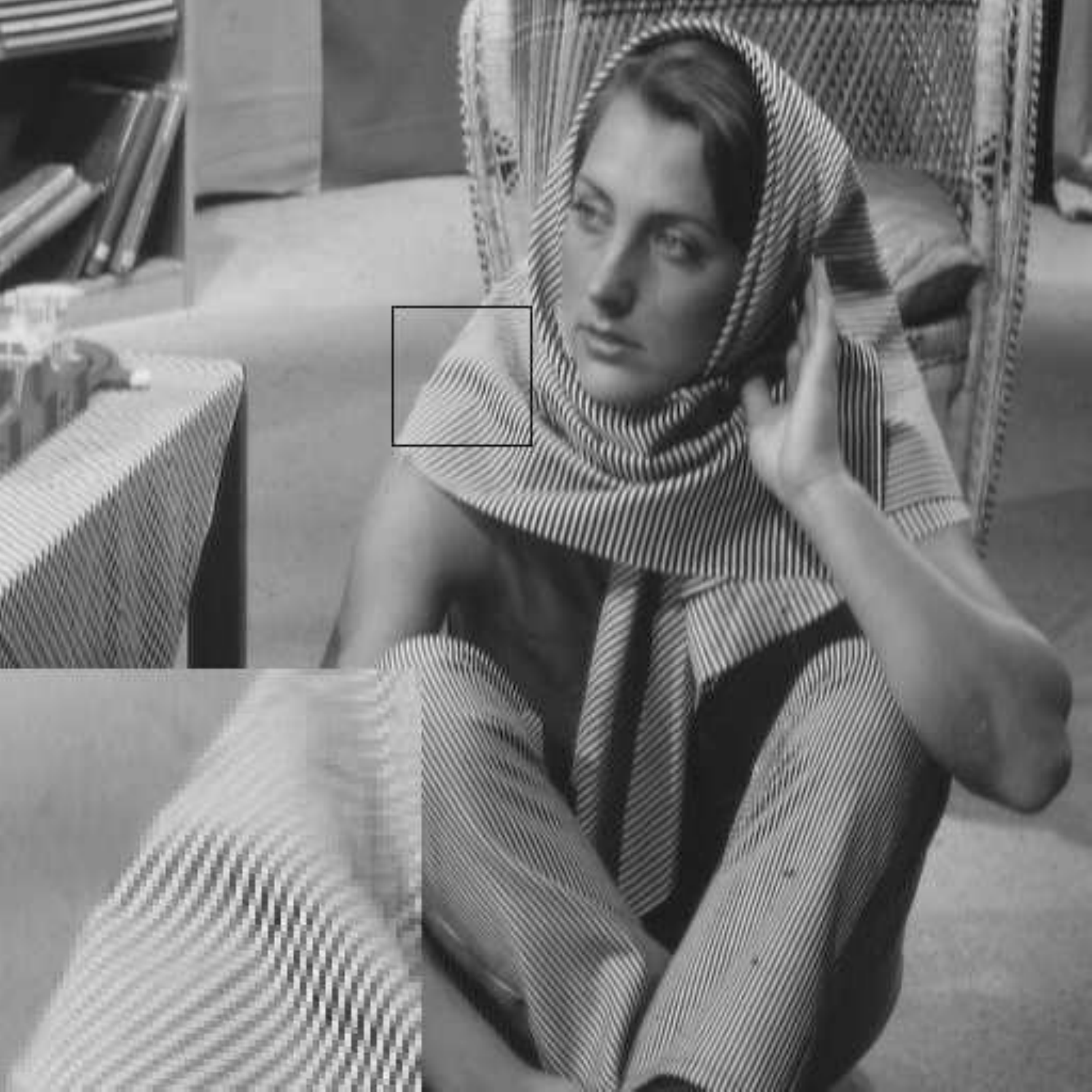}
&\includegraphics[width=0.30\linewidth]{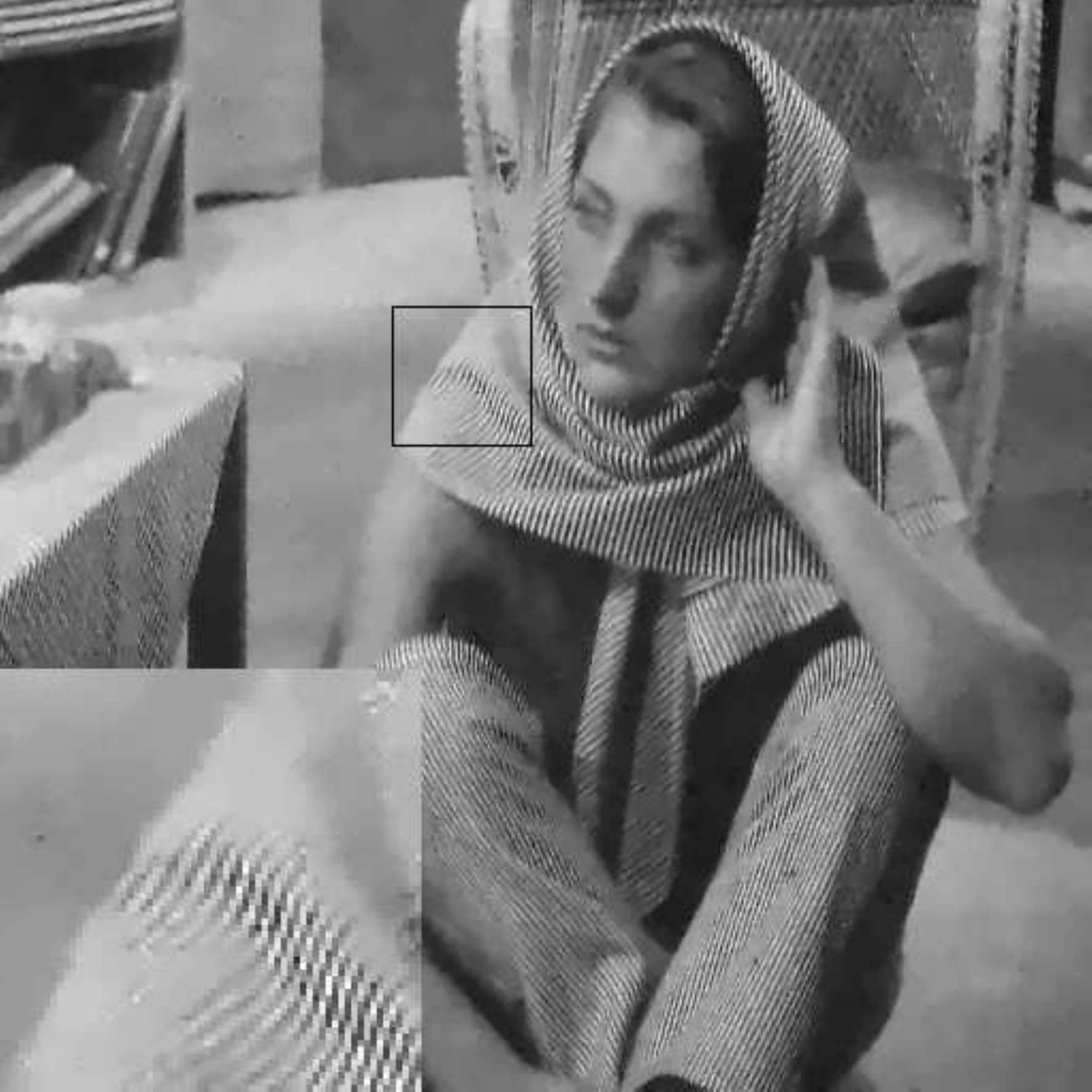}
&\includegraphics[width=0.30\linewidth]{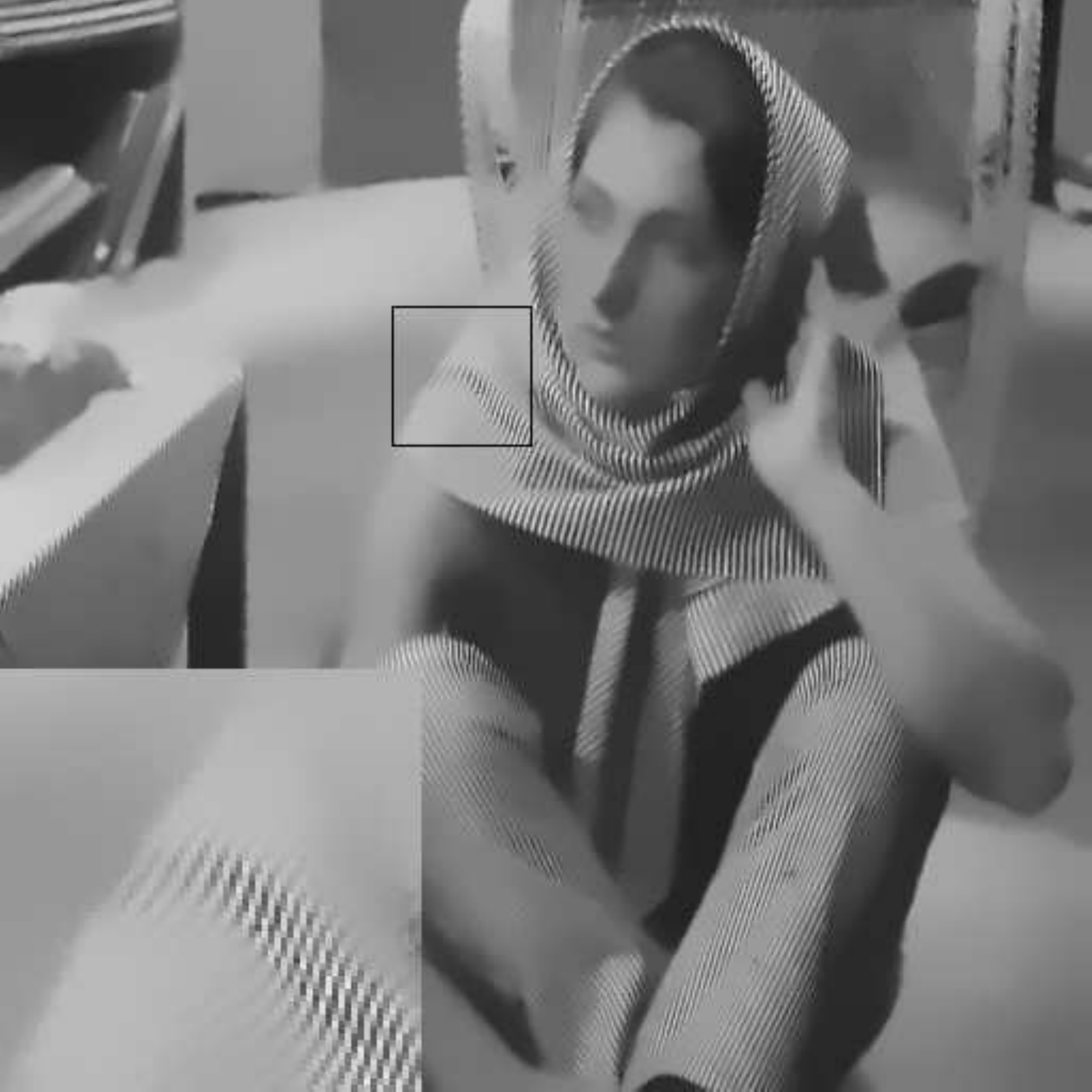}\\
\small{Original} & \small{$D=3$} & \small{$D=7$}\\
\end{tabular}
\caption{Denoised image by NLTV with\ $D=3, 7$ for Lena (top) and   Barbara (bottom) with $\sigma=20$.}\label{barnltvd}
\end{center}
\end{figure}

$\bullet$ We first study the influence of the search window size $D$. In Figure $\ref{barnltvd}$, we show the denoised images for different values of $D$. When
 $D$ increases, restored images look smoother globally, and more fine details disappear. However, in the areas near edges, there is more noise left. Though some regular textures are better recovered and the staircase artifact is less obvious for larger $D$, the global quality is deteriorated. Thus the proposed choice $D=21$  in \cite{lou2010image} in accordance with the choice in NL-means in \cite{buades2005review} is not an optimal choice for image denoising. Since a better choice for $D$ is a smaller value, the denoising process is faster. 

\begin{figure}
\begin{center}
\begin{tabular}{ccc}
\includegraphics[width=0.30\linewidth]{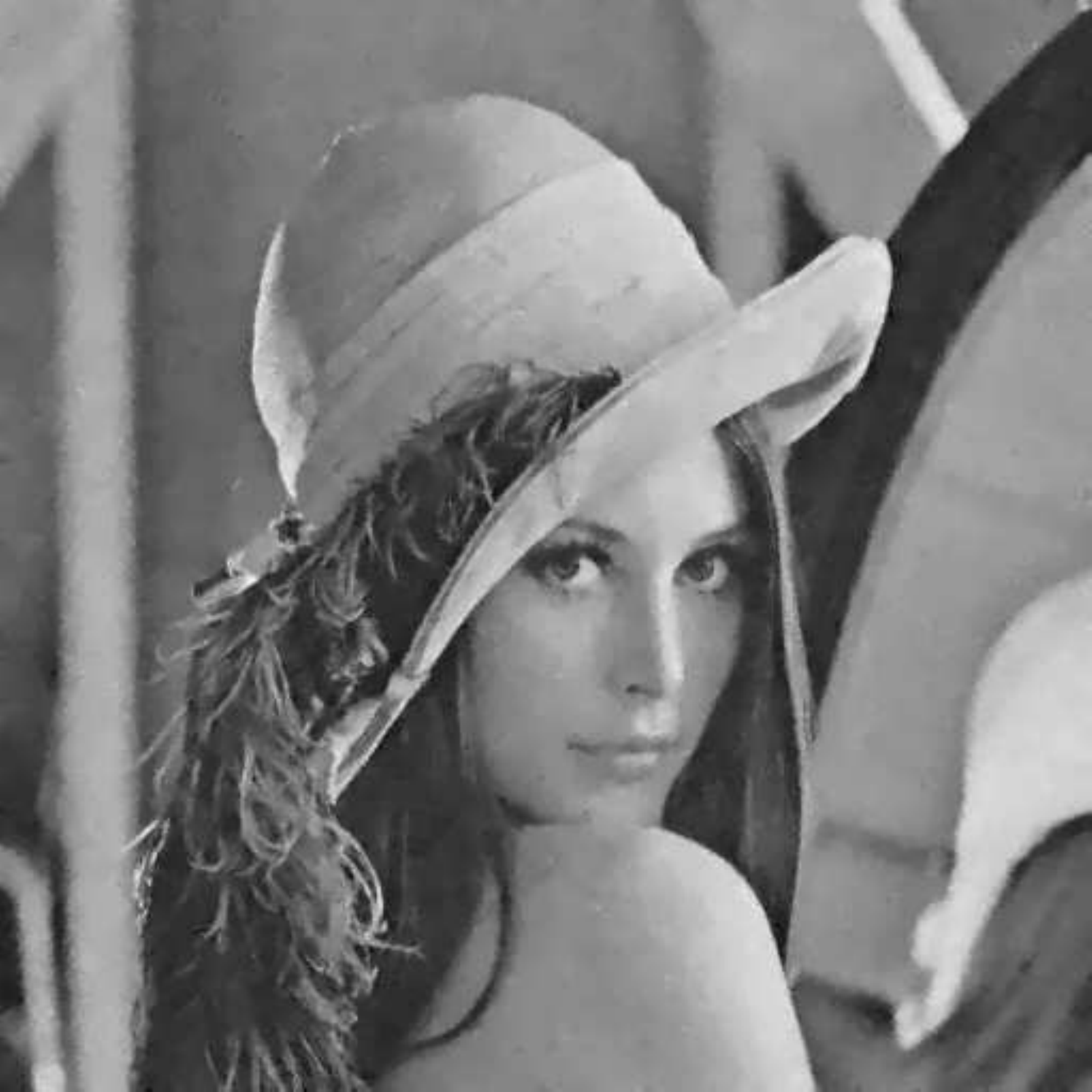} & 
\includegraphics[width=0.30\linewidth]{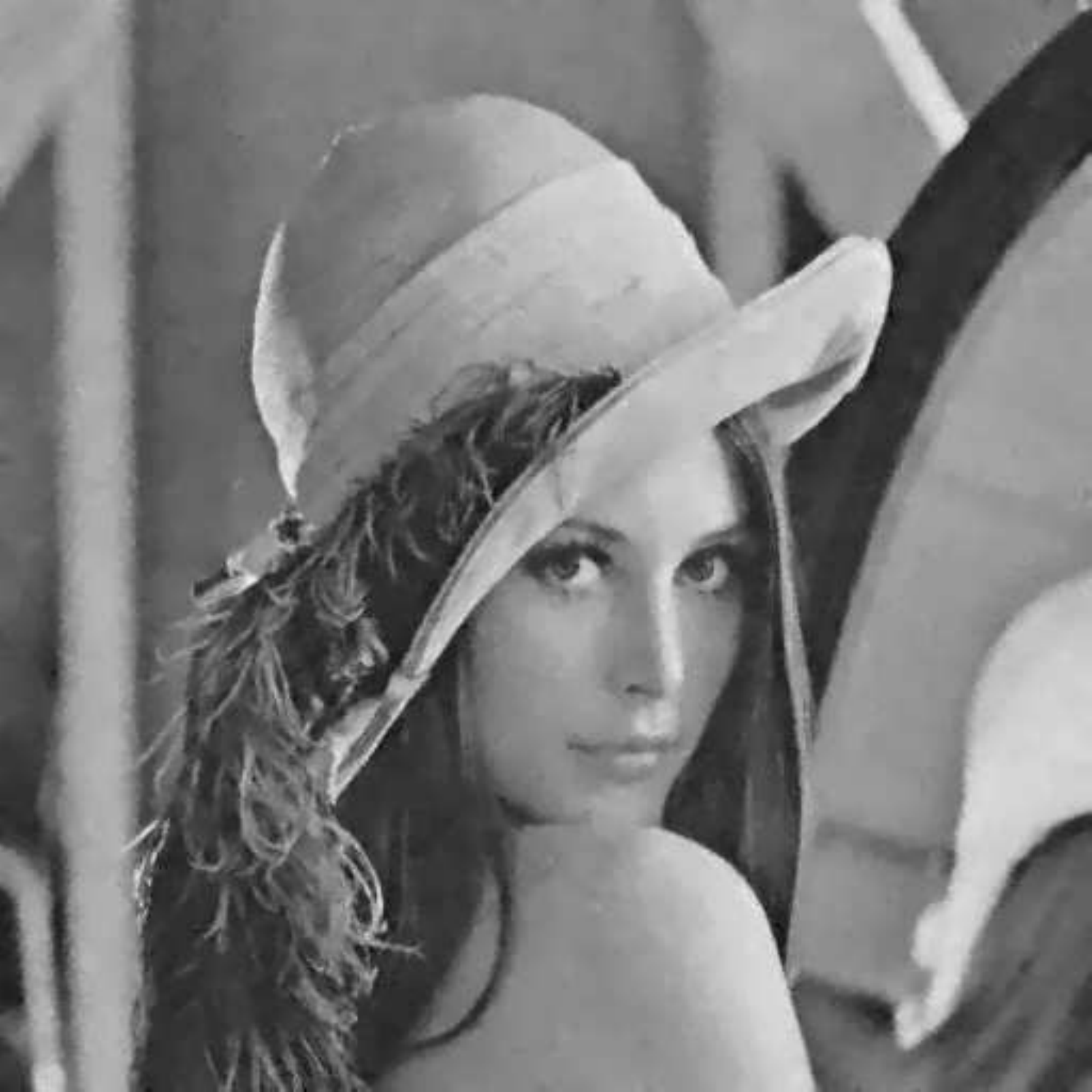}& 
\includegraphics[width=0.30\linewidth]{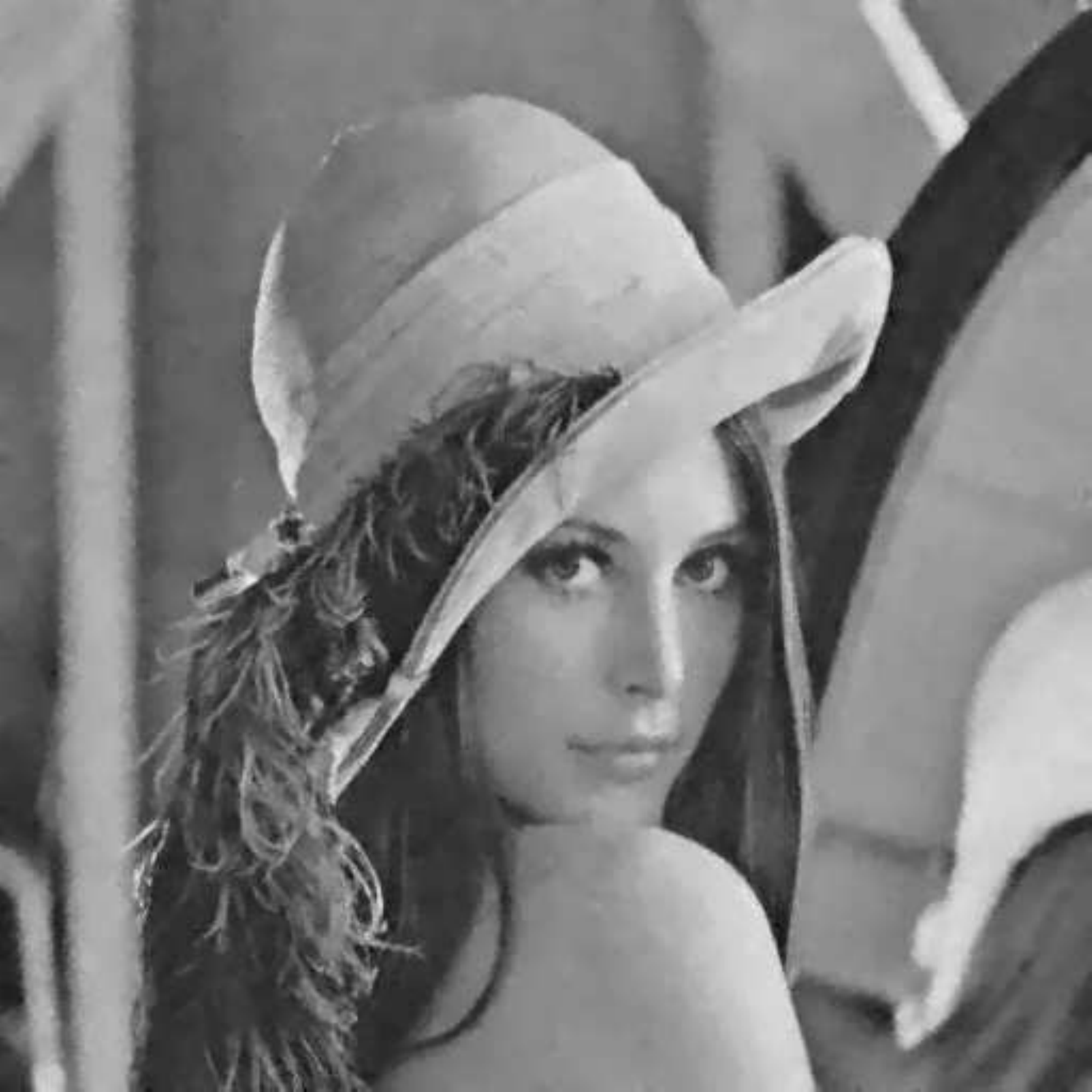} 
\end{tabular}
\caption{Denoised image by NLTV with\ $d=11, 17, 25 $ from left to right with $\sigma=20$.}\label{lenanltvs}
\end{center}
\end{figure}

$\bullet$ In Figure $\ref{lenanltvs}$, we compare the results for different $d$. We can see that there are not significant differences among the three images. In fact, as the patch size $d$ increases, in the limit case, all the nonzero weights $w(i,j)$ tend to be identical, thus close to ROF model if $D=3$.

$\bullet$ The choice of $\lambda$: the optimal $\lambda$ depends on the nonlocal total variation (thus depends on the other parameters $d, D, \sigma_r$) and on the noise level. In the above experiments, we always choose the optimal $\lambda$ for comparisons.

\subsection{SURE based NLTV}
In this section, we compute SURE for NLTV according to (\ref{sure}) in order to get an estimation of MSE.
Since there is no analytic expression for  NLTV model and it can be solved by gradient descent iterations, we calculate SURE for each iteration.
The core of SURE is the calculation of the divergence. For the proposed algorithms, the calculation of divergence of the gradient of $E(u)$ is crucial.
Since we use symmetric weights for NLTV, by ($\ref{W}$) and ($\ref{eulernltv}$), the gradient can be rewritten as
\begin{equation}
\nabla E(u)(i)=\lambda\sum_{j\in I}\left(u(i)-u(j)\right)w(i,j)(\frac{1}{|\nabla_{{{w}}}u(i)|}+
\frac{1}{|\nabla_{{{w}}}u(j)|})+u(i)-v(i). \label{eulernltv2}
\end{equation}
Note that the value at each pixel $i$ depends on its neighbors, so we calculate the partial derivative of  $\nabla E(u)(i)$
with respect to each pixel $l$ in the image. It can be easily obtained that
\begin{equation}
\frac{\partial \nabla E(u)(i)}{\partial v(l)}=\lambda(P_1(i,l)+P_2(i,l)+P_3(i,l))+\frac{\partial u(i)}{\partial v(l)}-\delta(i,l), \label{par_grad_e}
\end{equation}
with
\begin{equation}
P_1(i,l)=\sum_{j\in I}(\frac{\partial u(i)}{\partial v(l)}-\frac{\partial u(j)}{\partial v(l)})w(i,j)(\frac{1}{|\nabla_{{{w}}}u(i)|}+\frac{1}{|\nabla_{{{w}}}u(j)|}), \label{p1}
\end{equation}
\begin{equation}
P_2(i,l)=\sum_{j\in I}(u(i)-u(j))\frac{\partial w(i,j)}{\partial v(l)}(\frac{1}{|\nabla_{{{w}}}u(i)|}+\frac{1}{|\nabla_{{{w}}}u(j)|}),\label{p2}
\end{equation}
\begin{equation}
P_3(i,l)=\sum_{j\in I}\left(u(i)-u(j)\right)w(i,j)\left(\frac{\partial \frac{1}{|\nabla_{{{w}}}u(i)|}}{\partial v(l)}+\frac{\partial \frac{1}{|\nabla_{{{w}}}u(j)|}}{\partial v(l)}\right), \label{p3}
\end{equation}
and
\begin{equation}
\delta(i,l)=\left\{\begin{array}{ll} 1 & i=l\\
0 & i\neq l\end{array} \right.,
\end{equation}
in which
\begin{eqnarray}
\frac{\partial \frac{1}{|\nabla_{{{w}}}u(i)|}}{\partial v(l)}=&&\frac{-1}{2|\nabla_wu(i)|^3}\sum_{j\in I}[2(u(i)-u(j))(\frac{\partial u(i)}{\partial v(l)}-\frac{\partial u(j)}{\partial v(l)})w(i,j)\nonumber \\
&&+(u(i)-u(j))^2\frac{\partial w(i,j)}{\partial v(l)}]. \label{par_1grad}
\end{eqnarray}
Recall that
$$
|\nabla_{{w}}u(i)|:=\sqrt{\sum_{j\in I}(u(i)-u(j))^2 w(i,j)},
$$
where
\begin{equation}
w(i,j) = \left\{\begin{array} {ll}
e^{- ||v(\mathcal{N}_i)-v(\mathcal{N}_j )||_a^{2}/ (2\sigma_r^2)} &  \mbox{if} \quad j\in U_i(D) \\
0 & \mbox{else} \end{array}\right. , \label{nltvweight2}
\end{equation}
with
\begin{equation}
     ||v(\mathcal{N}_{i})-v(\mathcal{N}_{j})||_a^2 =\frac{ \sum_{k\in \mathcal{N}_i(d)} a(i,k) |v(k)-v(\mathcal{T}(k))|^2}{\sum_{k\in \mathcal{N}_i(d)} a(i,k)}. \nonumber
\end{equation}
Thus, when $j\in U_i(D)$,
\begin{eqnarray}
\frac{\partial w(i,j)}{\partial v(l)}=&&w(i,j)\frac{ \sum_{k\in \mathcal{N}_i(d)} a(i,k) 2(v(k)-v(\mathcal{T}(k)))(\frac{\partial v(k)}{\partial v(l)}-\frac{\partial v(\mathcal{T}(k))}{\partial v(l)}) } { -2\sigma_r^2\sum_{k\in \mathcal{N}_i(d)} a(i,k)} \nonumber \\
=&&w(i,j)\frac{ \sum_{k\in \mathcal{N}_i(d)} a(i,k) (v(k)-v(\mathcal{T}(k)))(\delta(k,l)-\delta(\mathcal{T}(k),l))} { -\sigma_r^2\sum_{k\in \mathcal{N}_i(d)} a(i,k)}. \nonumber
\end{eqnarray}

Recall that
\begin{equation}u^{k+1}(i)=u^k(i)-t_k\nabla E(u^k)(i), \quad k=0,1,2,\cdots \label{graddes2} \end{equation}
It follows that
\begin{equation} \frac{\partial u^{k+1}(i)}{\partial v(l)}=\frac{\partial u^k(i)}{\partial v(l)}-t_k \frac{\partial \nabla E(u^k)(i)}{\partial v(l)}, \quad k=0,1,2,\cdots \label{partial_graddes2} \end{equation}
In our experiments, $u^0=v$, so $\frac{\partial u^0(i)}{\partial v(l)}=\delta(i,l)$. Hence, by iteration, we can get the value of $\frac{\partial u^{k+1}(i)}{\partial v(l)}$ for $ k=0,1,2,\cdots $. Finally, we can obtain $\mbox{div}_v\{\bar{v}\}$, and then the value of SURE by
(\ref{surediv}) and (\ref{sure}).

\begin{figure}
\begin{center}
\begin{tabular}{ccc}
\includegraphics[width=1\linewidth]{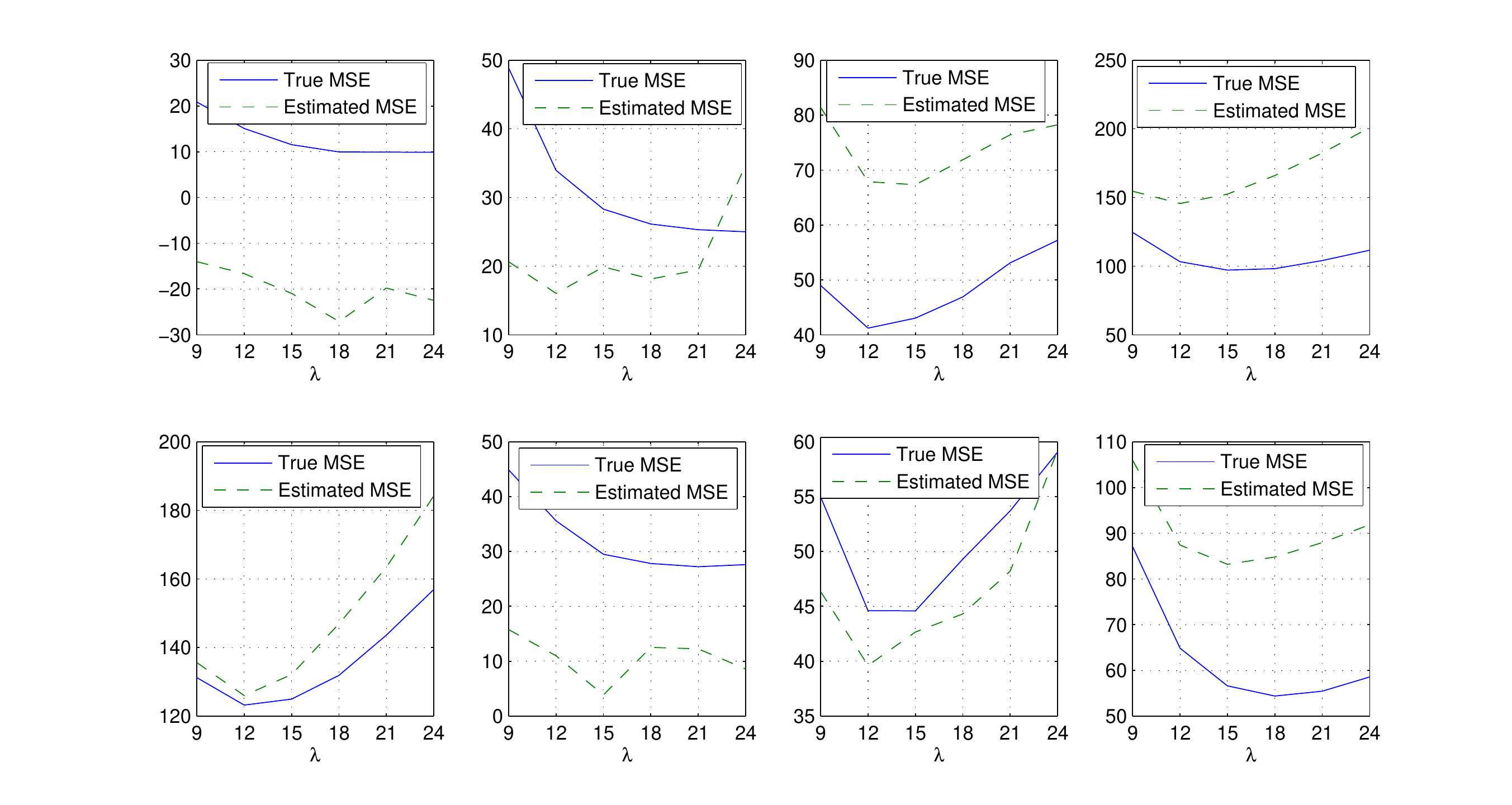}
\end{tabular}
\caption{Estimated MSE and True MSE for regions of size 16$\times$16 extracted from Lena image shown in Figure  \ref{lenapatches1632}.}
\label{estmse16}
\end{center}
\end{figure}

\begin{figure}
\begin{center}
\begin{tabular}{ccc}
\includegraphics[width=0.3\linewidth]{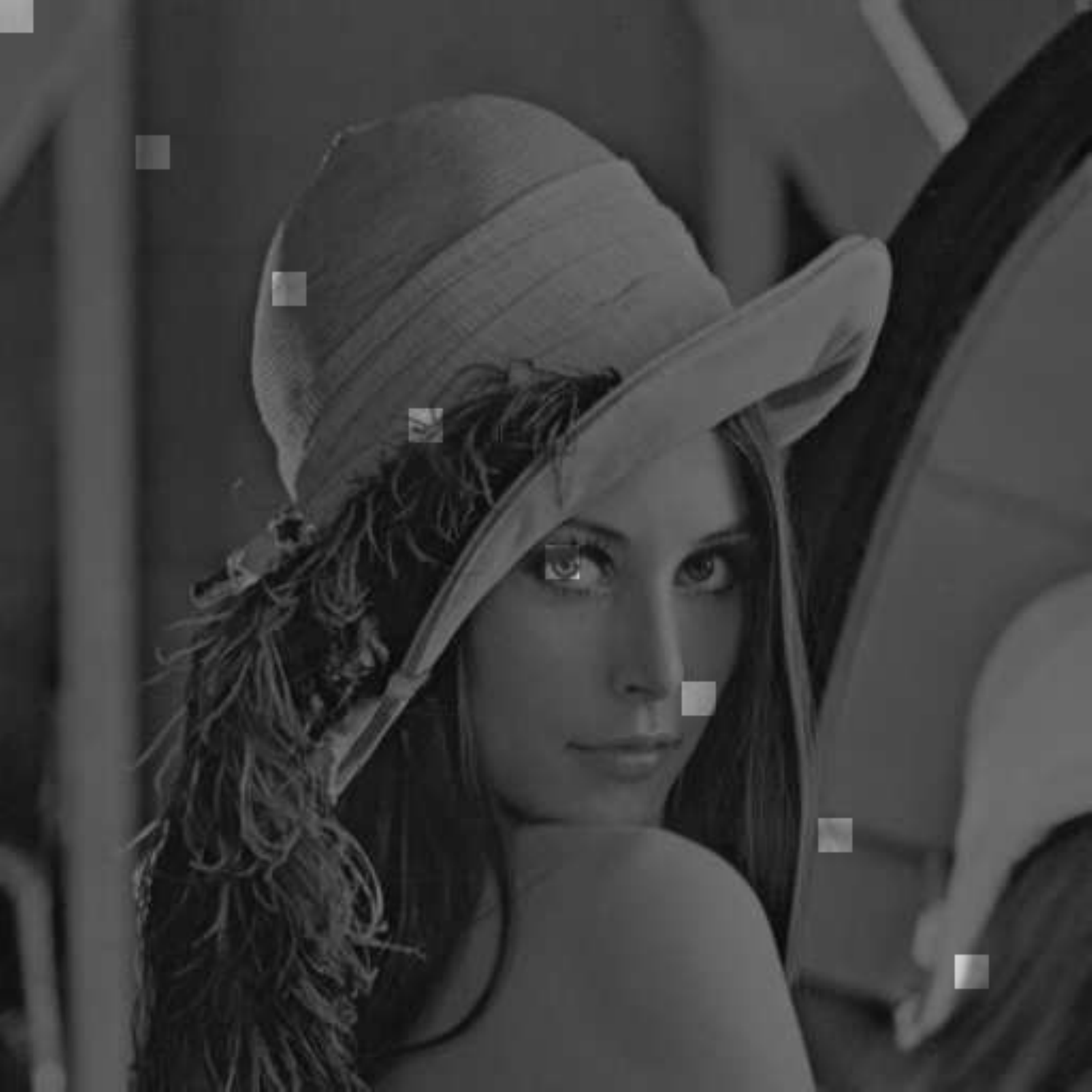}
\end{tabular}
\caption{Lena regions used to compare MSE and estimated MSE.}
\label{lenapatches1632}
\end{center}
\end{figure}

\begin{figure}
\begin{center}
\begin{tabular}{ccc}
\includegraphics[width=0.3\linewidth]{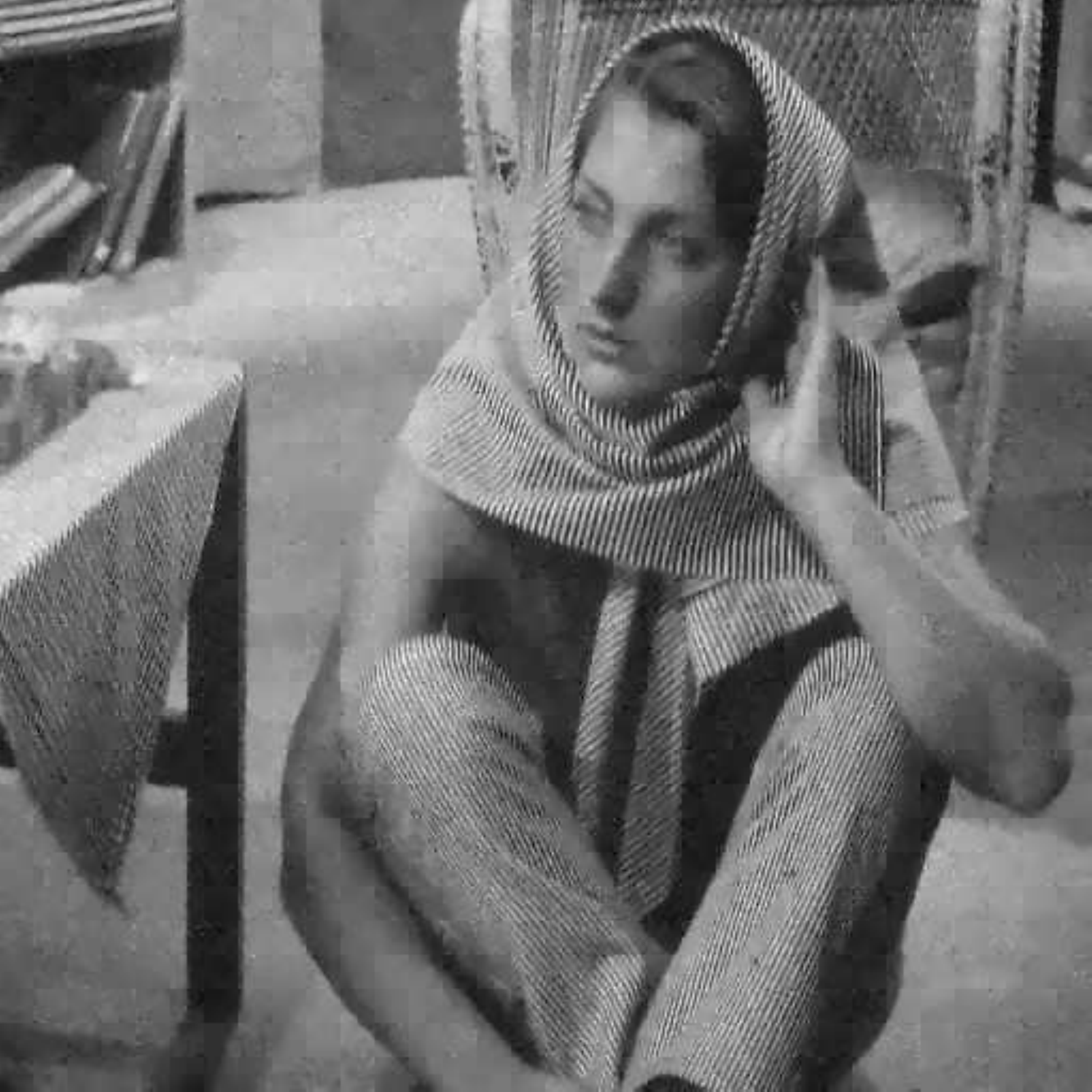}&\includegraphics[width=0.3\linewidth]{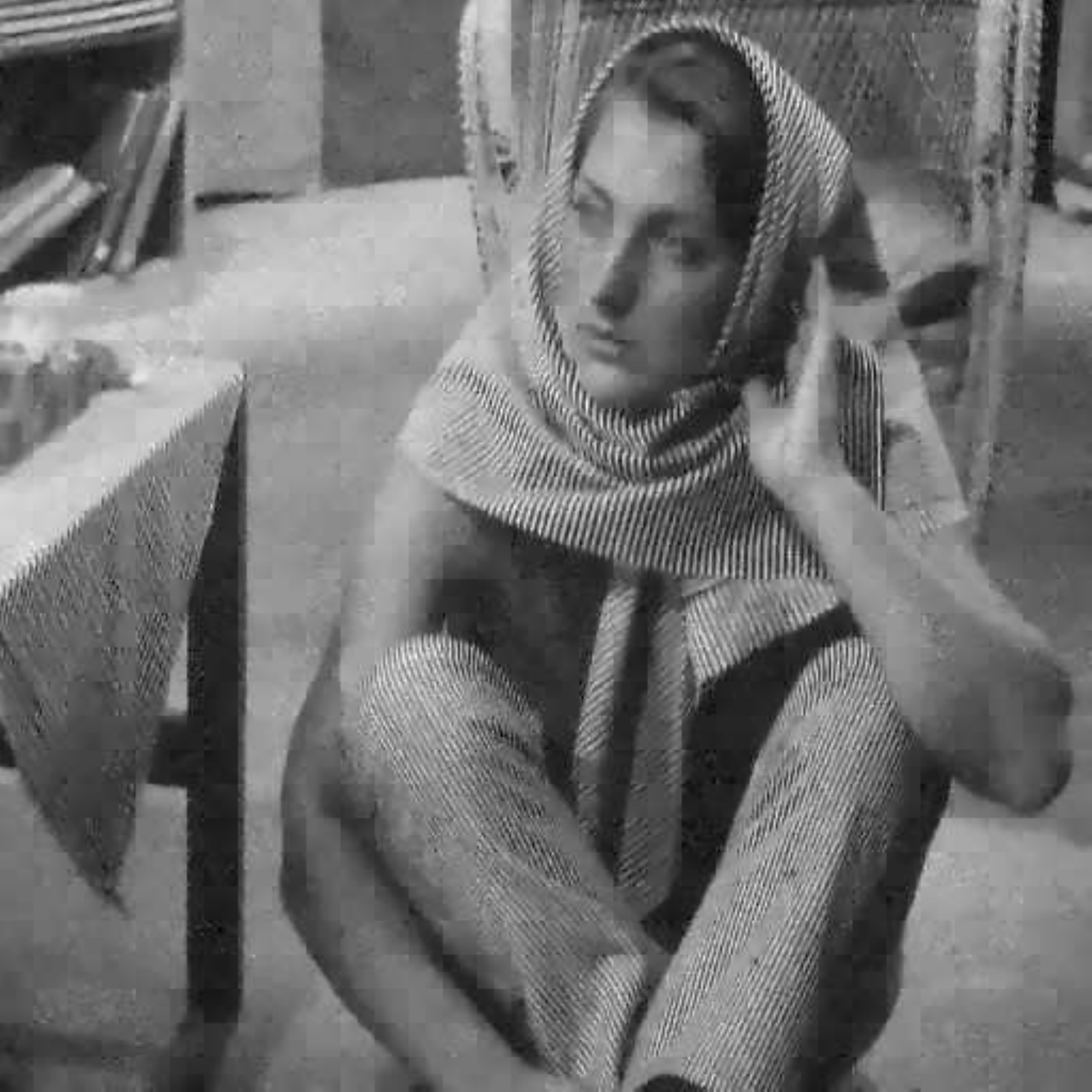}\\
\includegraphics[width=0.3\linewidth]{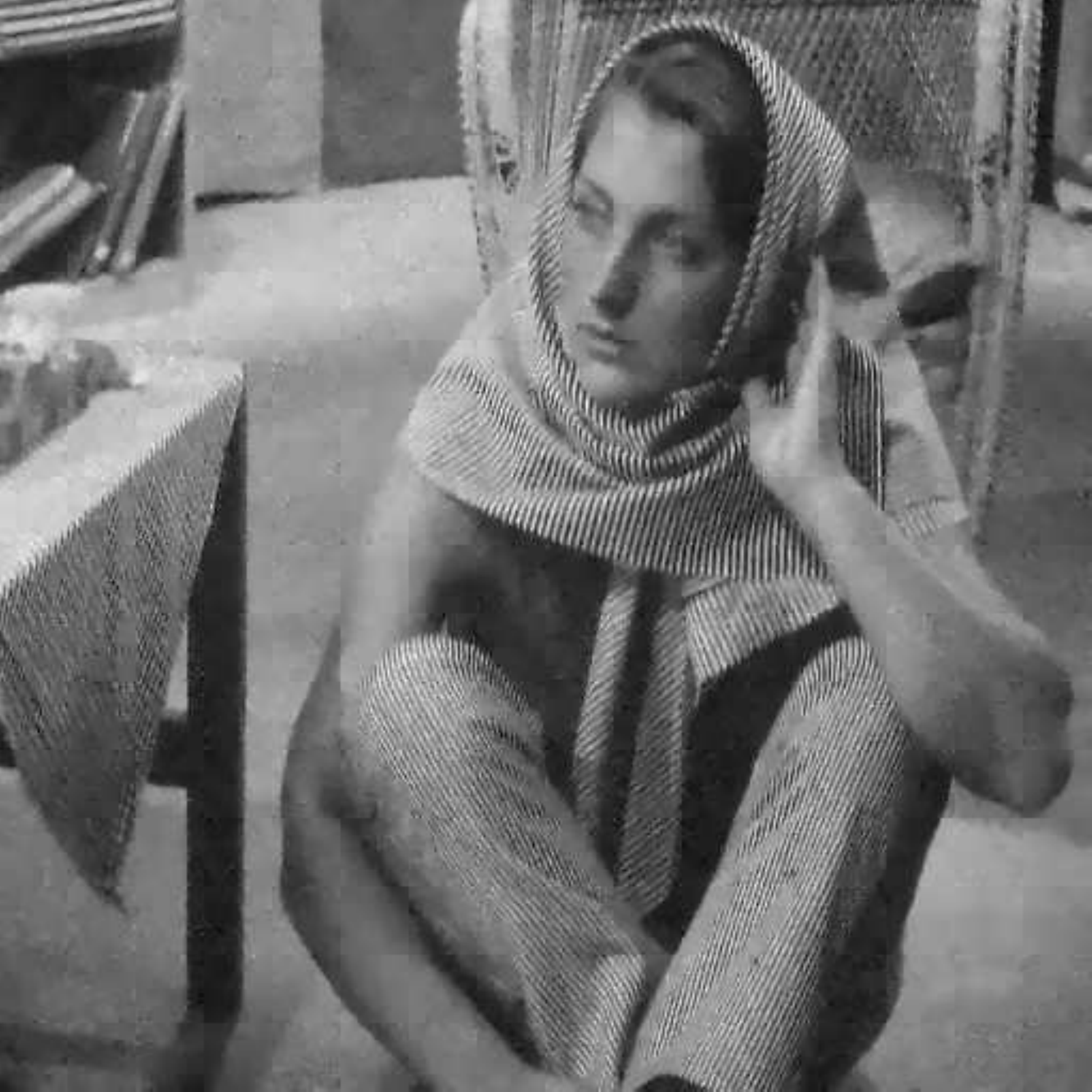}&\includegraphics[width=0.3\linewidth]{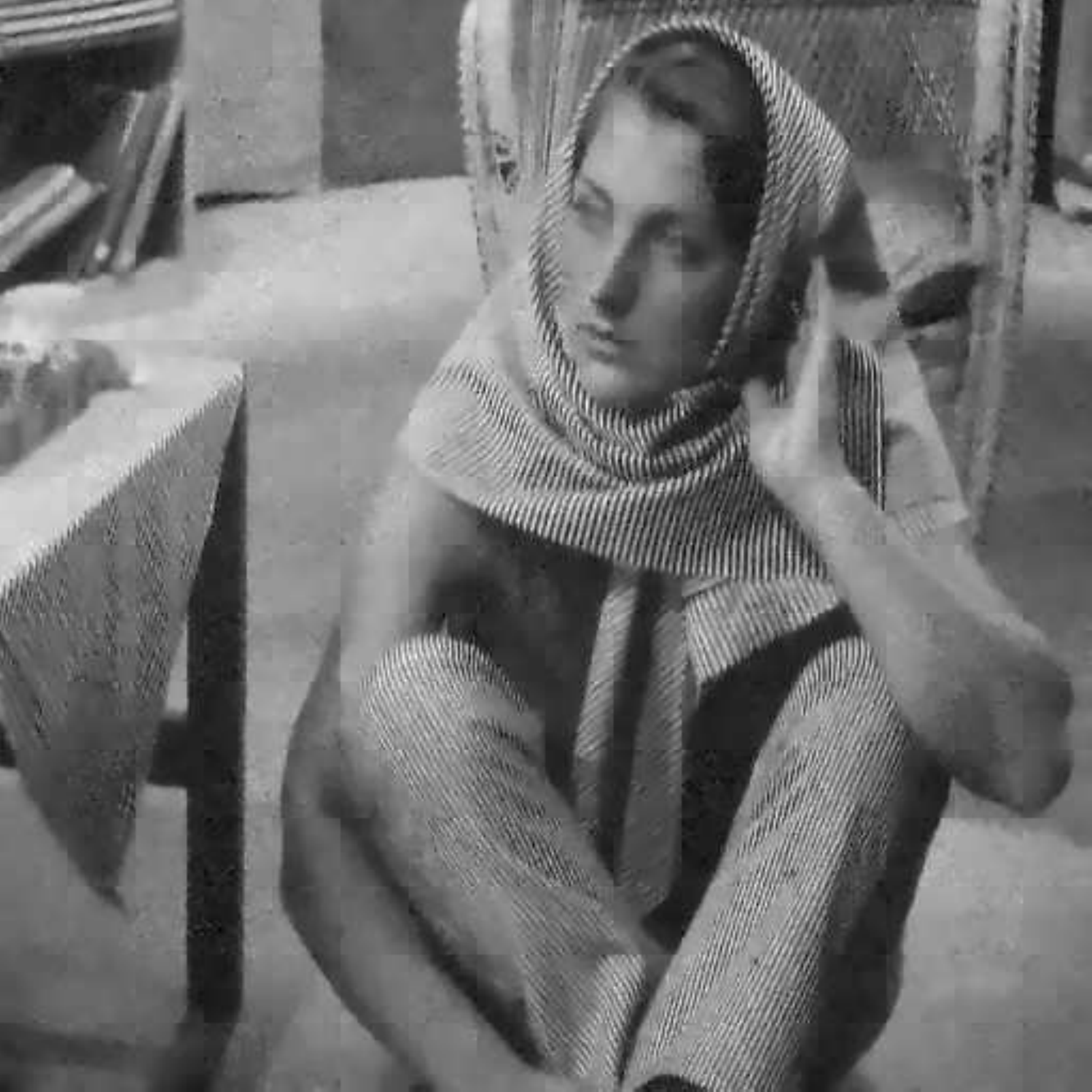}
\end{tabular}
\caption{Left: image denoised by choosing $\lambda$ with smallest estimated MSE; right: choosing  $\lambda$ randomly. Top: region size 16$\times$ 16; bottom region size 32$\times$ 32.}
\label{lenapatcheslam}
\end{center}
\end{figure}

We compare the estimated MSE with the true MSE for different parts of Lena image shown in Figure \ref{estmse16}. Since the estimated MSE values fluctuate around the true MSE values, it is possible that the estimated MSE values are negative when the true MSE values are low. The image parts are
shown in Figure  \ref{lenapatches1632}, which are of size 16$\times$16. The comparisons demonstrate that the estimated MSE are generally similar to the true ones.
To show the performance of estimated MSE, we consider the application to the choice of optimal $\lambda$. We test  the Barbara image and divide it into regions of size 16$\times$16 or of size 32$\times$32, then denoise the image for each region. The noise level $\sigma=20$ is utilized. We take into account the comparisons between  two cases:  choosing the optimal $\lambda$ from the set \{9, 12, 15, 18, 21, 24\} by estimated MSE, and choosing  $\lambda$ randomly in the same set. The results are displayed in Table \ref{nltv_rand_est}, which shows that choosing  $\lambda$ according to estimated MSE really works better than choosing  $\lambda$ at random.  The denoised Barbara images are displayed in Figure \ref{lenapatcheslam}. Since the regions are disjoint, the region boundaries are discernable in the images.

\begin{table}
\begin{center}
\caption{PSNR values with choosing $\lambda$ randomly (the first and third lines) and choosing $\lambda$ according to the estimated MSE (the second and the fourth lines ). The first two lines are with region size 16$\times$ 16; the last two lines are with region size 32$\times$ 32. }
\begin{tabular}{ccccccccccc} 
\hline
\noalign{\smallskip}
Image &Lena &Barbara& Peppers &Boats &Bridge & House & Cameraman \\
\noalign{\smallskip}
\hline
\noalign{\smallskip}
Random  &   30.49  & 27.73& 29.35  & 28.84  & 26.20 &  30.57 &  28.86 \\
Estimated MSE & 30.65 &  28.04  &    29.56 &  29.05 &  26.55 &   30.76&   29.11&\\
\noalign{\smallskip}
\hline
\noalign{\smallskip}
Random  &  30.78  &   27.90      &    29.72   &      29.02  &       26.27   &      31.07  &       29.14\\
Estimated MSE &30.98   &      28.24   &      29.88  &       29.24  &       26.67 &        31.10  &       29.31\\
\noalign{\smallskip}\hline
 \end{tabular}

\label{nltv_rand_est}
\end{center}
\end{table}

\section{Frequency domain nonlocal total variation (FNLTV) model}

The Fourier transform structures the image data in a totally different way from the standard basis.
For example, uniform areas generate low frequency values only and microtextures may be localized
in some narrow bands, whereas edges spread frequencies over a wide spectral band. In contrast to
these natural structures, Gaussian noise remains Gaussian noise. These remarks indicate that
denoising in the Fourier domain is likely to produce an effect complementary to that of  denoising
in the direct domain. As a result, additional denoising without artifact accumulation can be expected
by combining denoising in both domains.
Let us start by describing and analyzing the effect of denoising
applied in the Fourier domain only.

\subsection{The model}
For an image $u=\{u(i),i\in I\}$, denote its  discrete Fourier transform as $\hat{u}=\{\hat{u}(\omega),\omega \in I\}$, where we still use $I$ to represent the index set in the  frequency domain for simplicity and we use the discrete Fourier transform such that $\|\hat{u}\|_F=\|u\|_F$. In the frequency domain, the noise model becomes
\begin{equation}
  \hat{v}=\hat{u_0}+\hat{\eta}. \nonumber
\end{equation}
By the orthogonality of the discrete Fourier transform, $\hat{\eta}$ is also Gaussian noise with variance $\sigma^2$.
 Take into account the following energy functional called frequency domain nonlocal total variation (FNLTV) model  in the discrete form:
\begin{equation}
E(u)=\lambda_f\sum_{\omega \in I}|\nabla_{w_f}\hat{u}(\omega)|+\frac12\sum_{i \in I}(u(i)-v(i))^2,\label{eufnltv}
\end{equation}
where
\begin{equation}
|\nabla_{w_f}\hat{u}(\omega)|=\sqrt{\sum_{\xi \in I}|\hat{u}(\omega)-\hat{u}(\xi)|^2w_f(\omega,\xi)}. \nonumber
\end{equation}
The weight  $w_f(\omega,\xi)$ is taken as ($\ref{nltvweight}$) with $\hat{v}$ replacing $v$, and the parameters $d_f, D_f, \sigma_{rf}$ replacing $d, D, \sigma_r$ respectively. That is,
  \begin{equation}
w_f(\omega,\xi) = \left\{\begin{array} {ll}
e^{- ||\hat{v}(\mathcal{N}_{\omega})-\hat{v}(\mathcal{N}_{\xi} )||_a^{2}/ (2\sigma_{rf}^2)} &  \mbox{if} \quad {\xi}\in U_{\omega}(D_f) \\
0 & \mbox{else} \end{array}\right. , \label{nltvweightf}
\end{equation}
where
\begin{equation}
     ||\hat{v}(\mathcal{N}_{\omega})-\hat{v}(\mathcal{N}_{{\xi}})||_a^2 =\frac{ \sum_{k\in \mathcal{N}_{\omega}(d_f)} a(\omega,k) |\hat{v}(k)-\hat{v}(\mathcal{T}(k))|^2}{\sum_{k\in \mathcal{N}_{\omega}(d_f)} a(\omega,k)}. \label{nlmdisnormf}
\end{equation}
Since the Fourier transform is linear, the functional (\ref{eufnltv}) is still strictly convex. Thus we can also use the gradient descent algorithm to find the minimizer.

Let $J_f(u)=\sum_{\omega \in I}|\nabla_{w_f}\hat{u}(\omega)|$ be the frequency domain regularizer.
We will show that the gradient of $J_f(u)$ can be expressed as follows:
\begin{equation}
\nabla J_f(u)=\Re {f}, \mbox{with} \ \hat{f}(w)=\sum_{\xi \in I}(\hat{u}(\omega)-\hat{u}(\xi))\left(\frac{w_f(\omega,\xi)}{|\nabla_{{{w_f}}}\hat{u}(\omega)|}+\frac{w_f(\xi,\omega)}{|\nabla_{{{w_f}}}\hat{u}(\xi)|}\right), \label{gradfft}
\end{equation}
 where $\Re$ represents the real part of some complex number. 
In practice, as NLTV, $|\nabla_{{{w_f}}}\hat{u}(\omega)|$ is replaced by $\sqrt{|\nabla_{{{w_f}}}\hat{u}(\omega)|^2+\beta}$ with $\beta$ a small positive constant to avoid the division by 0.
It is easy to see that
\begin{eqnarray}
\frac d{dt}J_f(u+tv)|_{t=0}
&=&
\sum_{\omega \in I}\sum_{\xi \in I}[(\hat{u}(\omega)-\hat{u}(\xi))(\bar{\hat{v}}(\omega)-\bar{\hat{v}}(\xi))\nonumber \\
&+&
(\hat{v}(\omega)-\hat{v}(\xi))(\bar{\hat{u}}(\omega)-\bar{\hat{u}}(\xi))]w_f(\omega,\xi)\frac1{2|\nabla_{{{w_f}}}\hat{u}(\omega)|}. \label{granltvfft}
\end{eqnarray}
With a similar proof to ($\ref{eulernltv}$), we obtain
\begin{equation}
\sum_{\omega \in I}\sum_{\xi \in I}(\hat{u}(\omega)-\hat{u}(\xi))(\bar{\hat{v}}(\omega)-\bar{\hat{v}}(\xi))w_f(\omega,\xi)\frac1{2|\nabla_{{{w_f}}}\hat{u}(\omega)|}=\sum_{\omega \in I}\frac12\hat{f}(\omega)\bar{\hat{v}}(\omega). \label{granltvfft1}
\end{equation}
Then by ($\ref{granltvfft}$) and ($\ref{granltvfft1}$),
it follows that
\begin{eqnarray}
 \frac d{dt}J_f(u+tv)|_{t=0}&=&\sum_{\omega \in I}\frac12\hat{f}(\omega)\bar{\hat{v}}(\omega)+\overline{\sum_{\omega \in I}\frac12\hat{f}(\omega)
\bar{\hat{v}}(\omega)} \nonumber \\
&=&\sum_{i \in I}\frac12{f}(i)\bar{v}(i)+\sum_{i \in I}\frac12\bar{{f}}(i)\bar{v}(i)\nonumber\\
&=&\sum_{i \in I}\Re{f}(i)v(i). \nonumber 
\end{eqnarray}
Thus, we have the conclusion ($\ref{gradfft}$).

\begin{figure}
\begin{center}
\center 
\renewcommand{\arraystretch}{0.5} \addtolength{\tabcolsep}{0pt} \vskip3mm %
\fontsize{8pt}{\baselineskip}\selectfont
\begin{tabular}{ccc}
\includegraphics[width=0.30\linewidth]{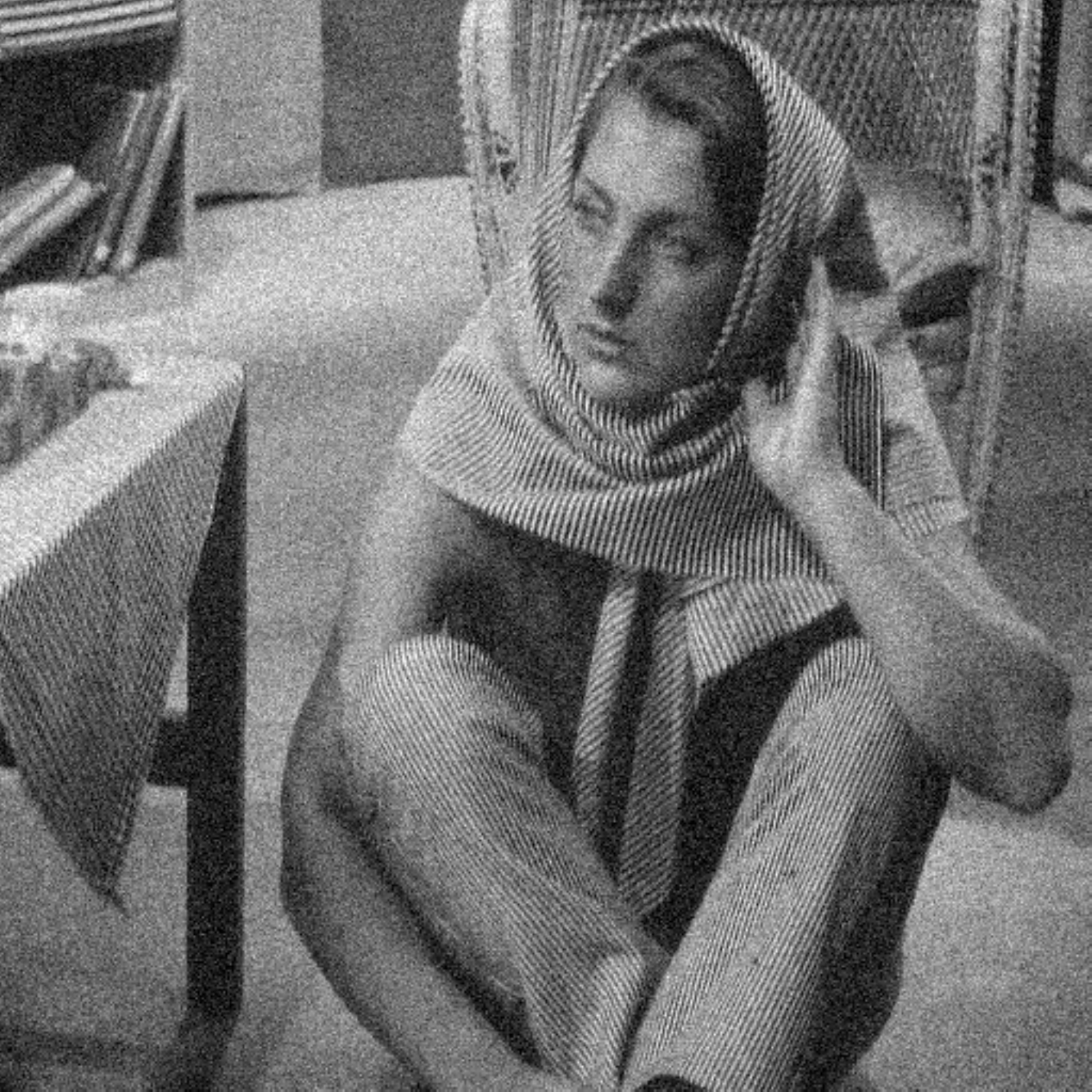}
&\includegraphics[width=0.30\linewidth]{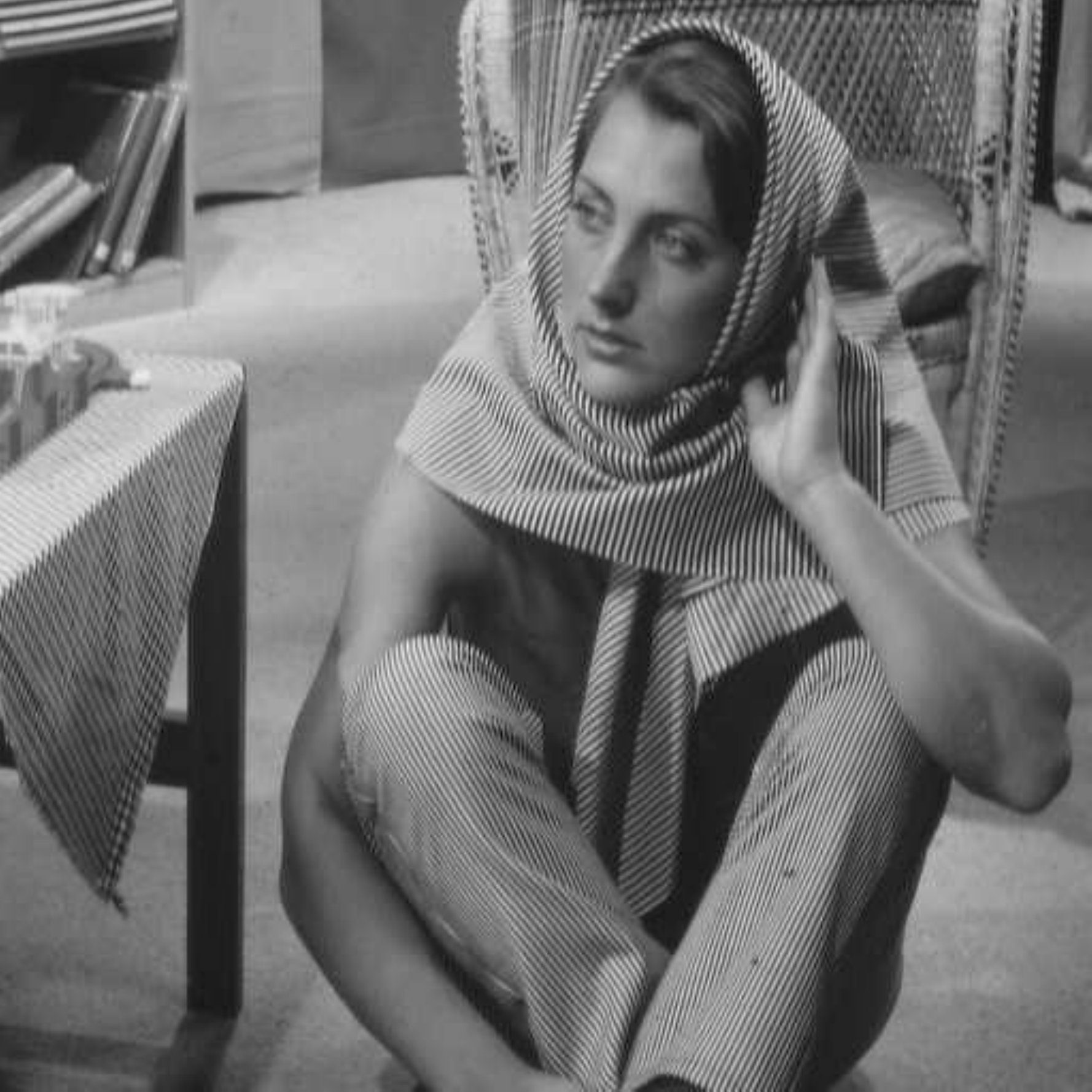}
&\includegraphics[width=0.3\linewidth]{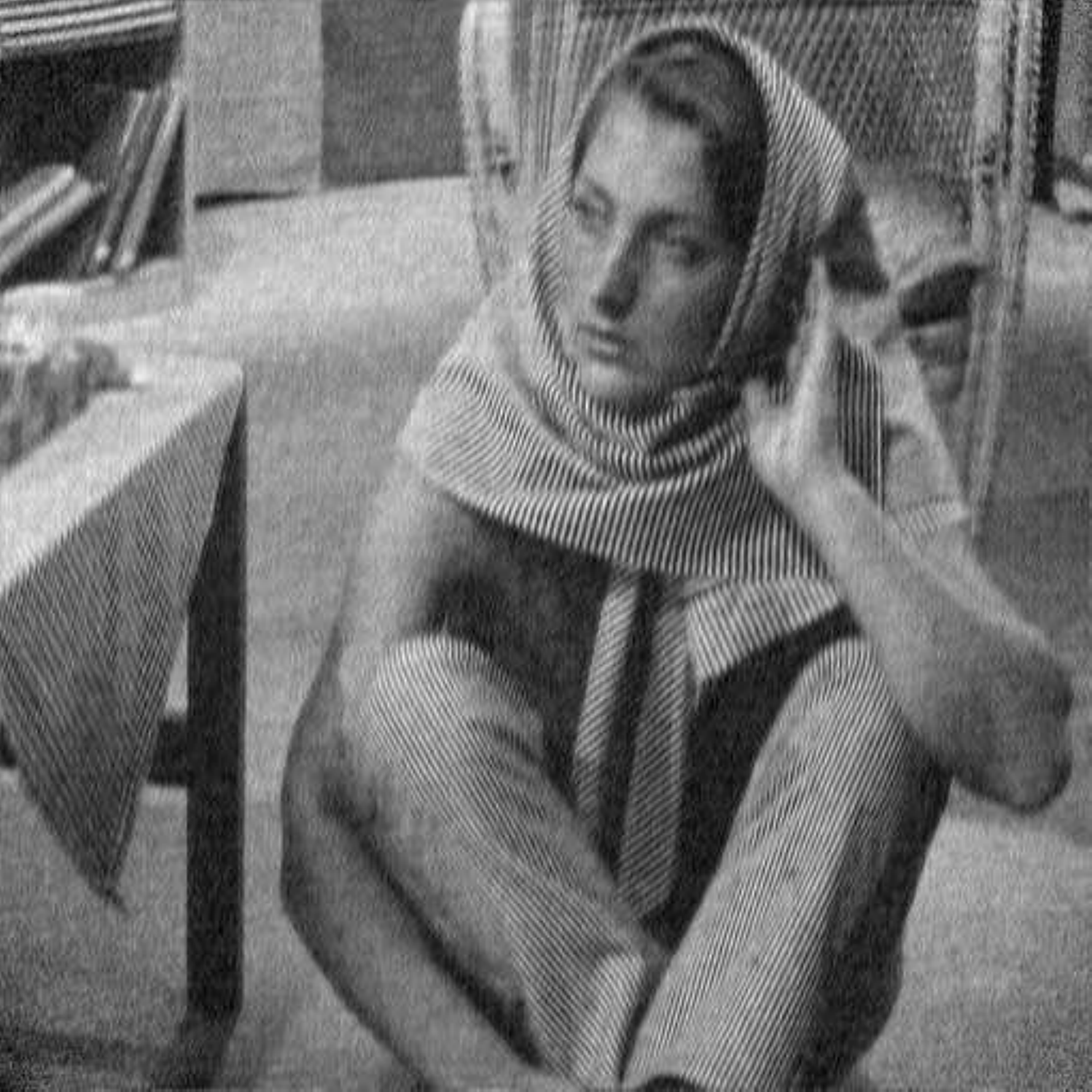}\\
Noisy &Original (noise-free)& FNLTV\\
\includegraphics[width=0.3\linewidth]{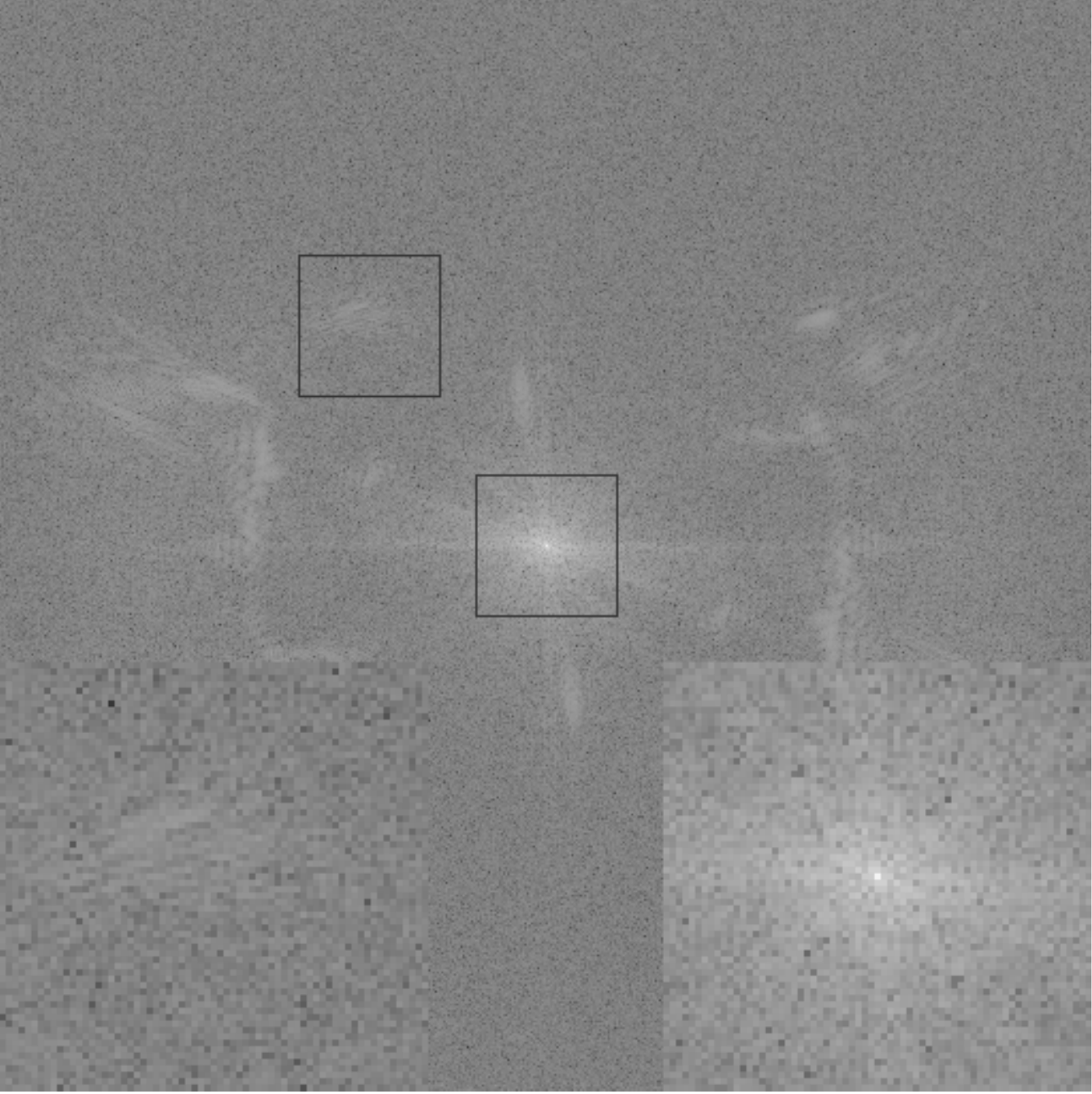}
&\includegraphics[width=0.3\linewidth]{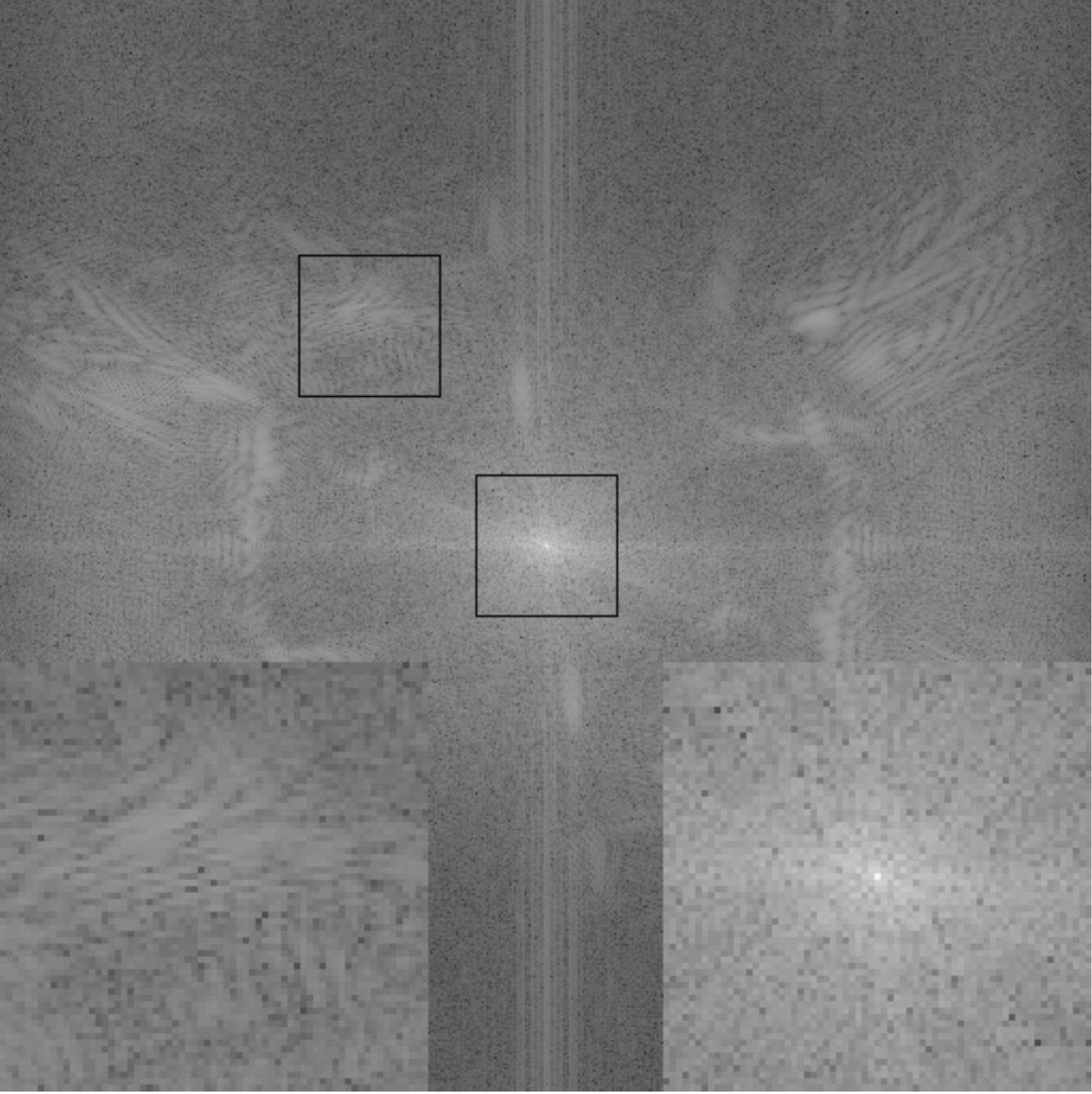}
&\includegraphics[width=0.3\linewidth]{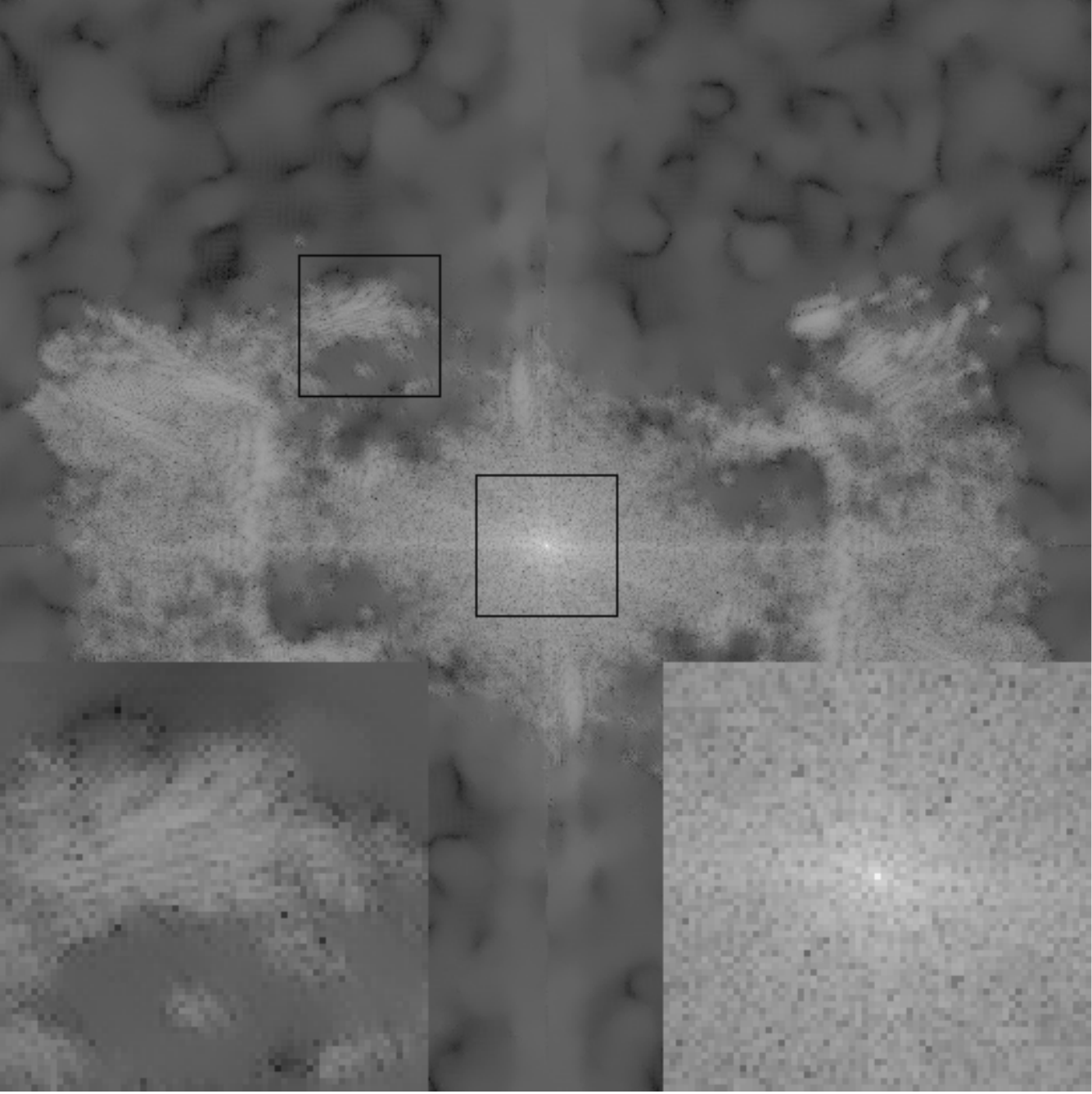}
\end{tabular}
\caption{Top: noisy image, original image and denoised image by FNLTV from left to right; Bottom: the Fourier transform of the top images, where the right one can also be considered as  denoised Fourier transform. }
\label{fft_fig}
\end{center}
\end{figure}

Note that applying FNLTV is equivalent to denoise the discrete Fourier transform  of the noisy image and then apply the inverse Fourier transform. For the structures with high values, e.g. low frequencies, since the values are very large, the weights are very small. Thus the low frequencies and other structures with high values can be hardly influenced.
In  Figure \ref{fft_fig}, the Fourier transforms of the original image, noisy image, and the denoised Fourier transform are shown. We can see that  FNLTV has really removed Gaussian noise in the Fourier domain, and the low frequencies and structures with high values stay unchanged.

\subsection{Parameters Analysis by Comparing NLTV and FNLTV} \label{secfnltvpar}
\begin{figure}
\begin{center}
\center 
\renewcommand{\arraystretch}{0.5} \addtolength{\tabcolsep}{0pt} \vskip3mm %
\fontsize{8pt}{\baselineskip}\selectfont
\begin{tabular}{ccc}
\includegraphics[width=0.3\linewidth]{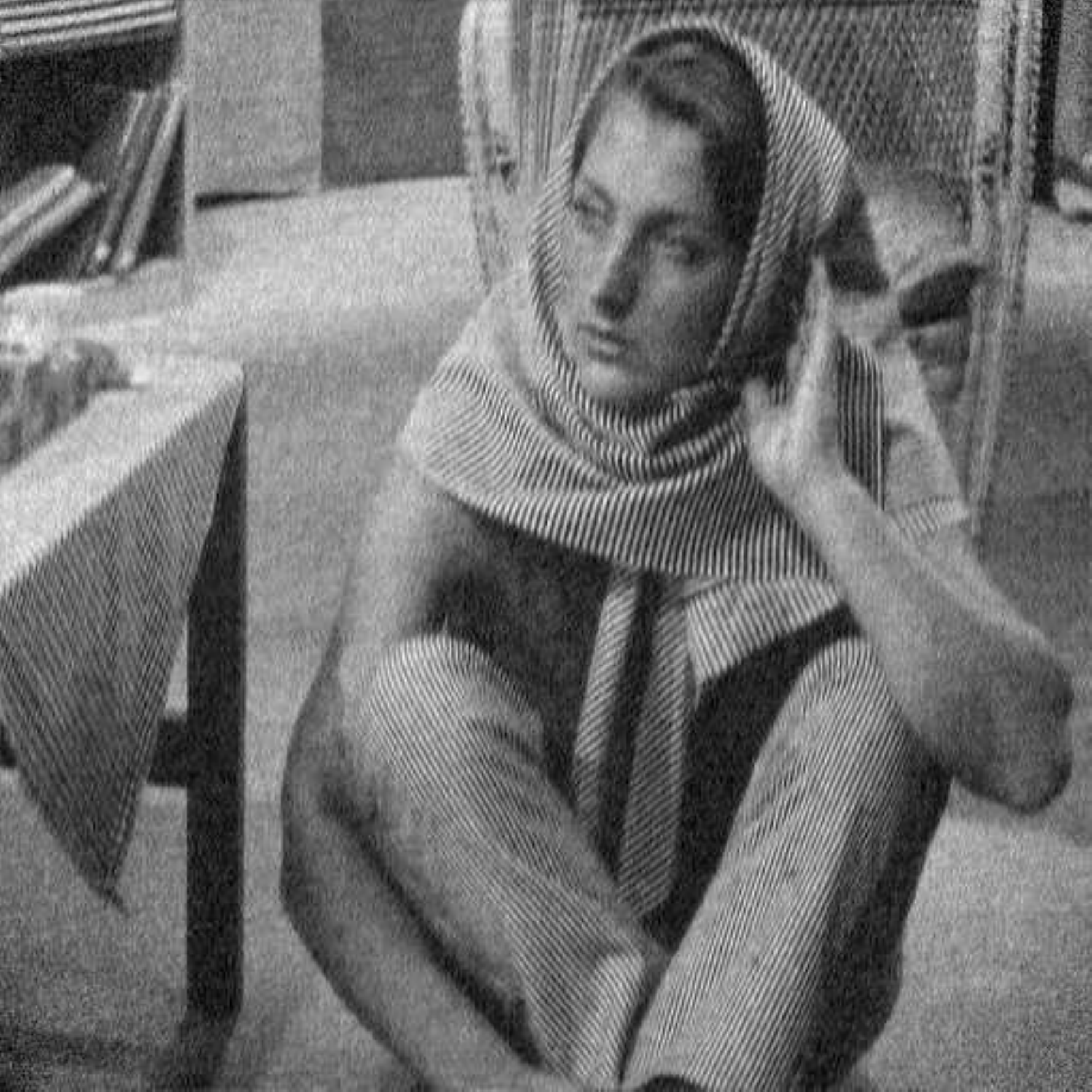}
&\includegraphics[width=0.3\linewidth]{f_nltv_lamf16_27_6455.pdf}
&\includegraphics[width=0.3\linewidth]{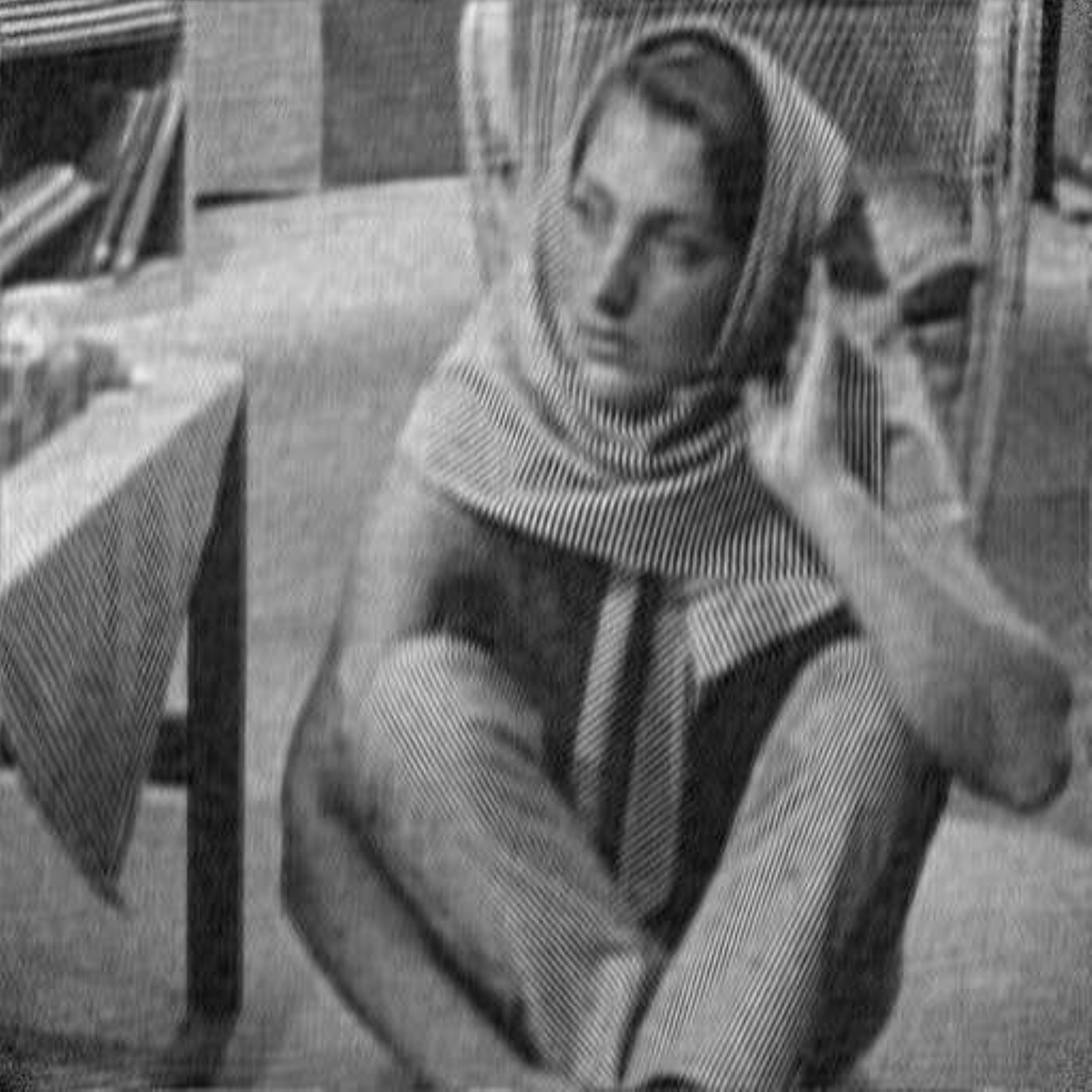}\\
$\lambda_f=10,$ PSNR=26.87
&$\lambda_f=16,$ PSNR=27.64
&$\lambda_f=50, $ PSNR=27.04\\
\includegraphics[width=0.3\linewidth]{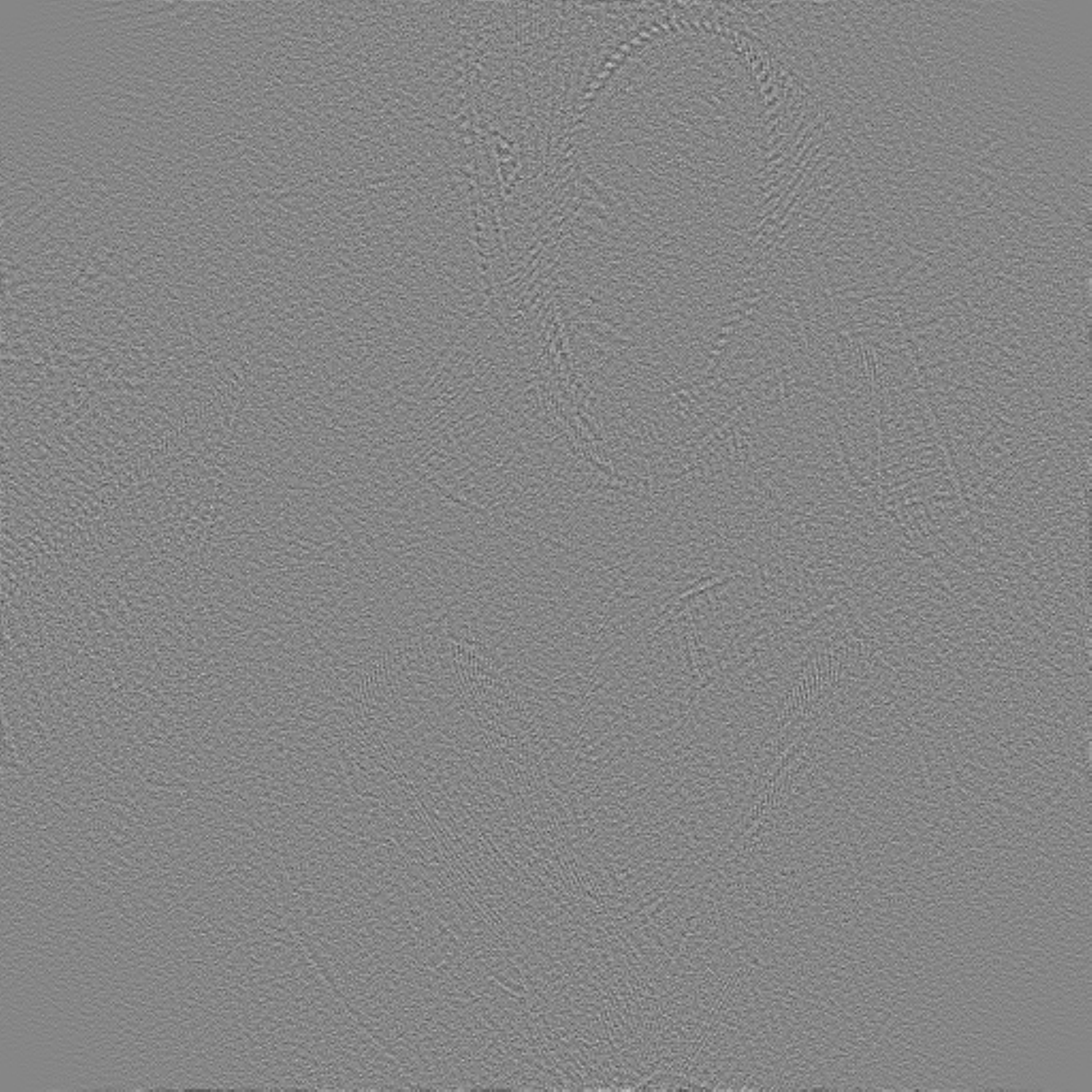}
&\includegraphics[width=0.3\linewidth]{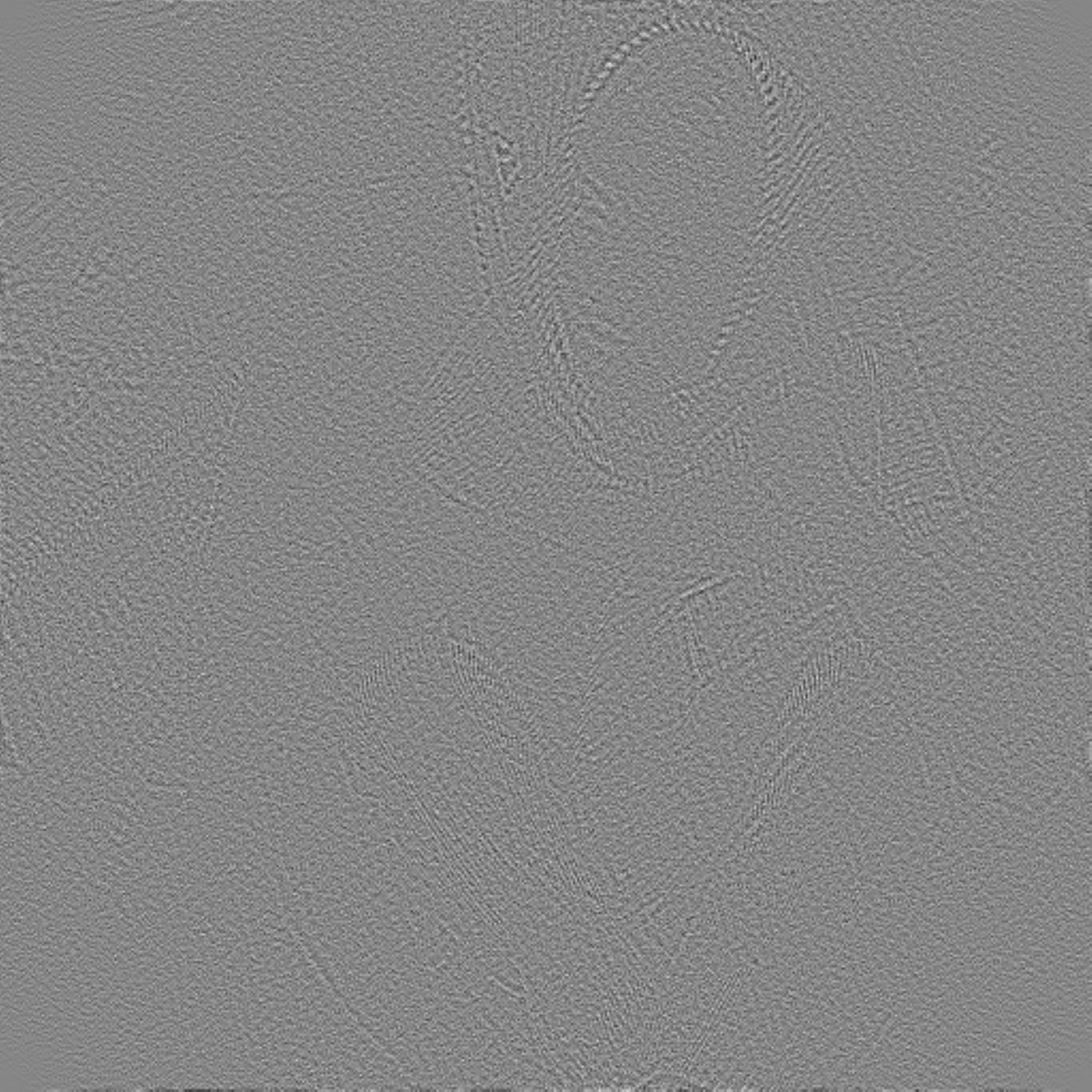}
&\includegraphics[width=0.3\linewidth]{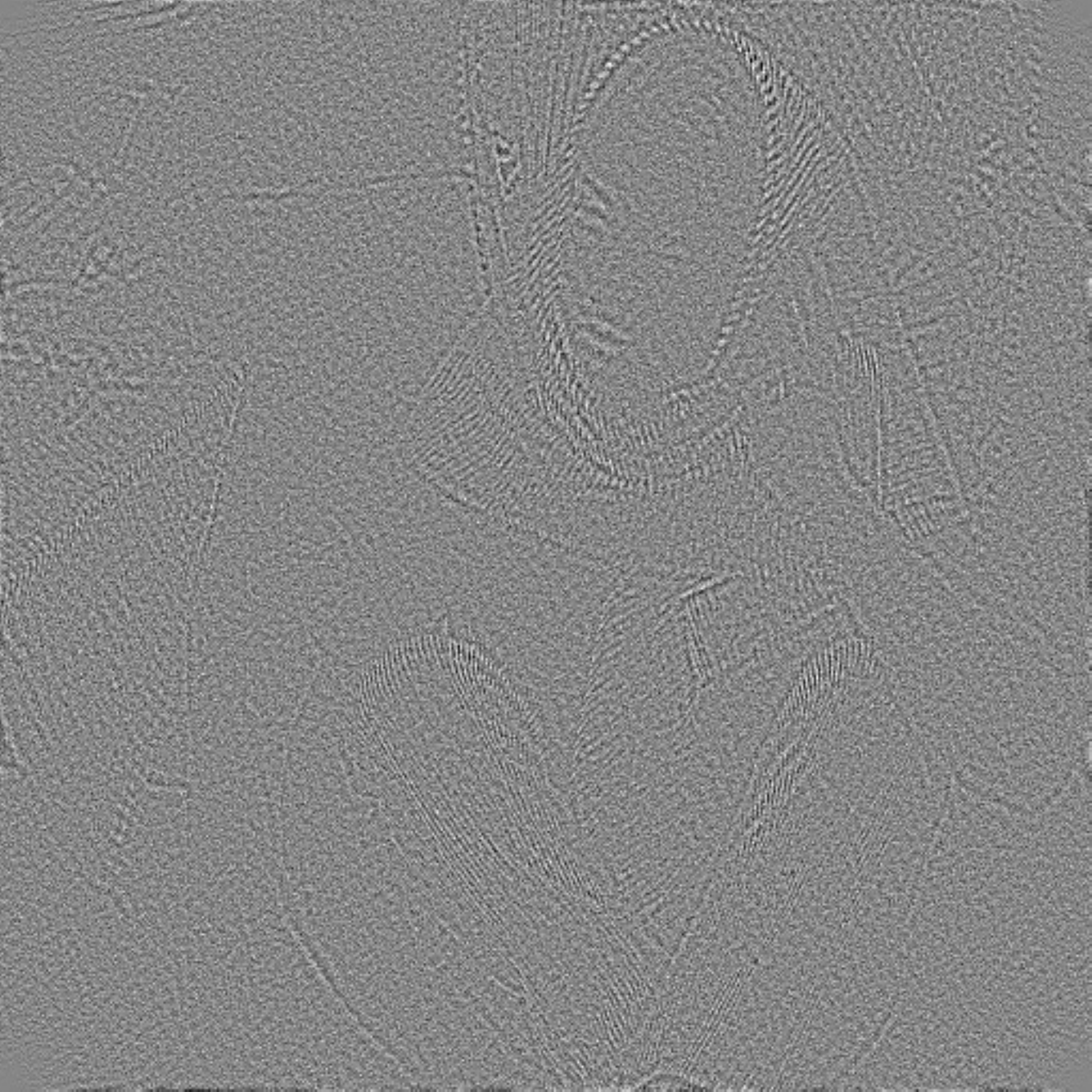}\\
$\lambda_f=10,$ Method noise
&$\lambda_f=16,$ Method noise
&$\lambda_f=50, $ Method noise\\
\includegraphics[width=0.3\linewidth]{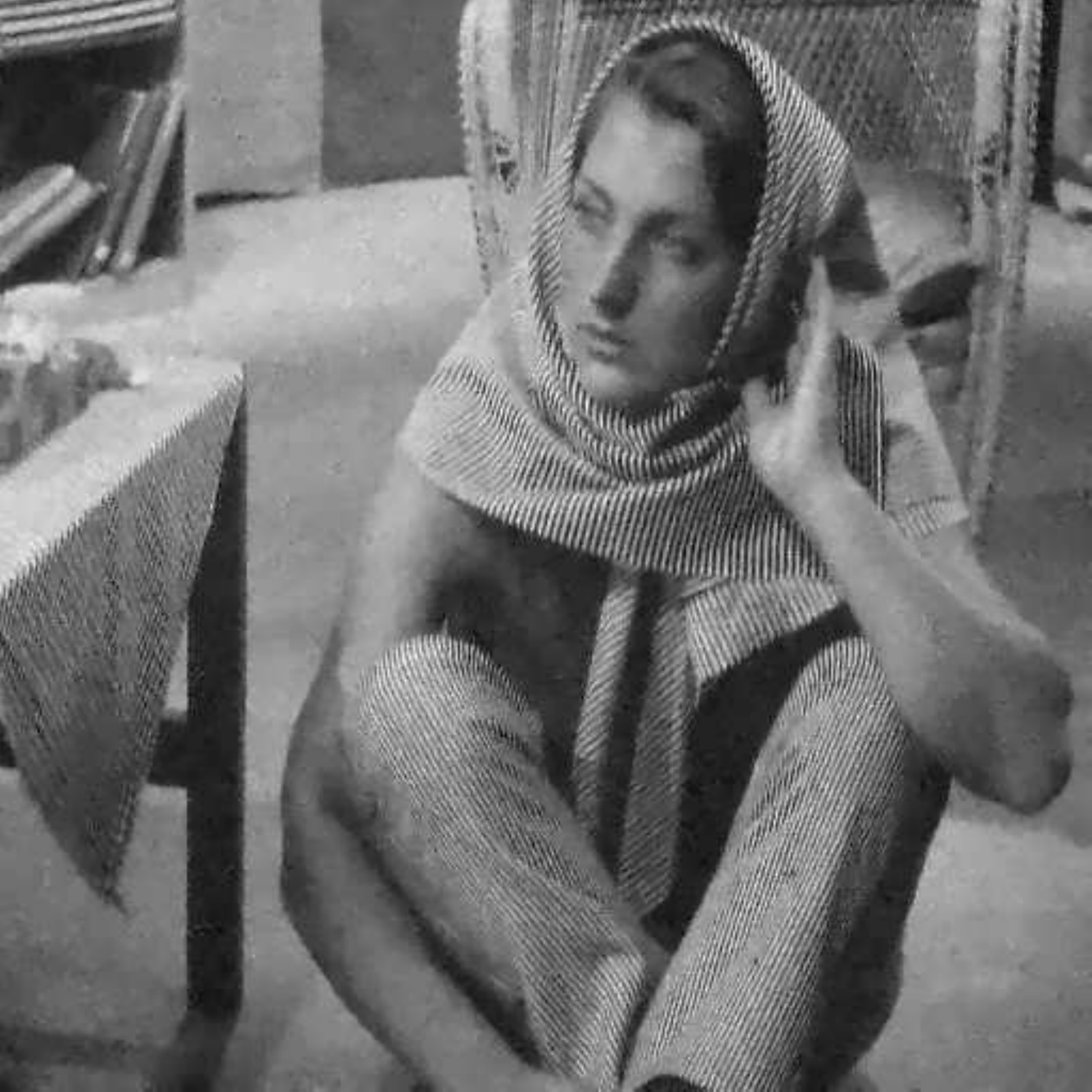}
&\includegraphics[width=0.3\linewidth]{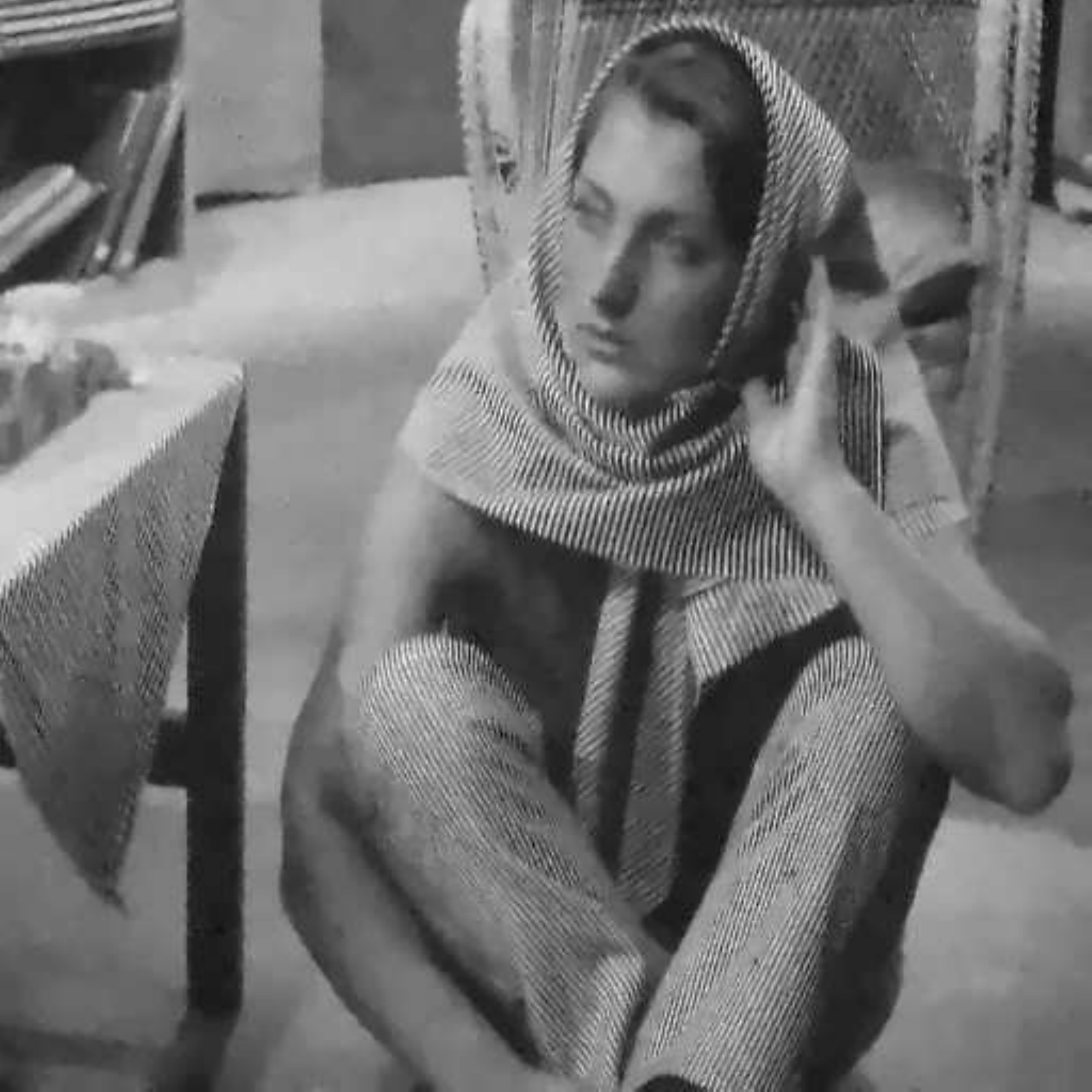}
&\includegraphics[width=0.3\linewidth]{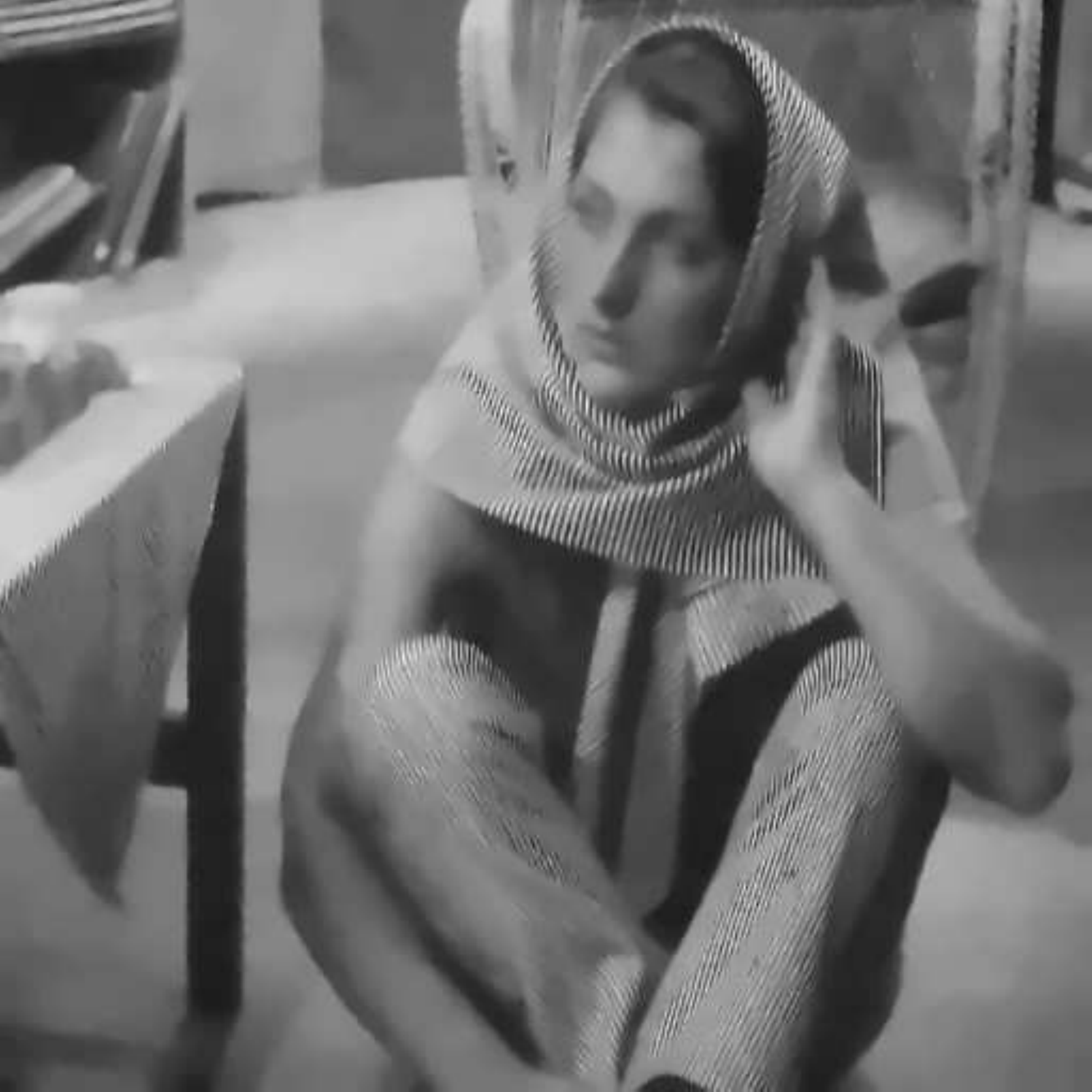}\\
$\lambda=10,$ PSNR=28.21
&$\lambda=16,$ PSNR=28.40
&$\lambda=50,$ PSNR=26.10\\
\includegraphics[width=0.3\linewidth]{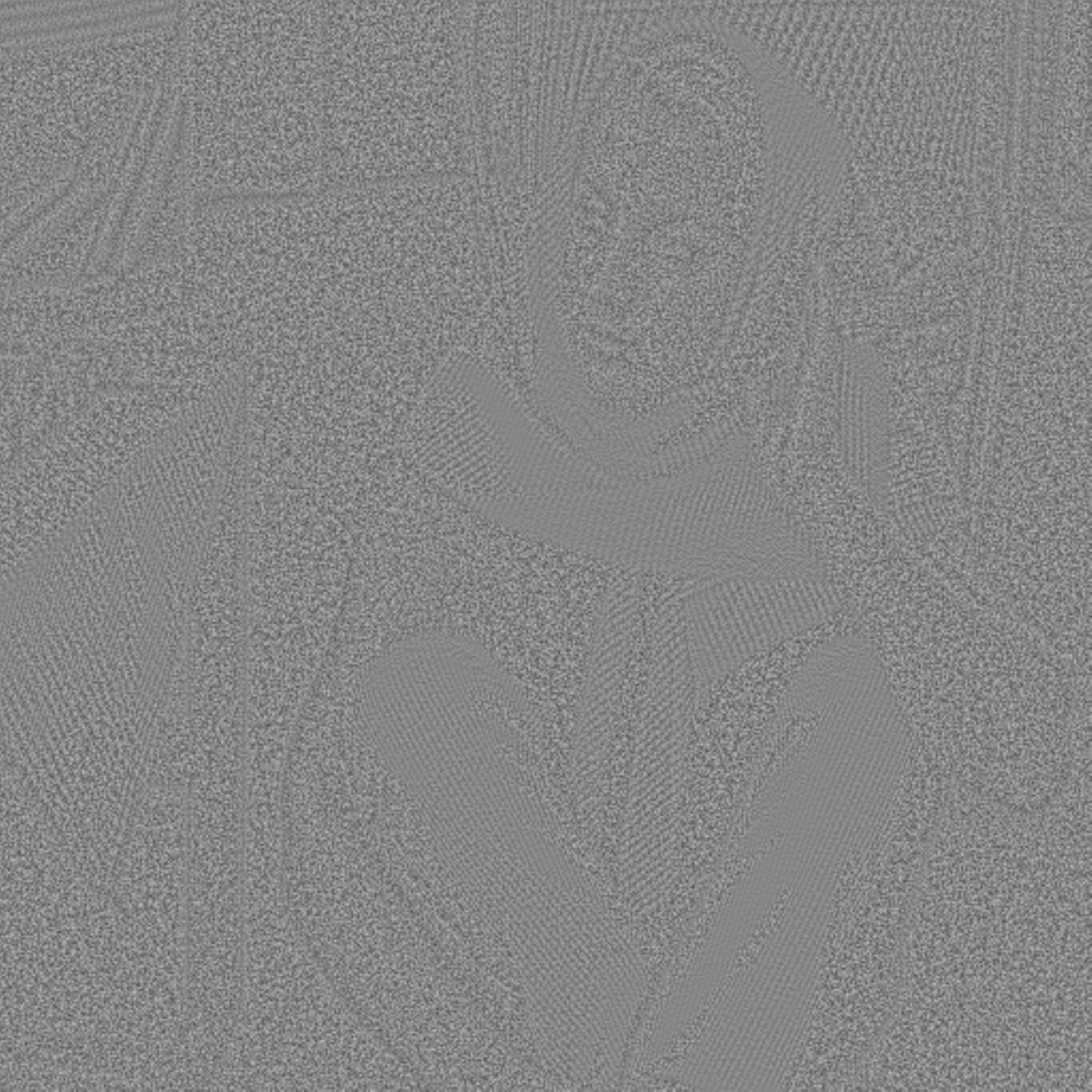}
&\includegraphics[width=0.3\linewidth]{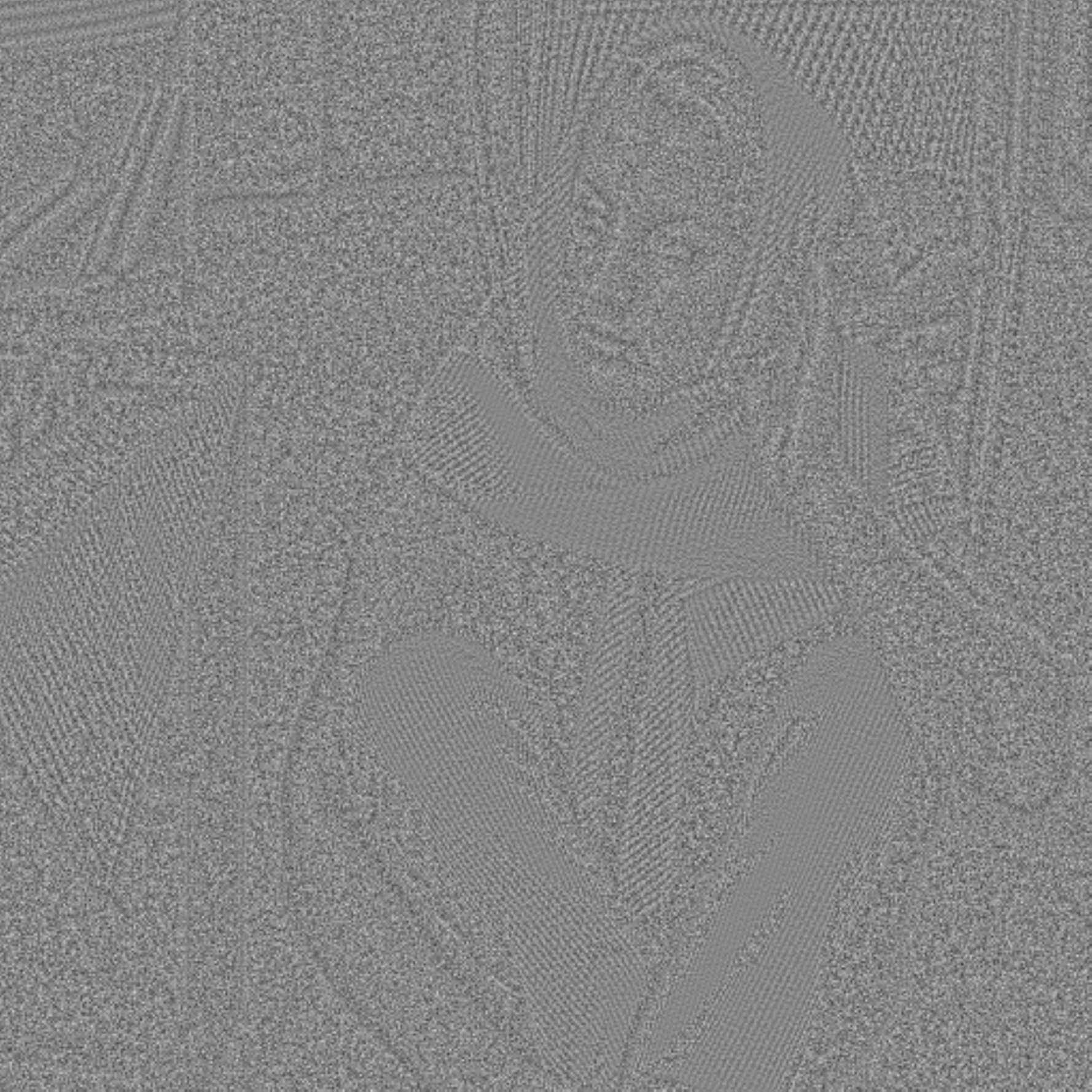}
&\includegraphics[width=0.3\linewidth]{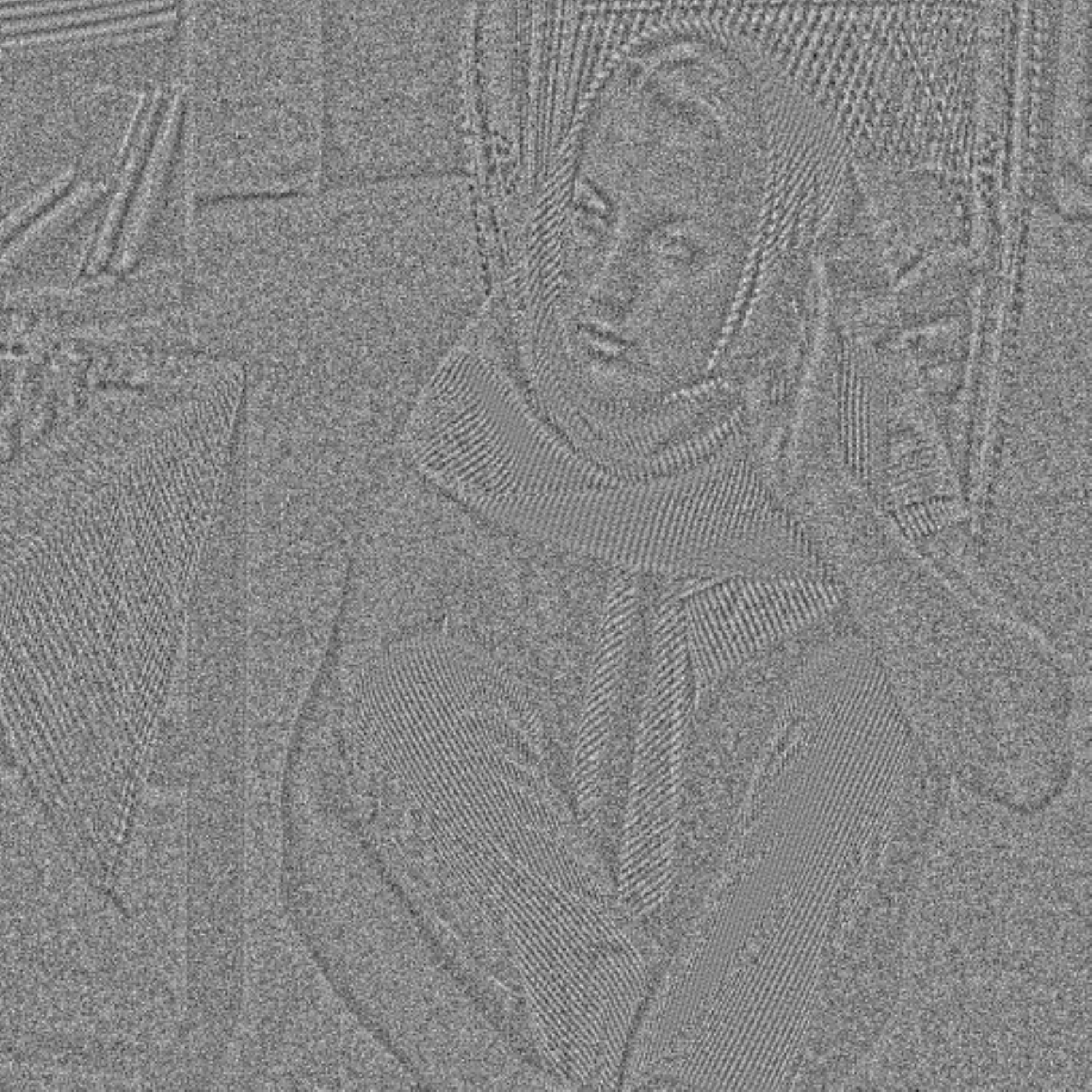}\\
$\lambda=10,$ Method noise
&$\lambda=16,$ Method noise
&$\lambda=50, $ Method noise\\
\end{tabular}
\caption{Top row: images denoised by FNLTV with different $\lambda_f$; the third row: images denoised by NLTV with different $\lambda$; the second row and bottom row: the corresponding method noise images of the top row and the third row. }
\label{lena_nltv_fnltv_lam}
\end{center}
\end{figure}

\begin{figure}
\begin{center}
\begin{tabular}{ccc}
\includegraphics[width=0.8\linewidth]{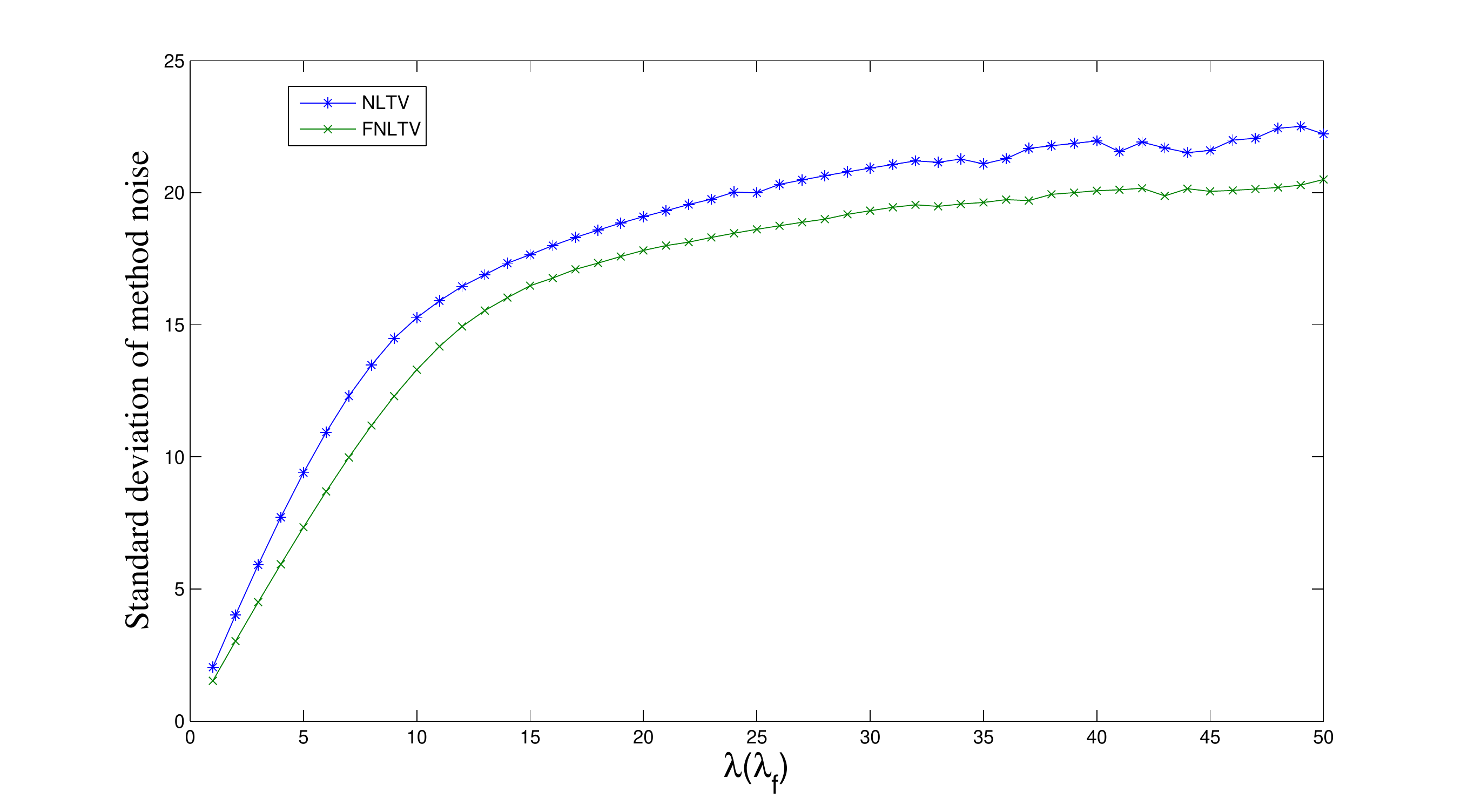}\\
\includegraphics[width=0.8\linewidth]{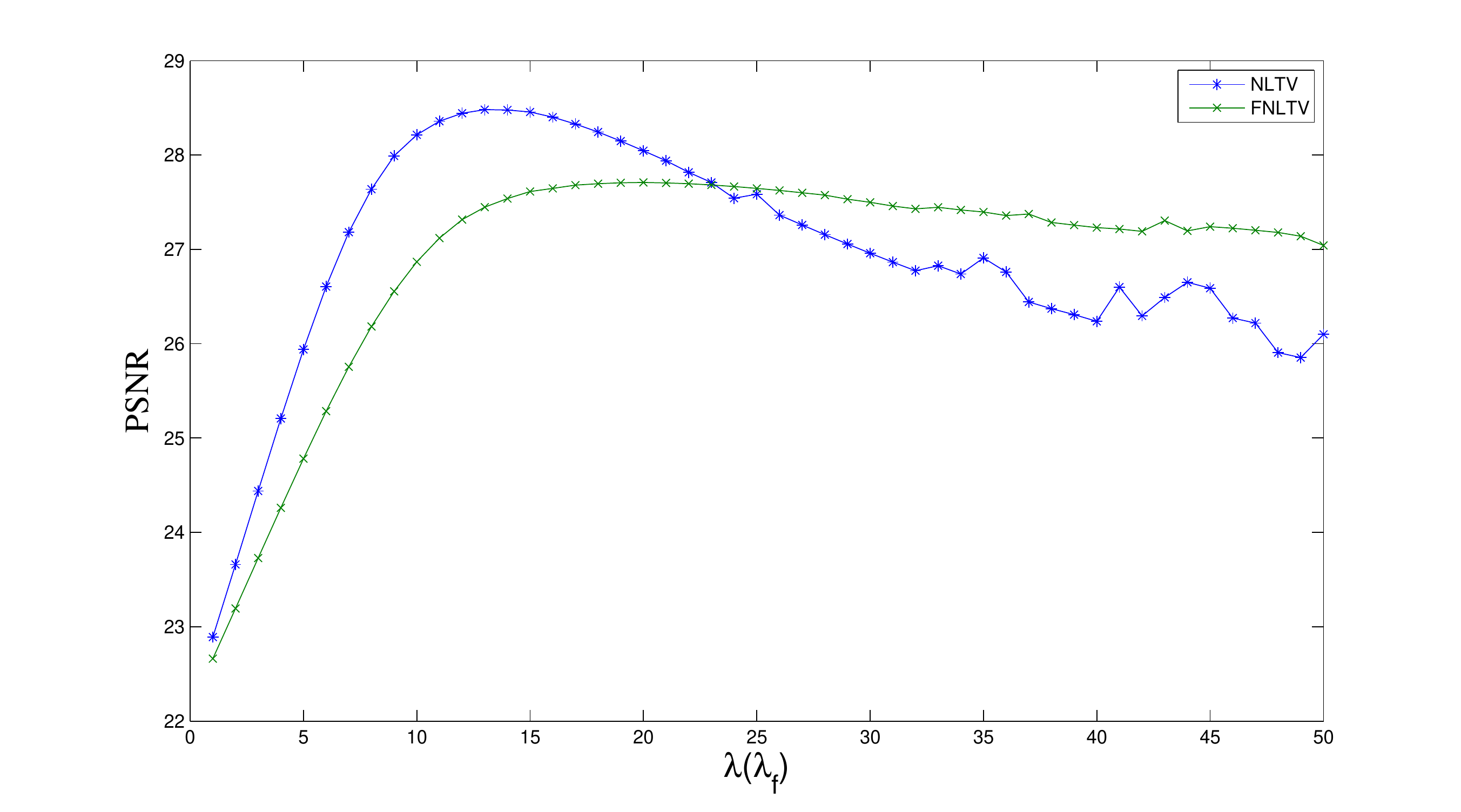}
\end{tabular}
\caption{Top: variances of method noise versus different $\lambda$ (for NLTV) or $\lambda_f$ (for FNLTV); Bottom: PSNR values versus different $\lambda$ (for NLTV)  or $\lambda_f$ (for FNLTV). }
\label{lena_nltv_fnltv_lam_curve}
\end{center}
\end{figure}

We first analyze the choice of regularization parameters by comparing the denoising performance and method noise (the difference of noisy image and denoised image $v-\bar{v}$) of NLTV and FNLTV. The regularization parameters  $\lambda$  and $\lambda_f$ are both taken as 10, 16, 50.
The visual results are shown in Figure \ref{lena_nltv_fnltv_lam}, where the six method noise images are scaled in the same way to show their differences.
The comparisons indicate that FNLTV model removes less image detail than NLTV model. Even when $\lambda_f$ is large, i.e. $\lambda_f=50$, the detail loss is not very evident. Note that when $\lambda$ is large,  much  image details are lost for NLTV. In Figure  \ref{lena_nltv_fnltv_lam_curve}, we  plot the variances of method noise and PSNR values\footnote{which is defined in Section \ref{sec_sim}.}  for more values of $\lambda$ and $\lambda_f$: $\lambda, \lambda_f=1,2,\cdots,50$. From the figures we can observe that when the regularization parameters $\lambda$ or $\lambda_f$ increase, the removing amount of noise are similar for two models, while the PSNR values vary less for FNLTV model than for NLTV model, especially when the regularization parameters are large.  Therefore,  the choice of the regularization parameter $\lambda_f$ is not very sensitive. Similar conclusions can be obtained for the search window size $D_f$.

\begin{figure}
\begin{center}
\center 
\renewcommand{\arraystretch}{0.5} \addtolength{\tabcolsep}{0pt} \vskip3mm %
\fontsize{8pt}{\baselineskip}\selectfont
\begin{tabular}{ccc}
\includegraphics[width=0.3\linewidth]{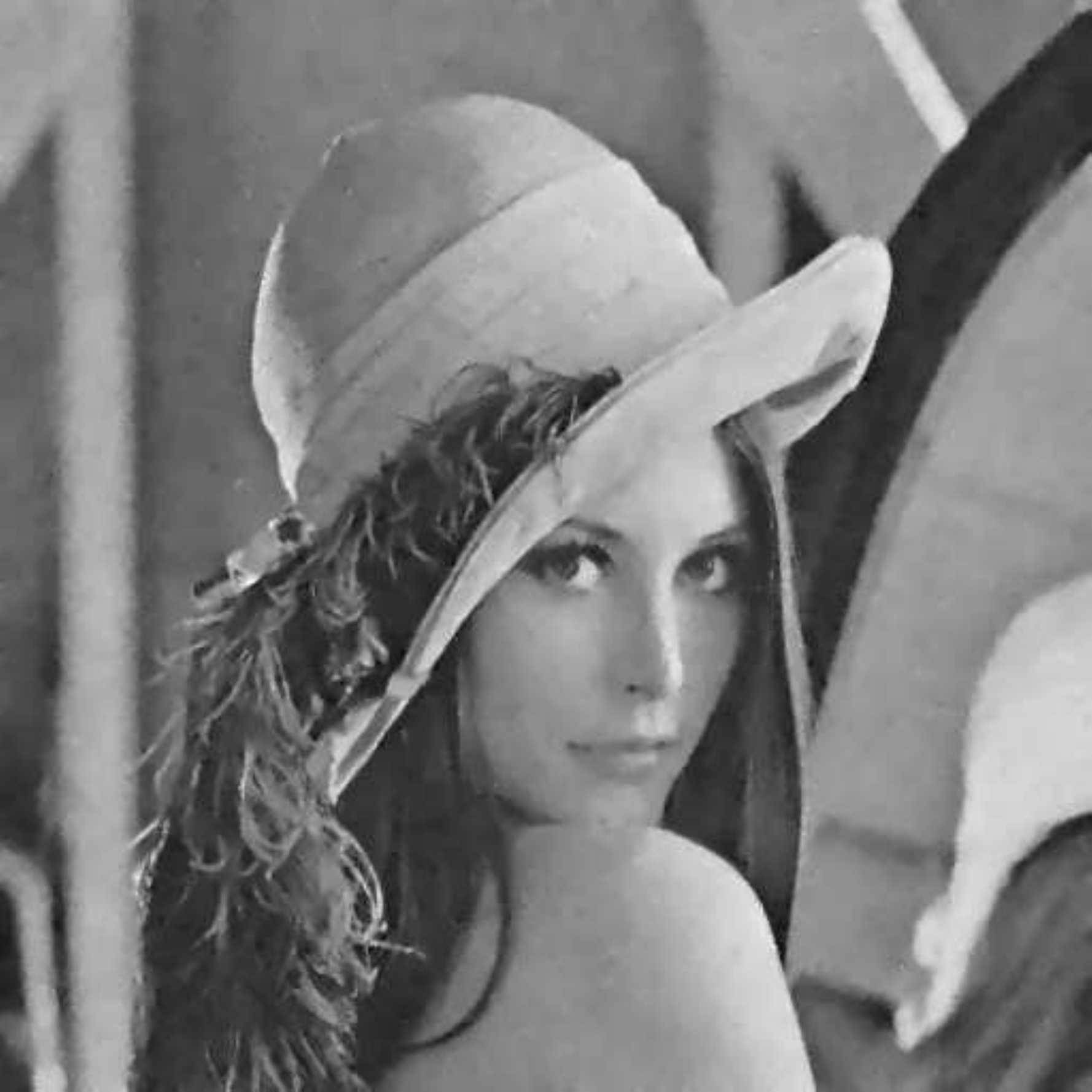}
&\includegraphics[width=0.3\linewidth]{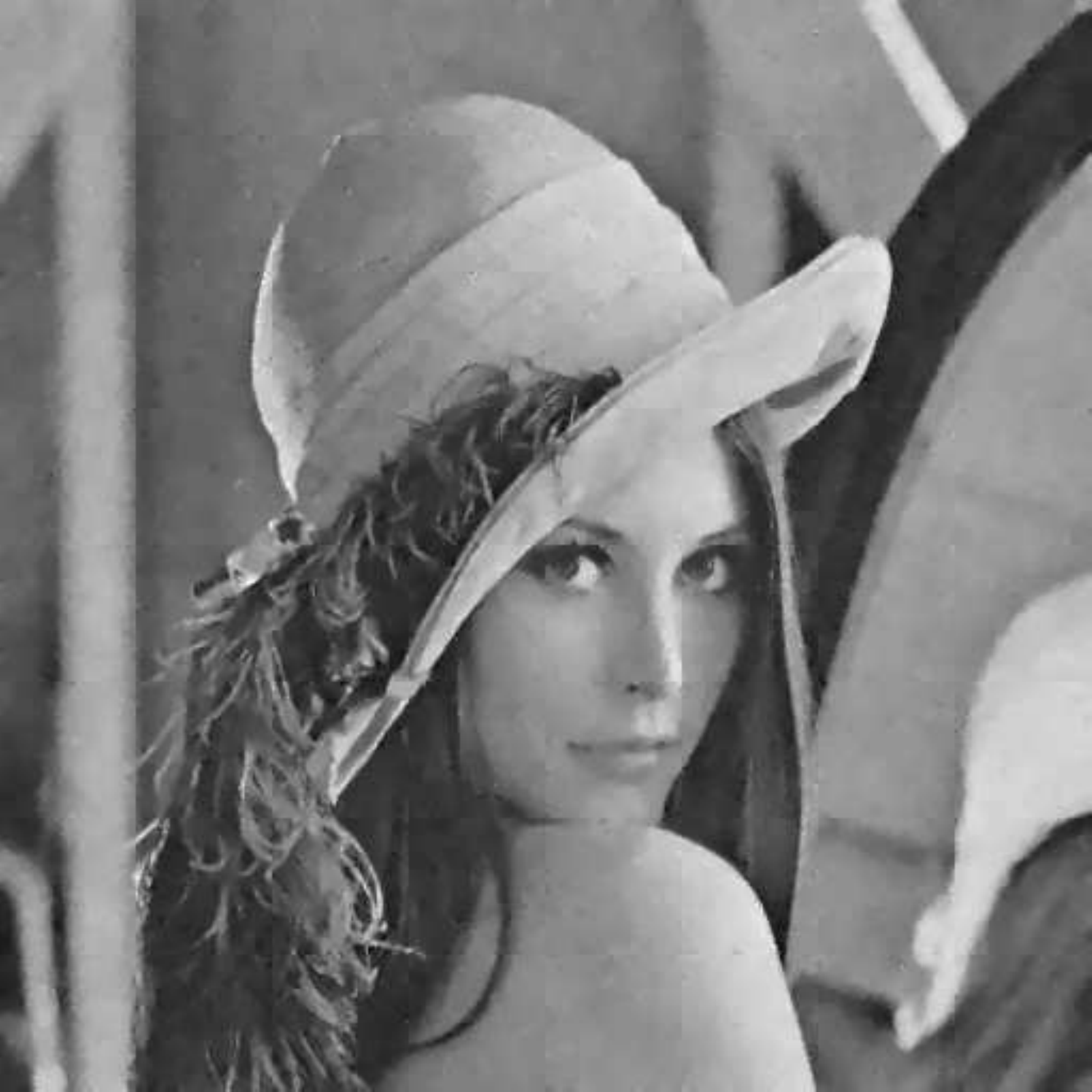}
&\includegraphics[width=0.3\linewidth]{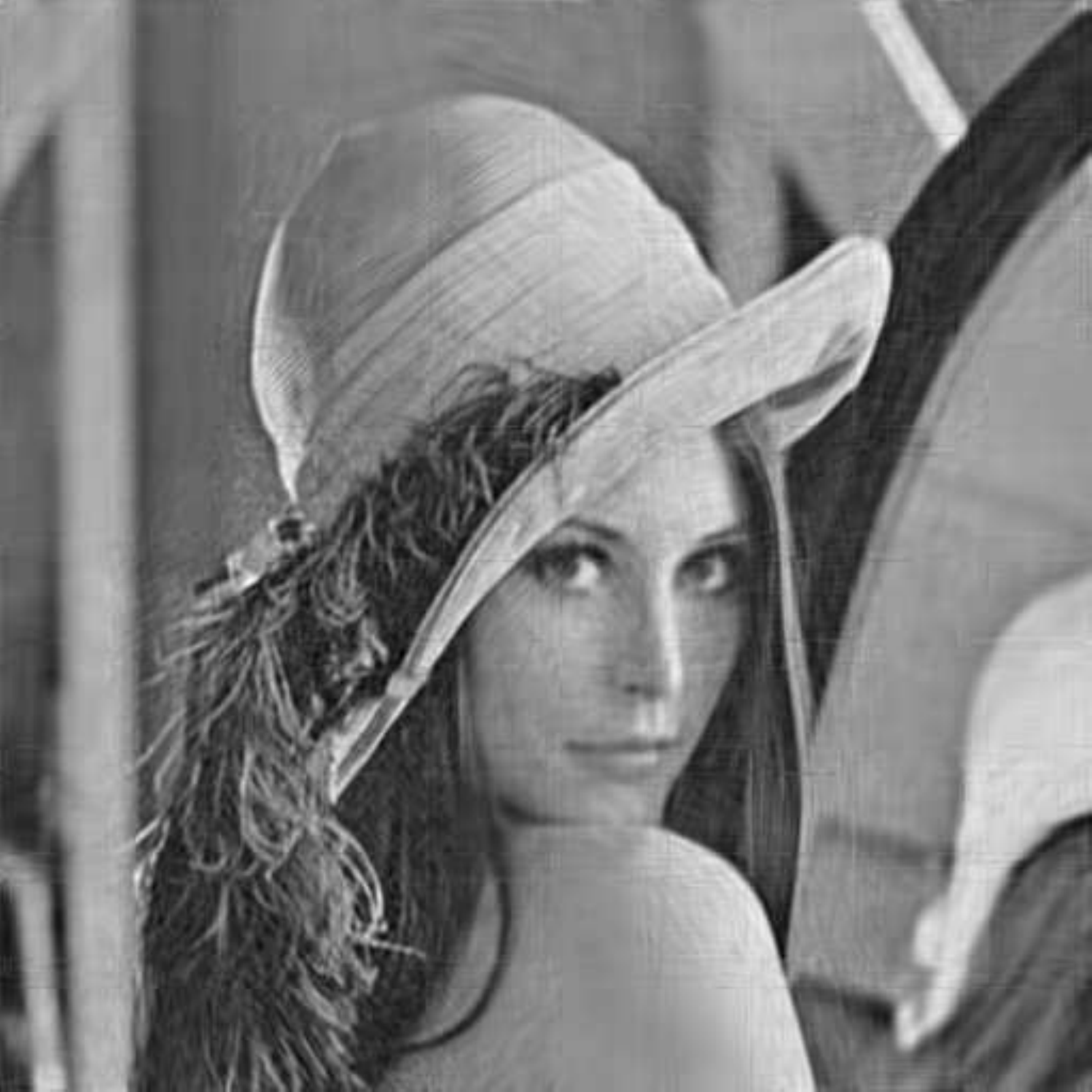}\\
NLTV &NLTV $S_r=64, n_s=64$ &FNLTV $ S_r=64, n_s=50$\\
\includegraphics[width=0.3\linewidth]{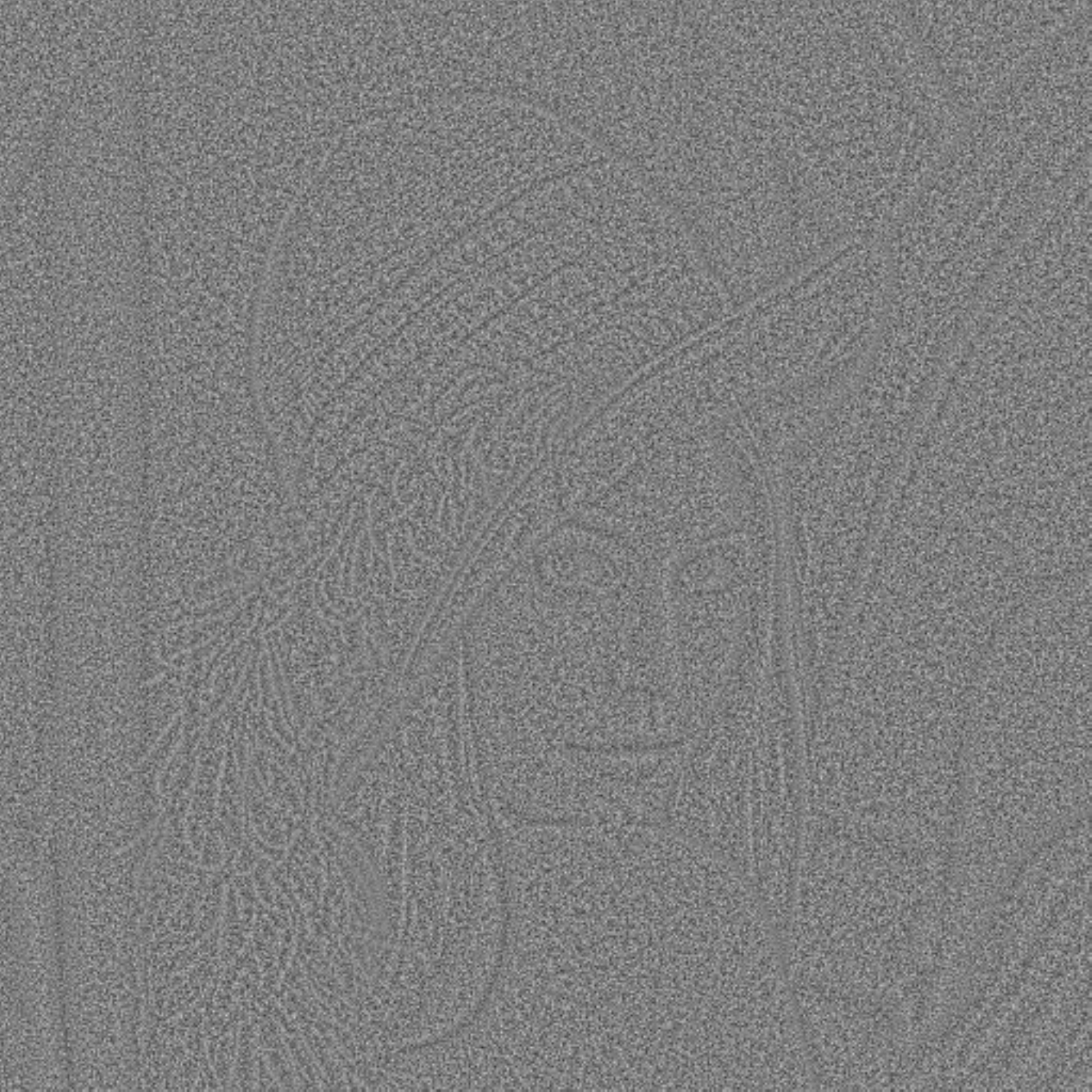}
&\includegraphics[width=0.3\linewidth]{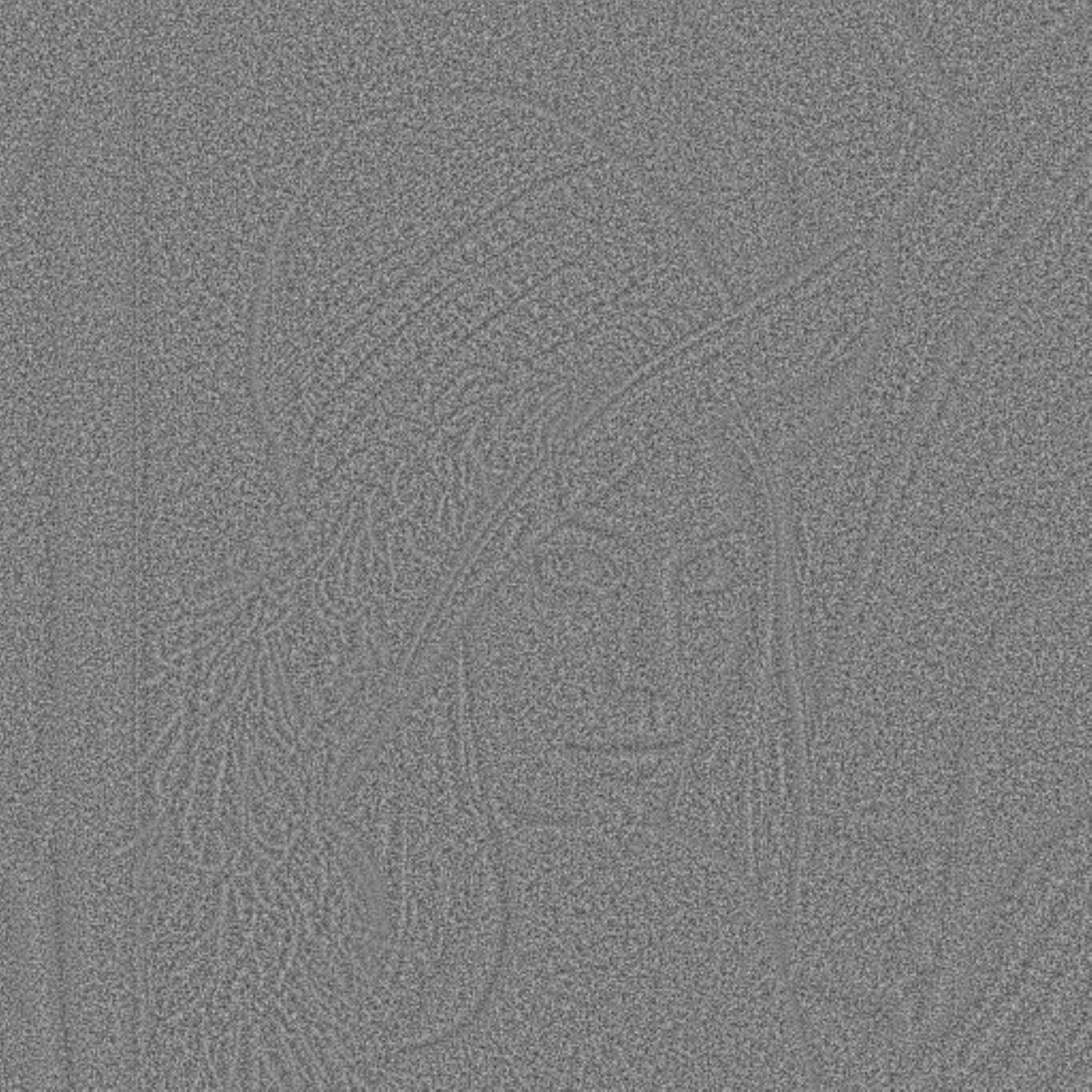}
&\includegraphics[width=0.3\linewidth]{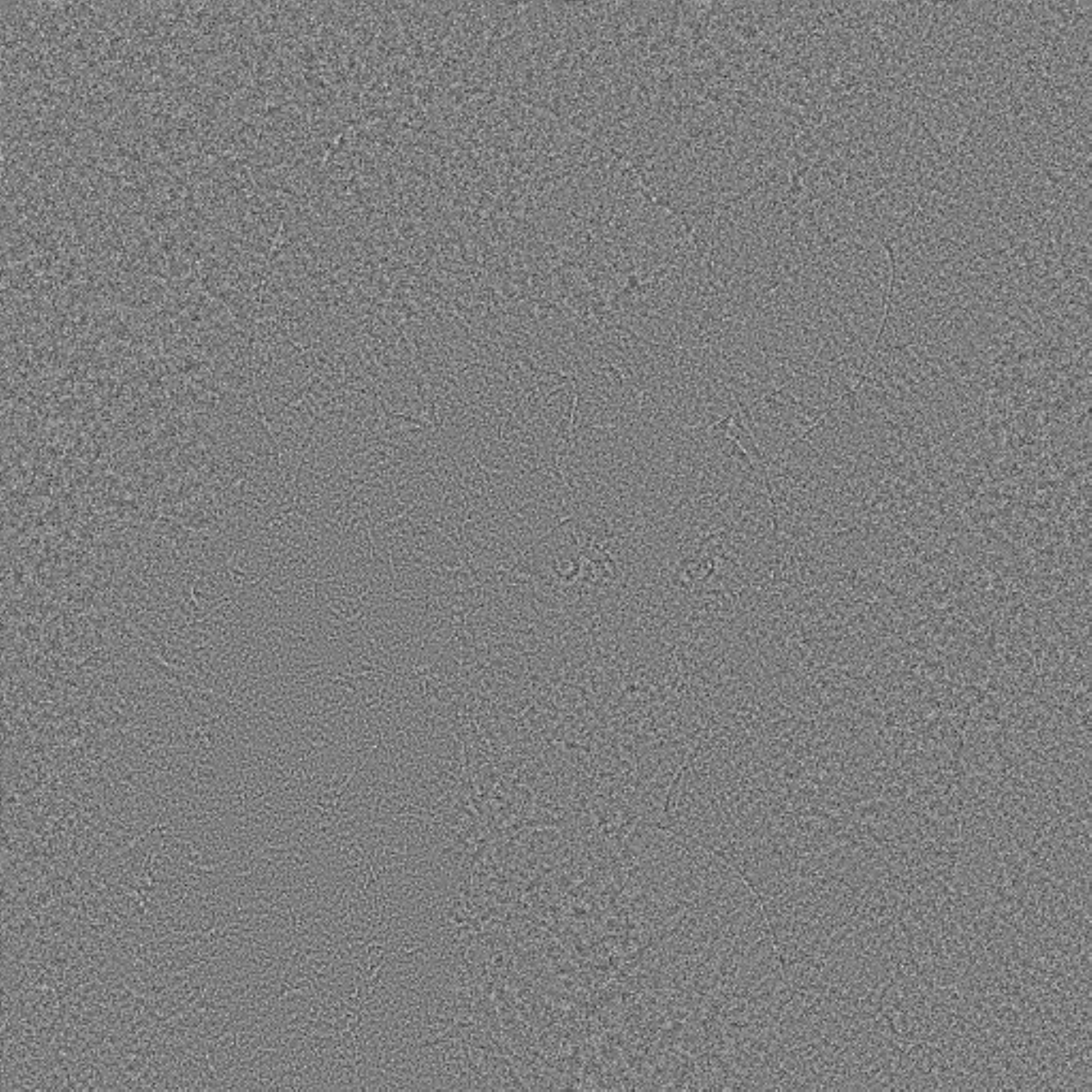}\\
\includegraphics[width=0.3\linewidth]{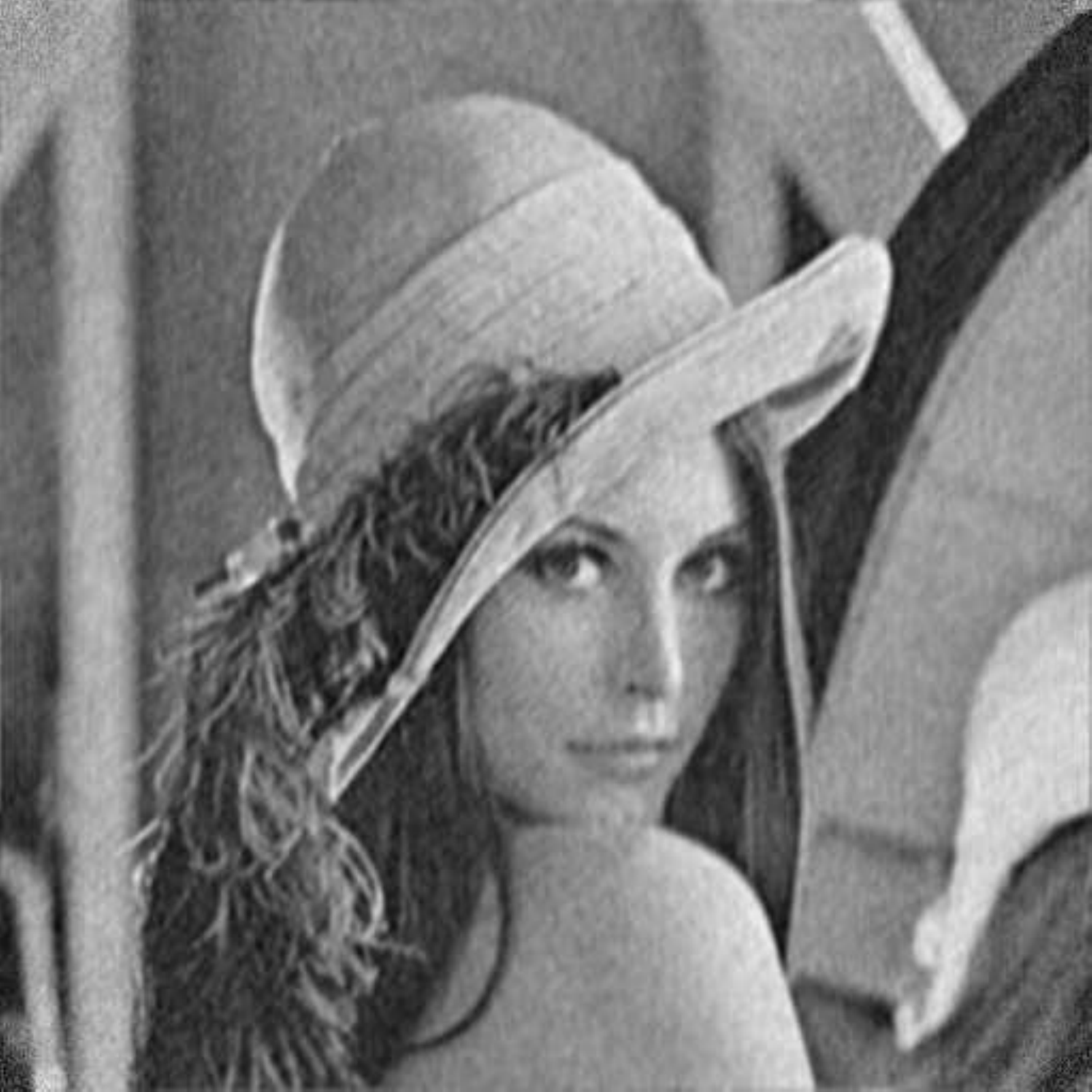}
&\includegraphics[width=0.3\linewidth]{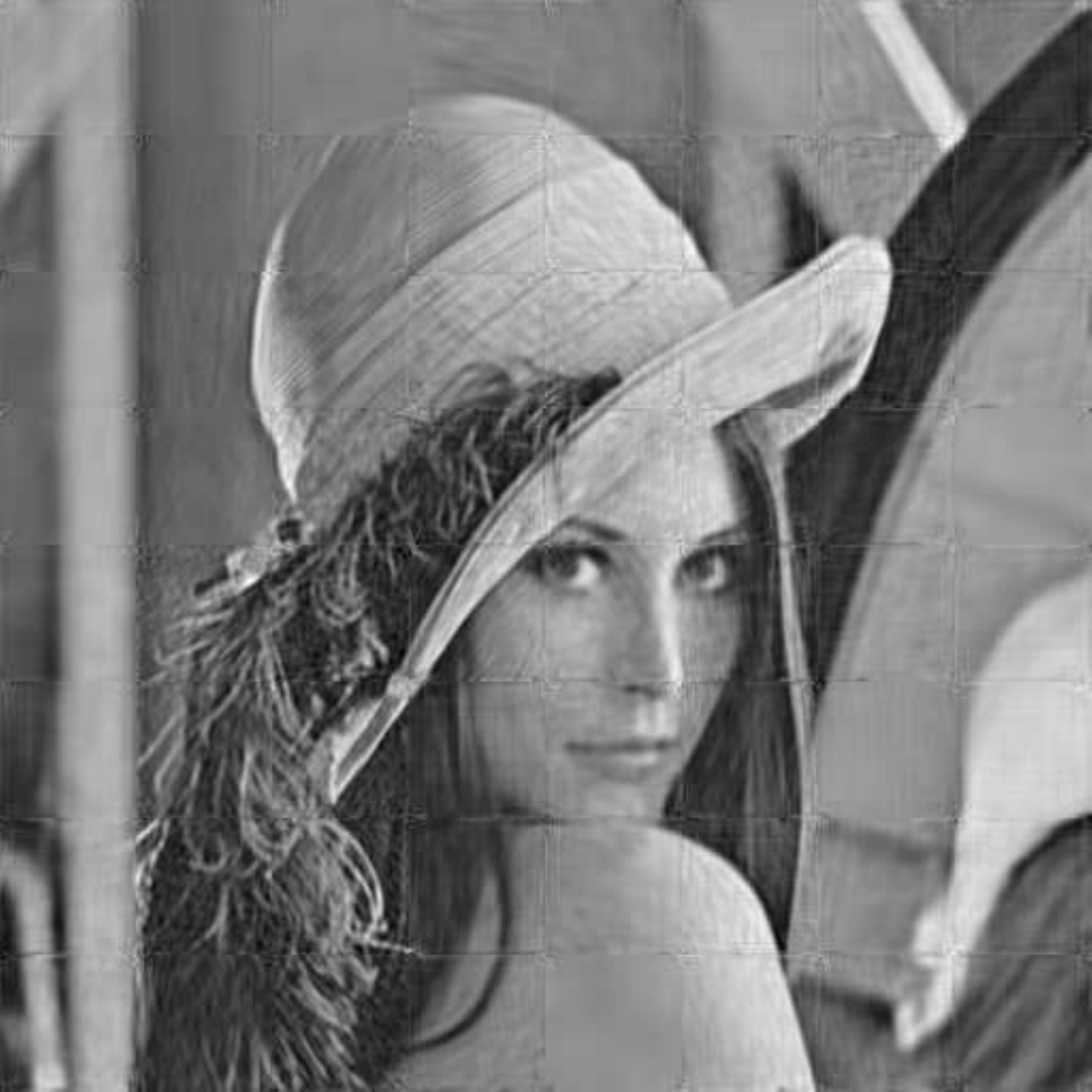}
&\includegraphics[width=0.3\linewidth]{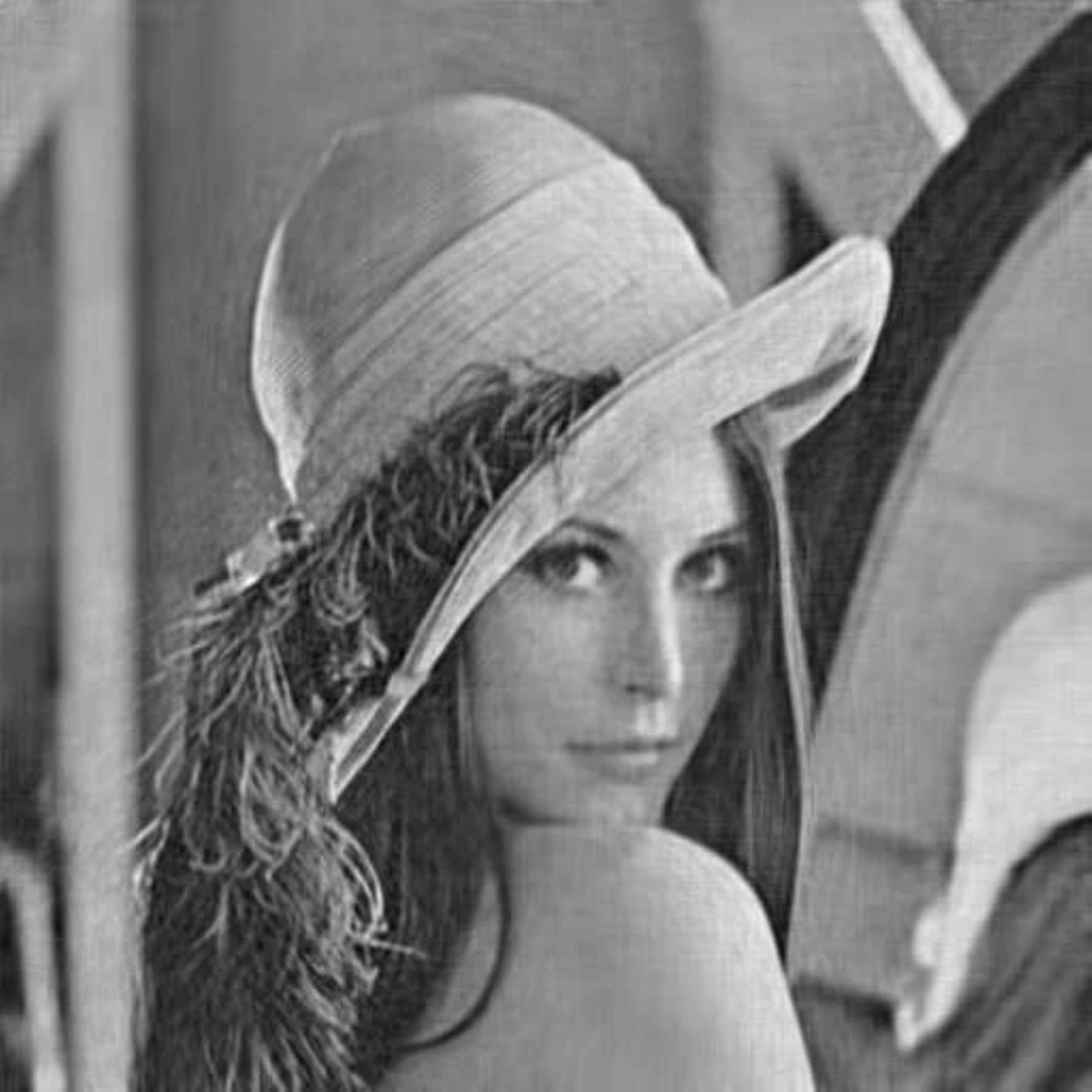}\\
FNLTV &FNLTV $S_r=64, n_s=64$ &FNLTV $ S_r=64, n_s=10$\\
\includegraphics[width=0.3\linewidth]{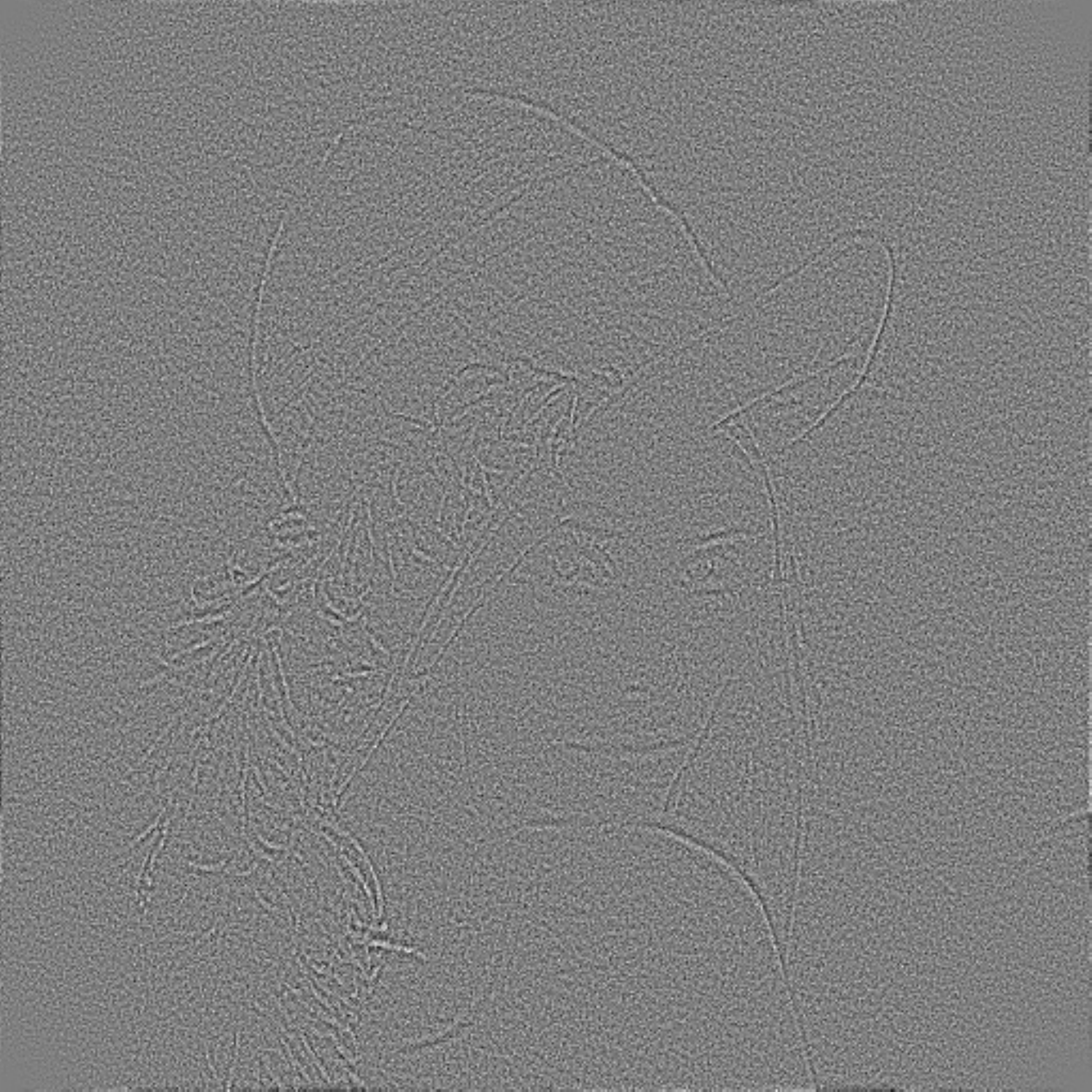}
&\includegraphics[width=0.3\linewidth]{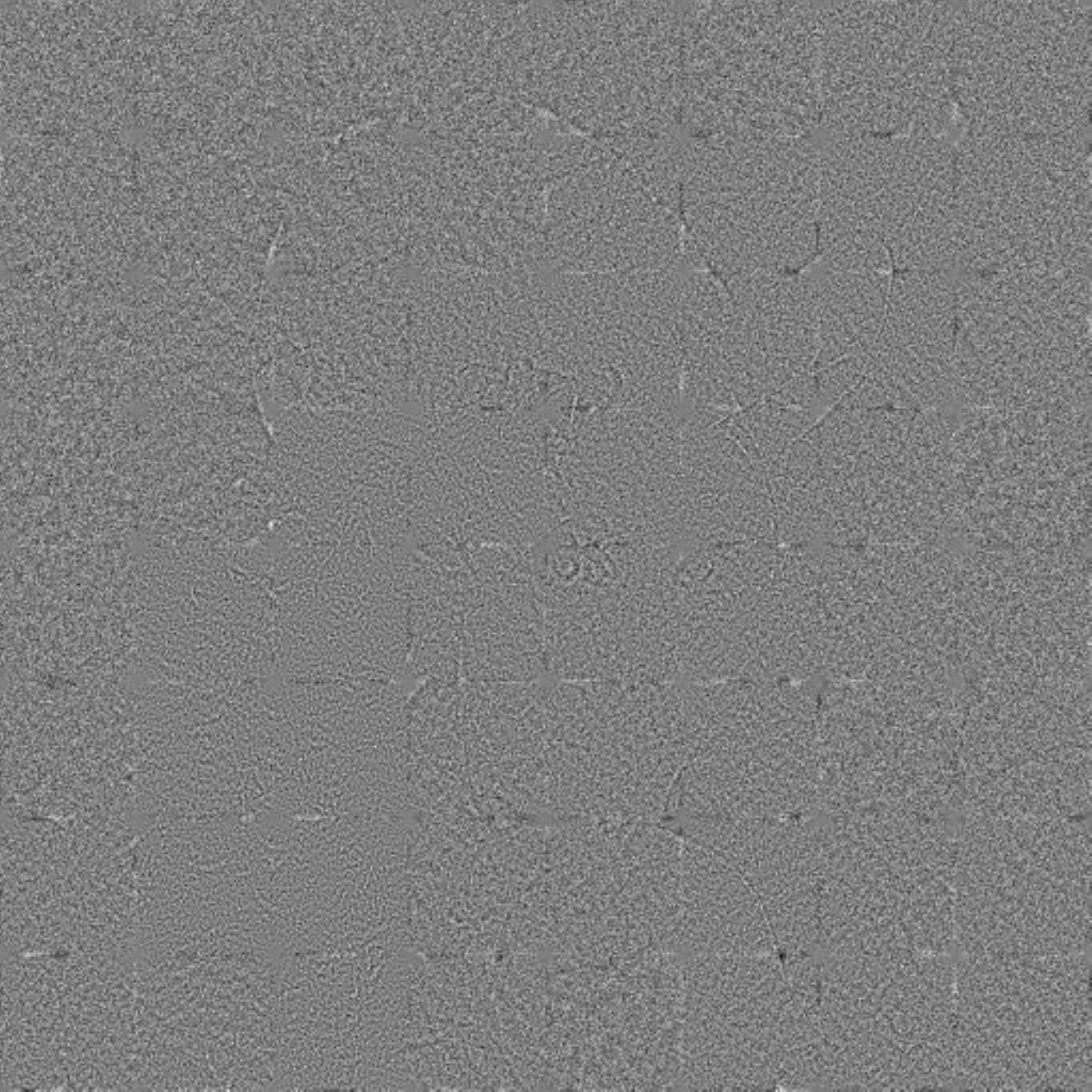}
&\includegraphics[width=0.3\linewidth]{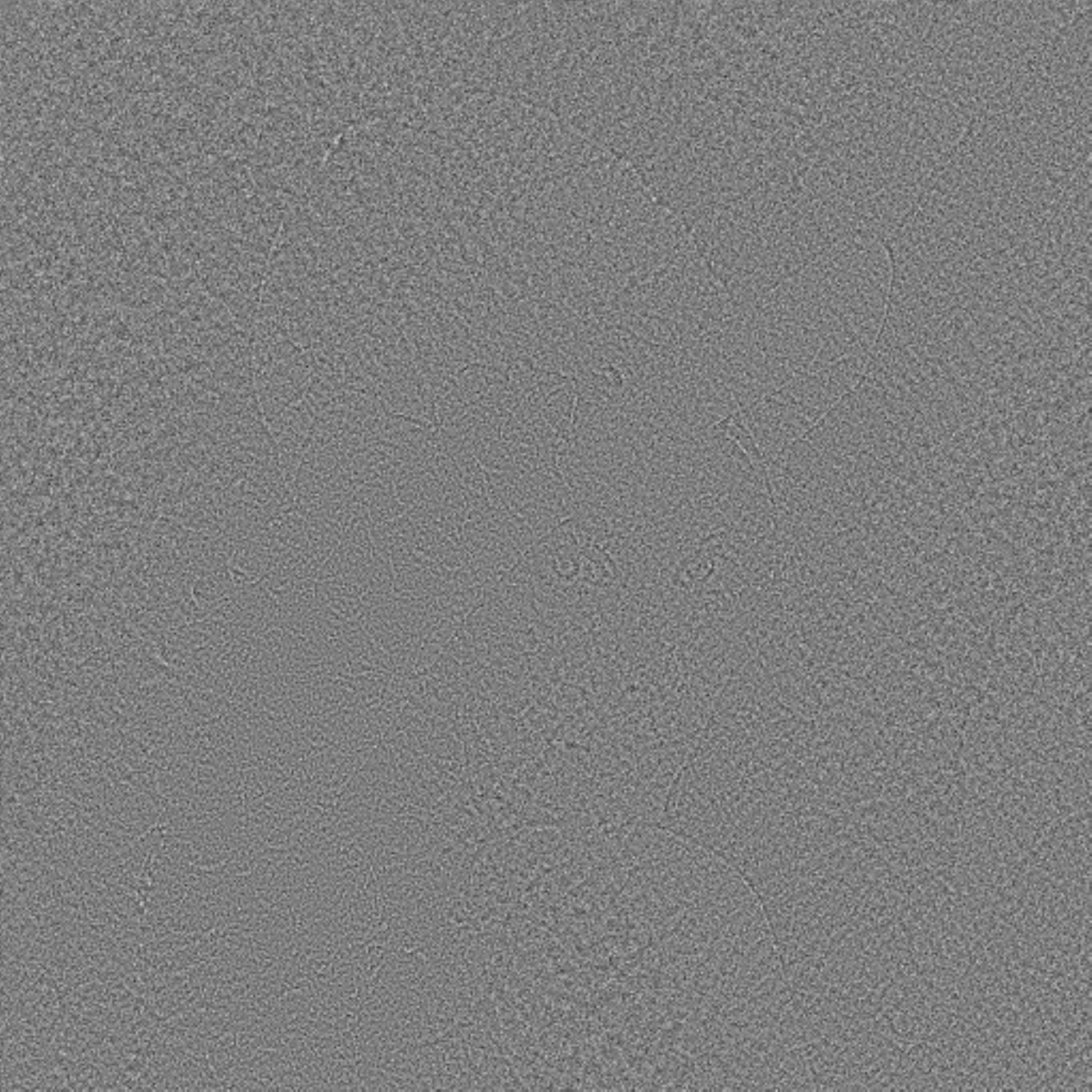}\\
\end{tabular}
\caption{Left: image denoised by NLTV and FNLTV globally; Middle: image denoised by NLTV and FNLTV locally with no-overlapping regions of  size 64$\times$64;    Right: image denoised by NLTV and FNLTV locally with overlapping regions of  size 64$\times$64 for moving step $n_s=50$(top) and $n_s=10$(third row). The second and bottom rows are the method noise images of the corresponding images of the top  and third rows.}
\label{figlocal}
\end{center}
\end{figure}

FNLTV demonstrates its advantage for retaining image detail as shown in Figure  \ref{lena_nltv_fnltv_lam}. At the other hand, FNLTV can not remove noise clearly in homogenous regions. Since Fourier transform is well suited for dealing with regular image patterns, and local image regions are more regular than entire images, we  then test the performance of local application of FNLTV. That is, we divide the image into local square regions, and apply  FNLTV for each local region respectively. The results are shown in Figure  \ref{figlocal}. Note that the local version of FNLTV yields  images which are more visually pleasant than global FNLTV does.  When the local regions are not overlapping, the region boundaries are evident. While  the
regions are overlapping the boundaries can disappear. In fact, as  DFT is a global operator,  the estimations from different local regions for same pixels are generally different.

Moreover, we also apply NLTV locally in the same manner as FNLTV for comparison. The results in Figure  \ref{figlocal}  show that the local application of NLTV does not make great difference, and the region boundary is hardly visible. This can be explained by the fact that NLTV is essentially  a neighborhood filter as stated in Section \ref{nltvneighbor}. When the local regions are overlapping, the resulting images are also similar, since all the estimations for  same pixels generally have similar values  as long as the local region is large enough.

\section{SFNLTV and Local SFNLTV}
Since NLTV model works better for homogenous region than FNLTV, and FNLTV retains more details than NLTV, we propose a model to combine the regularization terms of  NLTV and FNLTV to exploit the advantages of both models. It is called spatial-frequency domain nonlocal total variation (SFNLTV) model, which involve  minimizing the following energy functional:
  \begin{equation}
E(u)=\lambda \sum_{i \in I} |\nabla_{{w}}u(i)| +\lambda_f\sum_{\omega \in I}|\nabla_{w_f}\hat{u}(\omega)|+\frac12\sum_{i \in I}(u(i)-v(i))^2,\label{eunltvfft}
\end{equation}
recalling that
\begin{eqnarray}
|\nabla_{{w}}u(i)|&=&\sqrt{\sum_{j\in I}(u(i)-u(j))^2 w(i,j)} \label{juju}, \\
|\nabla_{w_f}\hat{u}(\omega)|&=&\sqrt{\sum_{\xi \in I}|\hat{u}(\omega)-\hat{u}(\xi)|^2w_f(\omega,\xi)}.
\end{eqnarray}
Since the two regularization terms are both convex with respect to $u$, SFNLTV energy functional is also strictly convex, and can be minimized by the gradient descent algorithm.
By (\ref{W}), (\ref{eulernltv}) and (\ref{gradfft}),
\begin{eqnarray}
\nabla E(u)(i)&=&\lambda\sum_{j\in I}\left(u(i)-u(j)\right)\frac{w(i,j)}{|\nabla_{{{w}}}u(i)|}+\frac{w(j,i)}{|\nabla_{{{w}}}u(j)|} \nonumber \\
&&+\lambda_f\Re\left\{\mathfrak{F}^{-1}\left[\sum_{\xi \in I}(\hat{u}(\omega)-\hat{u}(\xi))\left(\frac{w_f(\omega,\xi)}{|\nabla_{{{w_f}}}\hat{u}(\omega)|}+\frac{w_f(\xi,\omega)}{|\nabla_{{{w_f}}}\hat{u}(\xi)|}\right) \right]\right\} \nonumber \\
&&+u(i)-v(i), \label{grad_sfnltv}
\end{eqnarray}
where $\Re$ and $\mathfrak{F}^{-1}$ represent the real part and the inverse discrete Fourier transform.

Moreover, as demonstrated in Section \ref{secfnltvpar}, local application of FNLTV can improve FNLTV significantly and local application of NLTV is similar to NLTV. We now take account of applying SFNLTV locally. Concretely, an image is divided into overlapping regions of size $S_r \times S_r$. The
horizontal and vertical steps of moving regions are equal and  denoted as $n_s$. SFNLTV is then applied to each region. Since regions are overlapping,  some pixels  belong to several regions, and thus have several estimations. The final estimation of each pixel is the average  of  all the estimations.
In addition, to alleviate the edge effects caused by the Fourier transform, for each local region, the edges with the size of one pixel are removed from the denoised region.
The algorithm is called local SFNLTV (L-SFNLTV). Note that there are two differences between L-SFNLTV and SFNLTV, which make L-SFNLTV better: firstly, local implementation replaces global implementation; secondly  the average of several estimations is taken as the final estimation. Moreover, the local application methods can be easily implemented in parallel, which makes L-SFNLTV a fast method. The  algorithm is summarized in Algorithm \ref{algo1}.
\begin{CJK*}{UTF8}{gkai}
\begin{algorithm}
        \caption{Image denoising by L-SFNLTV}
        \begin{algorithmic}[1] 
        \Require $ v$
        \State Divide $ v$ into  overlapping regions of size  $S_r \times S_r$. The horizontal and vertical steps of moving regions are equal and  denoted as $n_s$
        \For  {each region}
        \State SFNLTV is applied.
        \EndFor
        \State Return each region to the corresponding original location and take averages for repeated estimates.
         \Ensure $ {\bar{v}}$
         \end{algorithmic}
         \label{algo1}
    \end{algorithm}
\end{CJK*}

\section{Simulations} \label{sec_sim}
We use  the Peak Signal-to-Noise Ratio (PSNR) to measure the
quality of restored images:
\begin{equation}
\mbox{PSNR } (\bar {v}) = 10\log_{10} \frac{255^2}{\mbox{MSE}}, \nonumber
  \end{equation}
  where MSE is defined in (\ref{mse}).
In our experiments we  use ($512\times 512$) images,  Lena, Barbara,  Bridge, Boats and ($256\times256$) images, Peppers, House, Cameraman.\footnote{ They were all downloaded online. Lena, Peppers,  Boats and House:

 http://decsai.ugr.es/$\sim$javier/denoise/test\_images/index.htm

Bridge: www.math.cuhk.edu.hk/$\sim$rchan/paper/dcx/

Barbara, Cameraman: www.dcs.qmul.ac.uk/$\sim$phao/CIP/Images/.} 
The level of noise is supposed to be known, otherwise there are methods to estimate it; see e.g. \cite{johnstone1997wavelet}.
\subsection{Performance of NLTV model and SFNLTV} \label{sec6_nltv}
We test our images in the case $\sigma=20$, and search for the best PSNR value with $\sigma_r=20$ fixed and other parameters varying in some ranges. We find that except the Bridge image, one can use the same parameters $D=3, d=9,  \lambda=15$. %
The PSNR value of each image is close to the best one (that we have tested) with a difference less than 0.1. So the optimal parameters are not sensitive to general images.
Since the image Bridge has many irregular fine details, the optimal parameters are a little different: $D=3, d=15, \lambda=11 $. 


{ NLTV} model is based on the idea of  ROF model \cite{rudin_nonlinear_1992} and NL-means \cite{buades2005review}. We now compare NLTV with them.
Recall that the denoised image by NL-means is
$$
\bar{v}(i)=\frac{\sum_{j\in U_i(D)}w(i,j)v(j)}{\sum_{j\in U_i(D)}w(i,j)},
$$
where $w(i,j)$ is defined in (\ref{nltvweight}), and ROF model is obtained by $(\ref{Eudis})$ and (\ref{ju})   with $w(i,j)$ defined in (\ref{weighttv}).
For NL-means, we use $D=5, d=3, \sigma_r=24$ for image Bridge, and  $D=11, d=7, \sigma_r=18 $ for other images.
For ROF model, $\lambda=11$ is used for the images Barbara and Bridge, and $\lambda=16$ for other images.

For SFNLTV model, $w(i,j)$ is taken as ($\ref{nltvweight}$) in NLTV model, and $w_f(\omega,\xi)$ is taken as ($\ref{nltvweightf}$) in FNLTV model. Note that we have now eight parameters $\lambda$, $d$, $D$, $\sigma_r$, $\lambda_f$, $d_f$, $D_f$, $\sigma_{rf}$. We use the same choices of  parameters in $w(i,j)$ in NLTV model for general natural images, i.e. $d=9, D=3, \sigma_r=20$, and  test the other parameters in the case $\sigma=20$.
We find that $\lambda=11, \lambda_f=2, d_f=9, D_f=5, \sigma_{rf}=16$ is a good choice for all the images. Therefore, despite  more parameters than NLTV model, the choice of parameters for SFNLTV model is less sensitive to images, which happens because the choice of the frequency domain regularizer coefficient $\lambda_f$ is not very sensitive.

\begin{figure} 
\begin{center}
\renewcommand{\arraystretch}{0.5} \vskip3mm {
\fontsize{8pt}{\baselineskip}\selectfont
\begin{tabular}{ccc}
\includegraphics[width=0.30\linewidth]{bar.pdf}
&\includegraphics[width=0.30\linewidth]{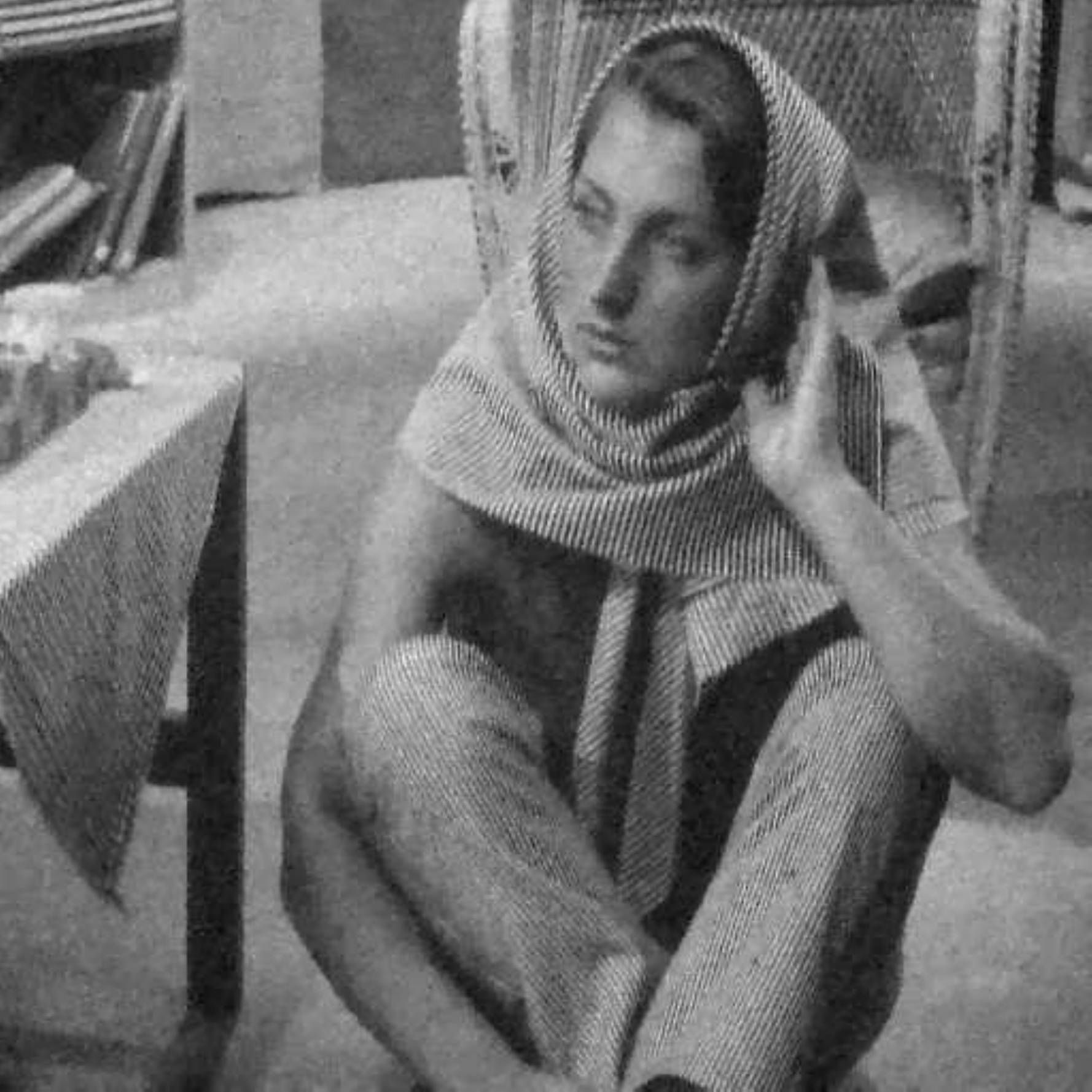}
&\includegraphics[width=0.30\linewidth]{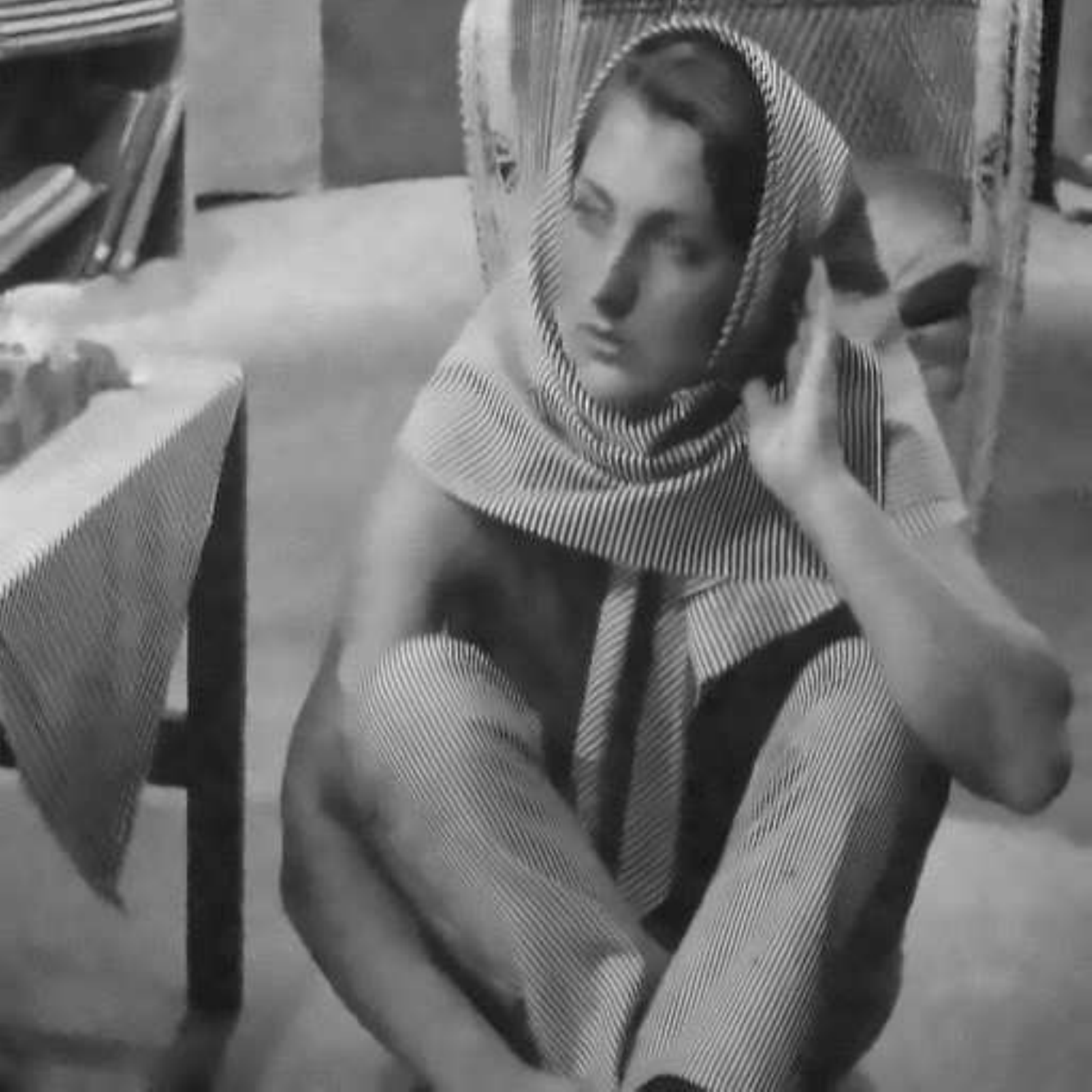} \\
Original (noise-free) &ROF: PSNR=27.10& NL-means: PSNR=29.68\\
 \includegraphics[width=0.30\linewidth]{bar_noisy.pdf}
&\includegraphics[width=0.30\linewidth]{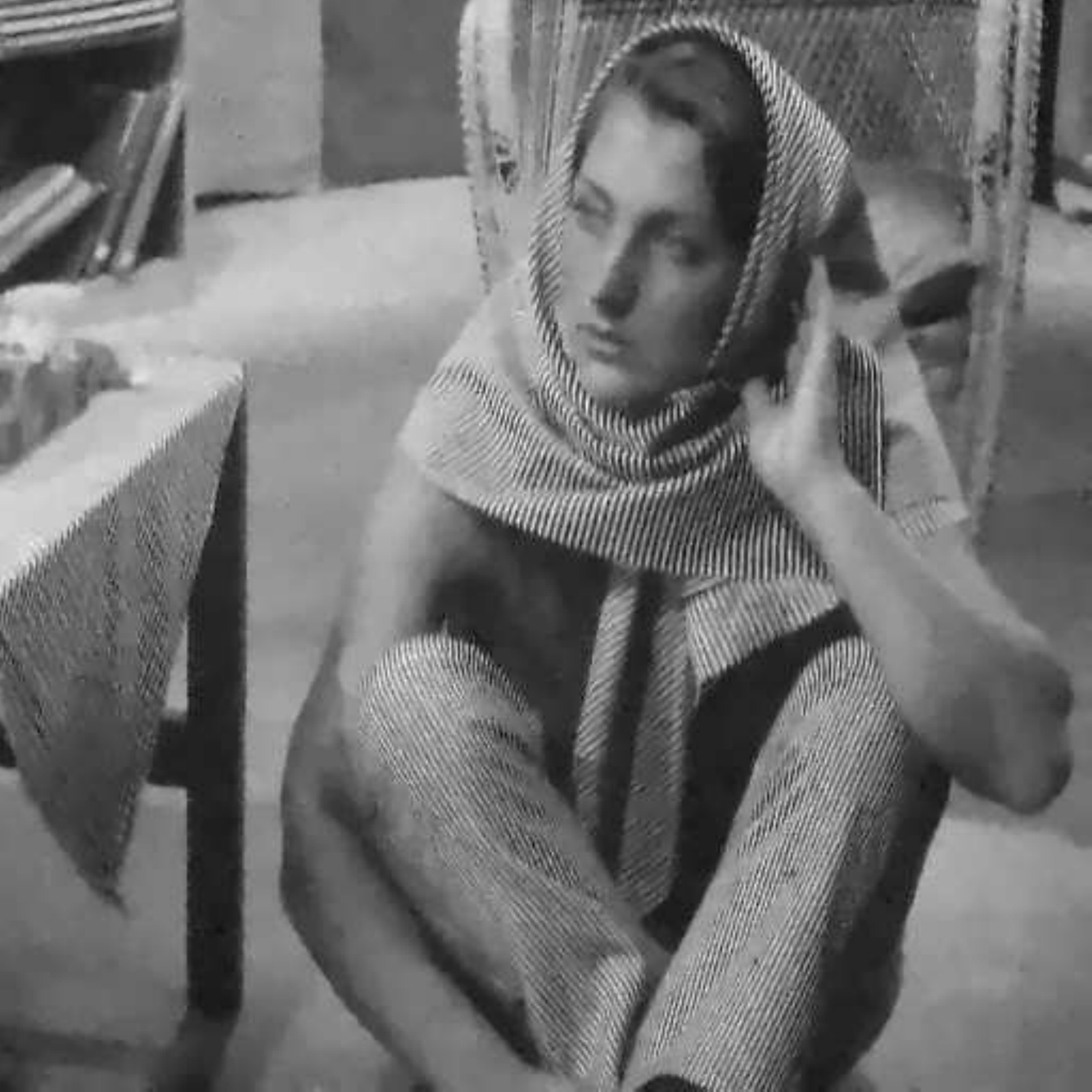}
&\includegraphics[width=0.30\linewidth]{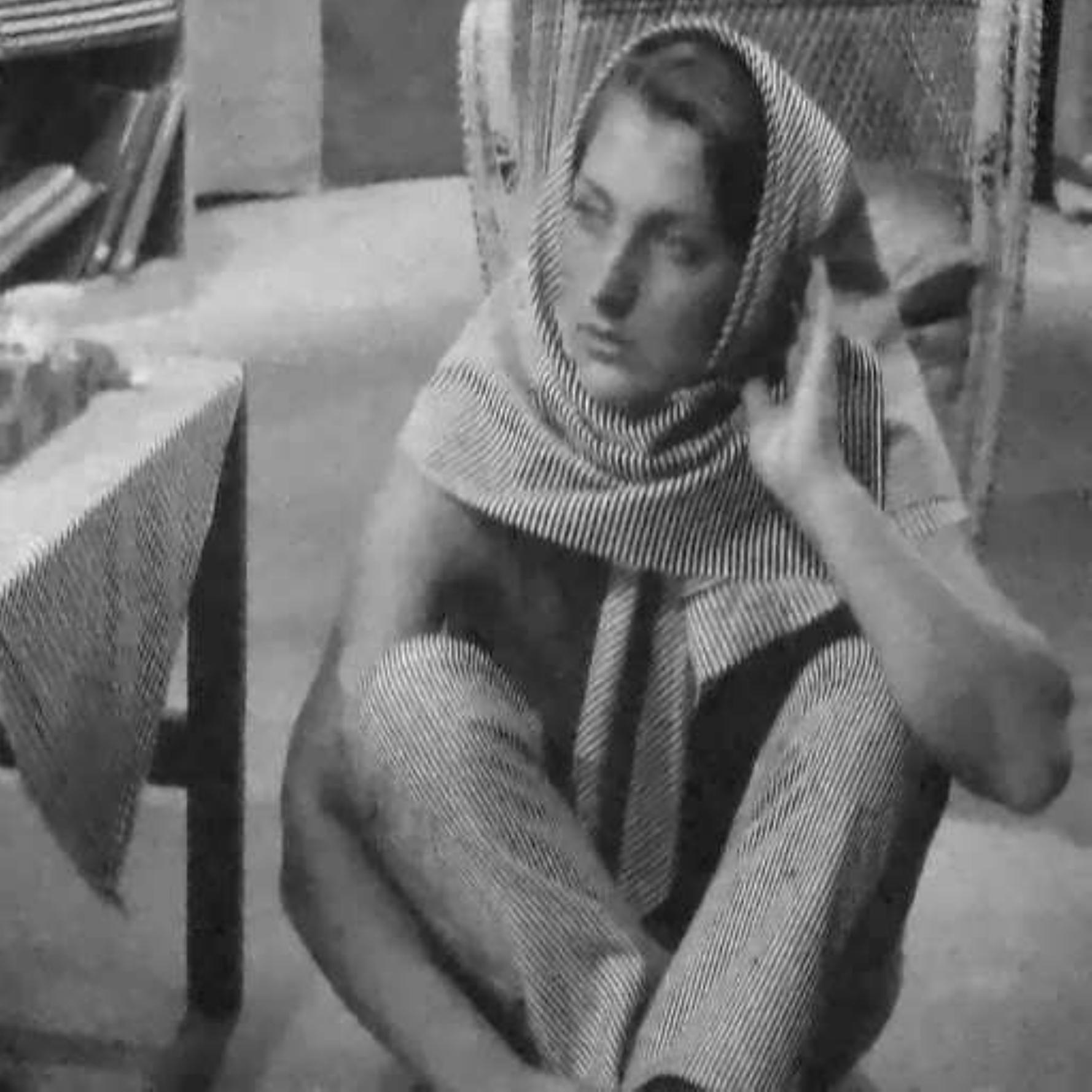} \\
 Noisy image & NLTV: PSNR=28.46 & SFNLTV: PSNR=29.19
\end{tabular}
} \vskip1mm
\par
\rule{0pt}{-0.2pt}
\par
\vskip1mm
\caption[Denoised images by NLTV model, NL-means  and ROF model for  Barbara.]{Denoised images  by ROF model, NL-means, NLTV model and SFNLTV model for  Barbara.} 
\label{nltv_nl_tv3}
\end{center}
\end{figure}

\begin{figure}
\begin{center}
\begin{tabular} {ccc}
\includegraphics[width=0.30\linewidth]{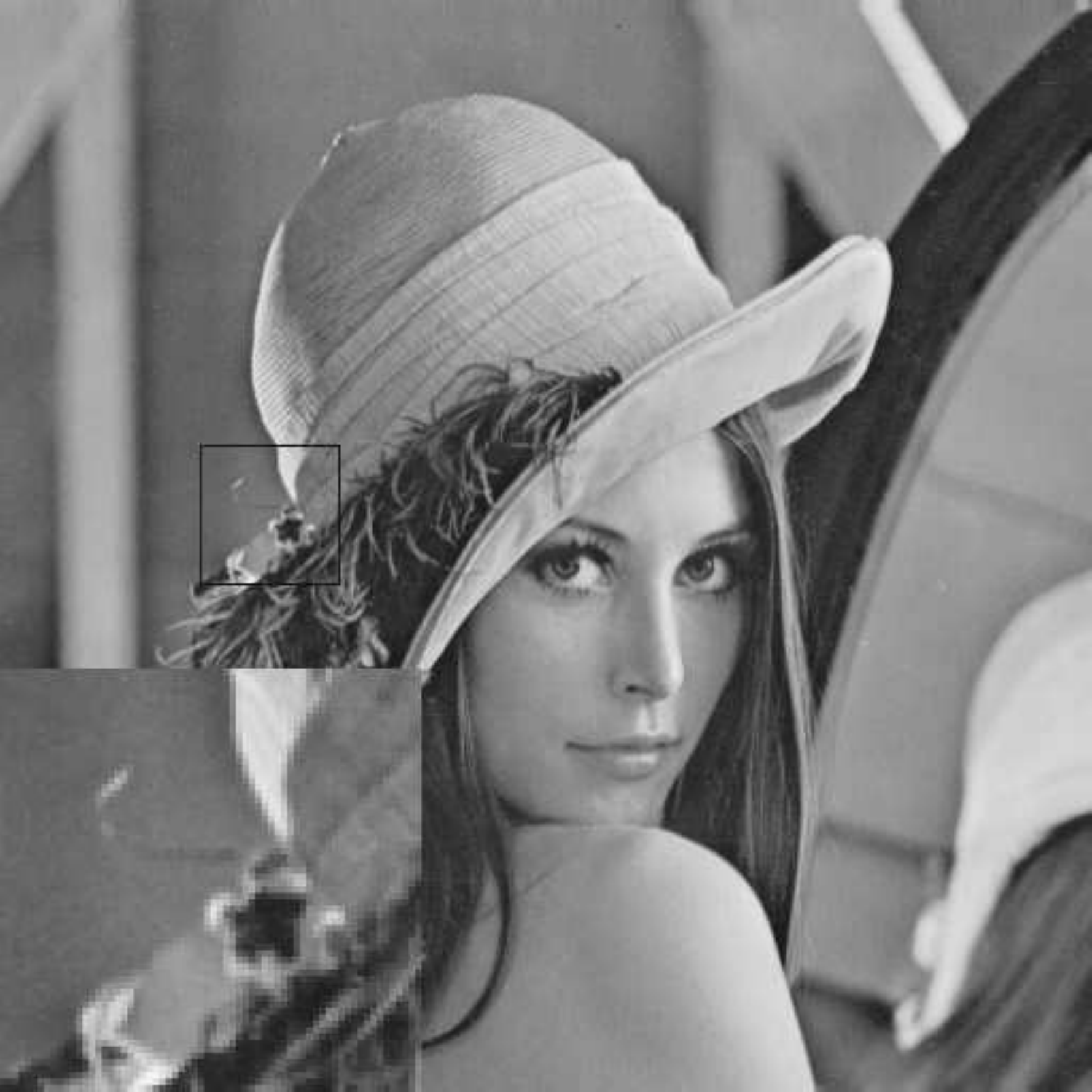} &
\includegraphics[width=0.30\linewidth]{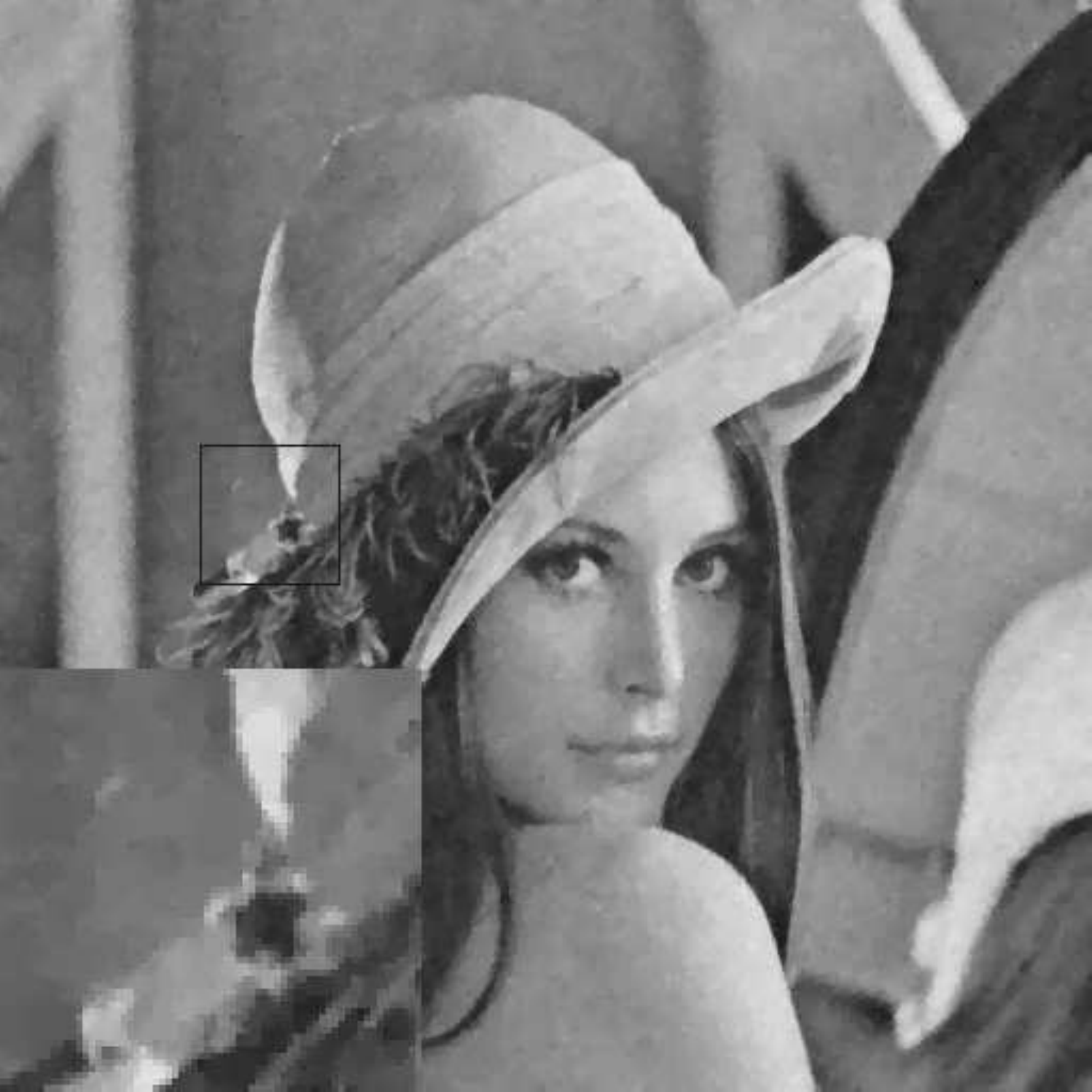} 
&\includegraphics[width=0.30\linewidth]{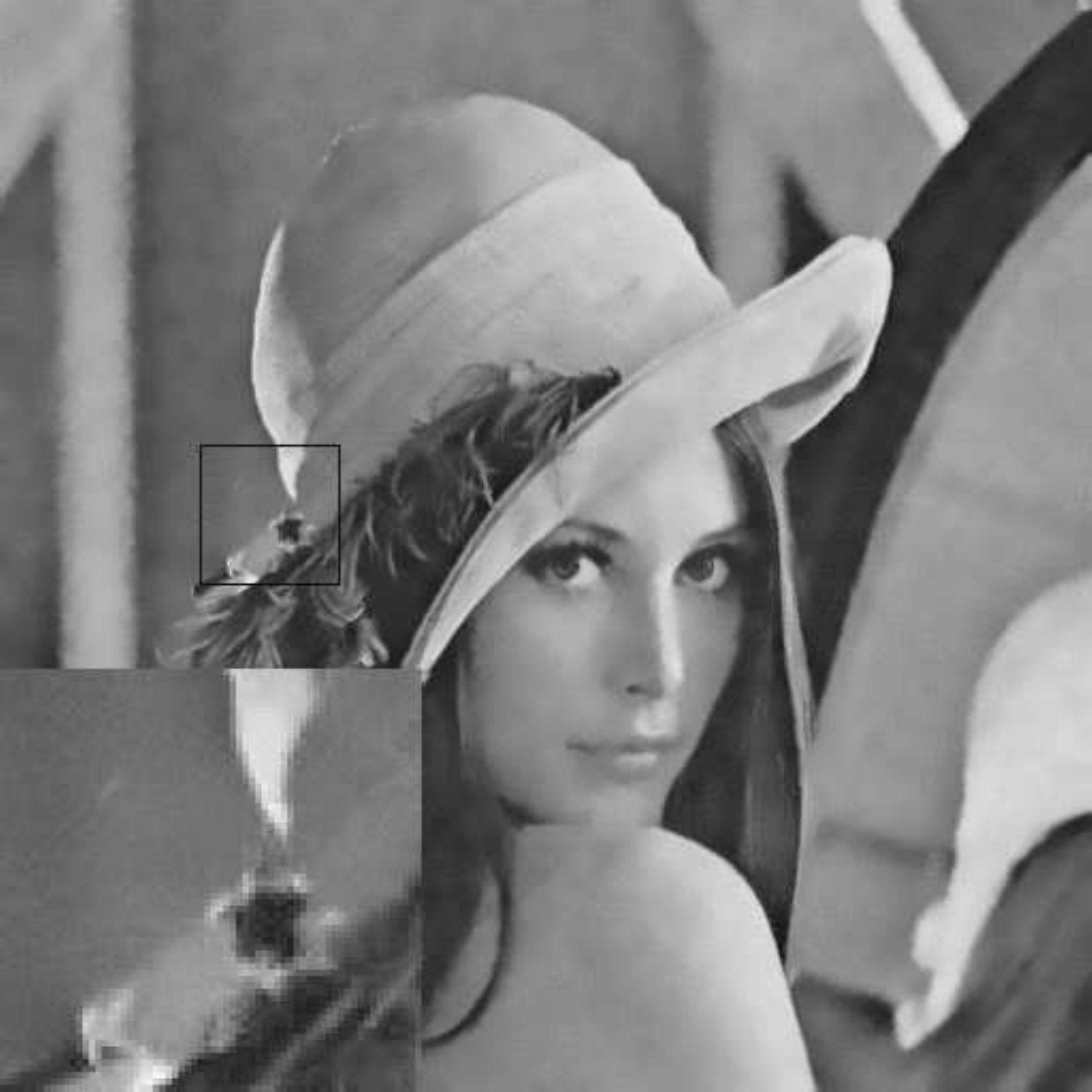}\\
\footnotesize{Original} & \footnotesize{ROF: PSNR=31.07}&\footnotesize {NL-means: PSNR=31.56}\\
\includegraphics[width=0.30\linewidth]{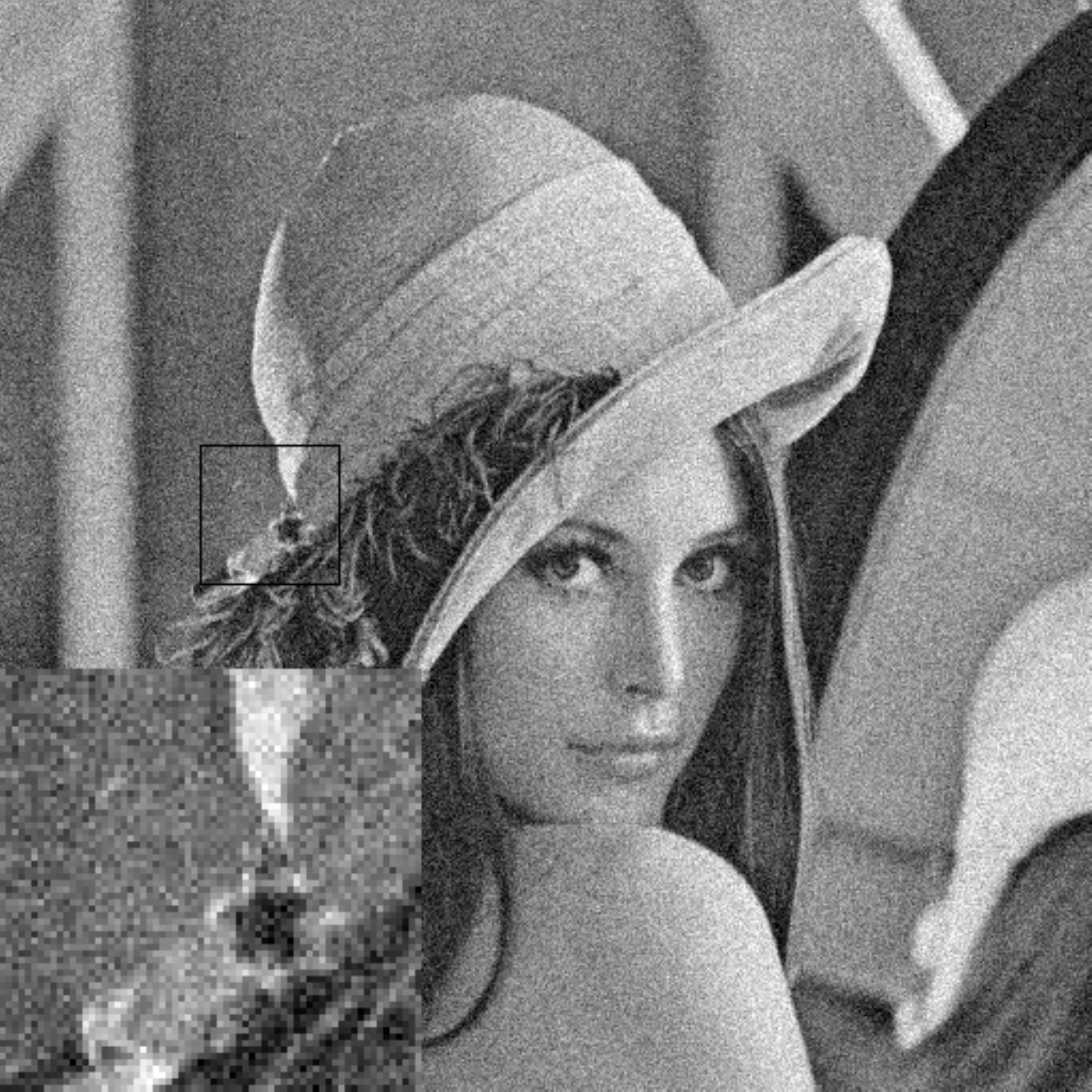}&
\includegraphics[width=0.30\linewidth]{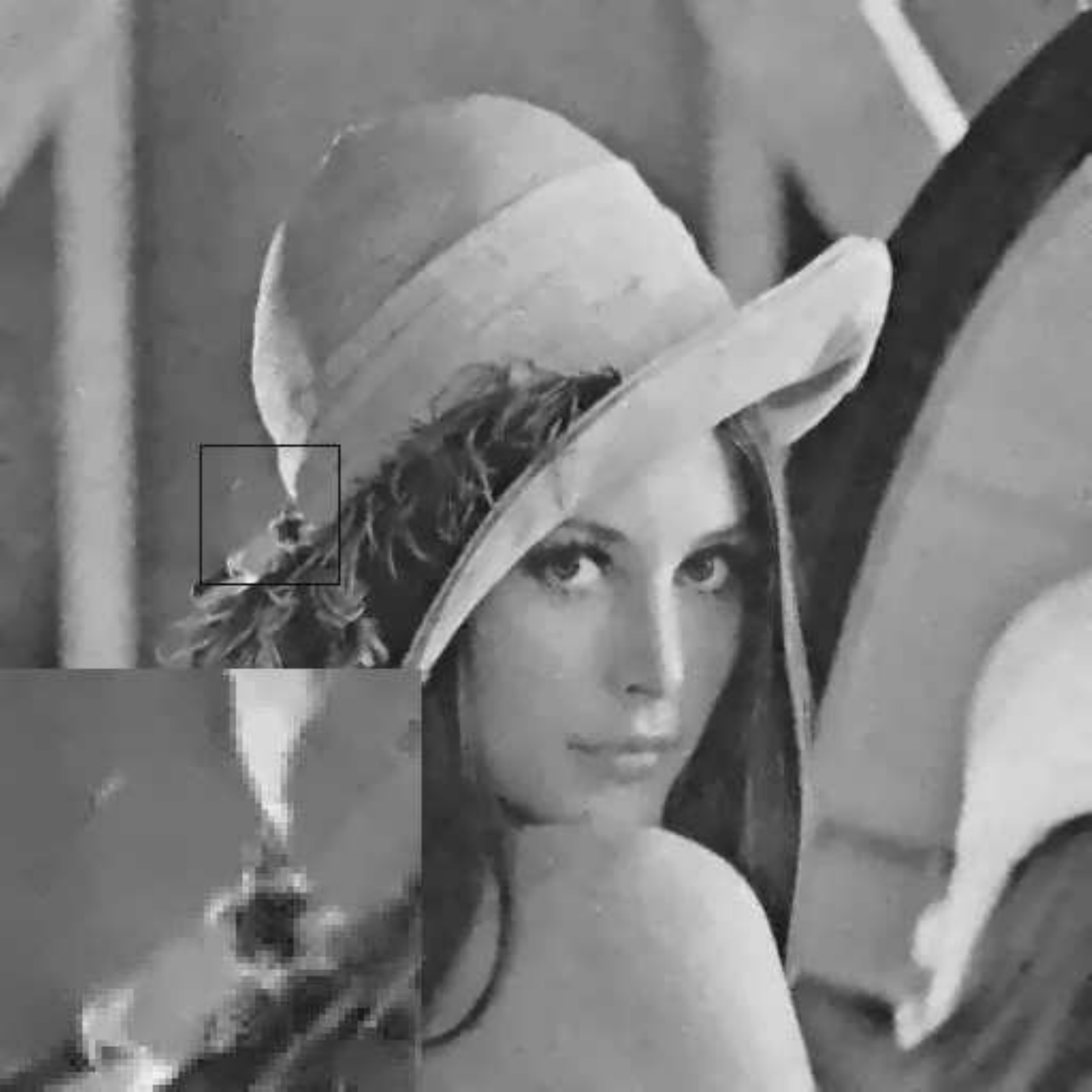}&
\includegraphics[width=0.30\linewidth]{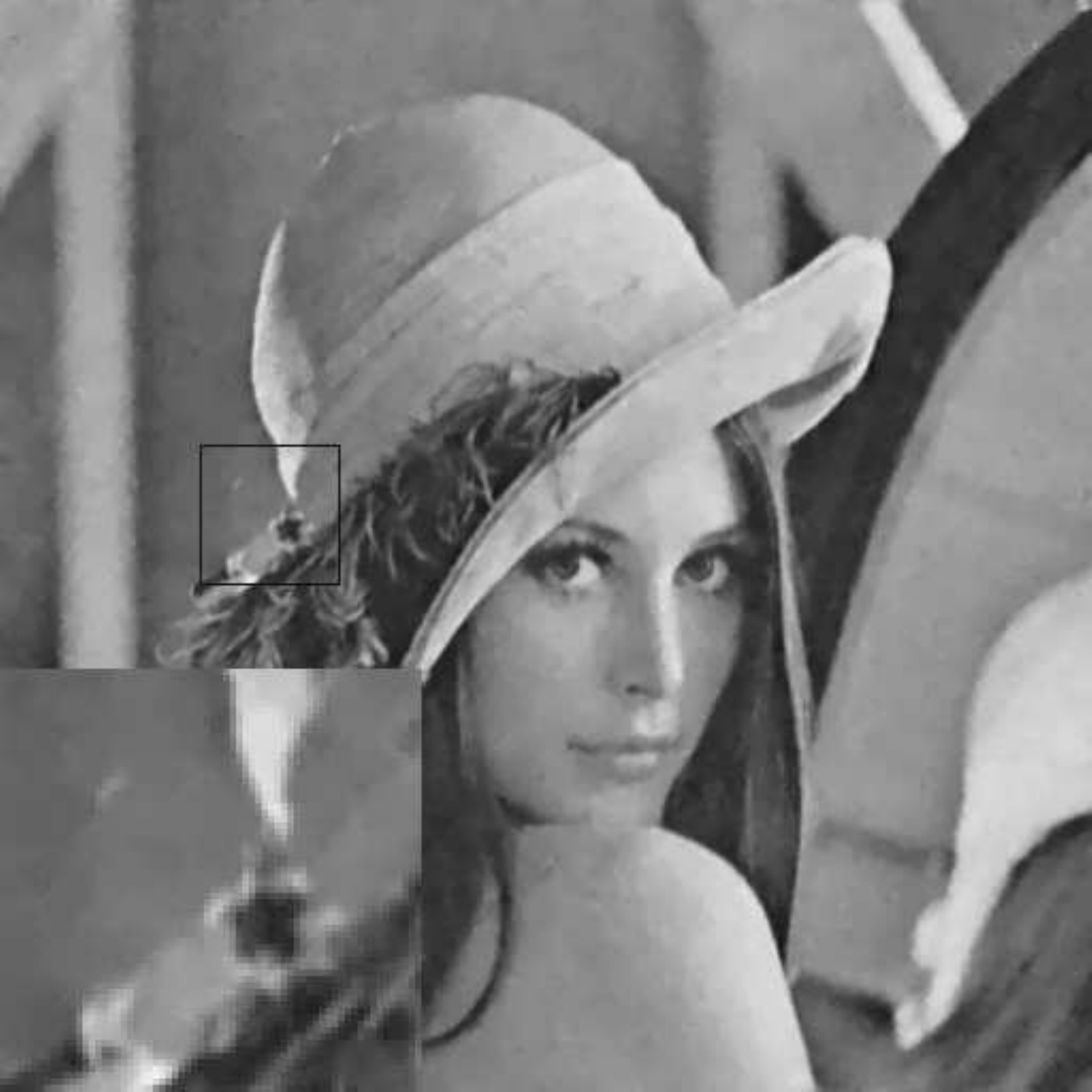}\\
\footnotesize {Noisy ($\sigma=20$)} &\footnotesize {NLTV: PSNR=31.56}&\footnotesize{SFNLTV: PSNR=31.77}
\end{tabular}
\caption[Zoom in the left area of the Lena's hat. In order to enhance visual artifacts,
the same sharpen filter has been applied to all displayed images.]{Denoised images  by ROF model, NL-means, NLTV model and SFNLTV model for Lena.}
\label{lenaext}
\end{center}
\end{figure}

 The comparisons of PSNR values are shown in  Table $\ref{nltv_nl_tvtab}$, which shows that NLTV model is better than ROF model for all the images, and  is similar to NL-means except the image Barbara.  SFNLTV model is better than NLTV model for all the images, even for Bridge image, and  it is the best for almost all the tested images among all the methods.
 Examples of  denoised images are shown in Figures.$\ref{nltv_nl_tv3}$ and \ref{lenaext}. We can see that  NLTV model is better than NL-means for isolated pixels (rare patches); the denoised images with NLTV model  have less staircase artifacts than ROF model, and have less noise left. SFNLTV model is better than NLTV model for fine details, but a little noisier in homogeneous regions.



\begin{table}
\begin{center}
\caption{PSNR values for different images with NLTV model, NL-means, ROF model, and SFNLTV model in the case $\sigma=20$. 
The values marked with * are obtained by optimal parameters for the corresponding images different from other images.
For SFNLTV model, we use uniform parameters for all the images.}
\begin{tabular}{ccccccccccc} 
\hline
\noalign{\smallskip}
Image &Lena &Barbara& Peppers &Boats &Bridge & House & Cameraman \\
\noalign{\smallskip}
\hline
\noalign{\smallskip}
NLTV    & 31.56 &28.46 &30.21 &29.49 &26.81* &31.74 &29.45 \\
NL-means    & 31.56 & $\mathbf{29.68}$  &30.18 &29.32 &26.81* &31.92 &29.35 \\

ROF &31.07 &  27.10* &  29.65&   29.15&  26.69*&   31.18&   28.70\\
SFNLTV & $\mathbf{31.77}$ &  29.19 &   $\mathbf{30.29}$ &   $\mathbf{29.89}$ &  $\mathbf{ 26.92}$ &  $\mathbf{ 32.14}$ &   $\mathbf{29.64}$ \\
\noalign{\smallskip}\hline
 \end{tabular}

\label{nltv_nl_tvtab}
\end{center}
\end{table}

\subsection{Performance of L-SFNLTV}

To test the performance of our algorithm L-SFNLTV, we have done experiments with $\sigma=10,20,30,50$. The region  size $16\times 16$,  the moving step $n_s=6$, the search window size $D=D_f=3$, the control parameters of the weight functions $\sigma_r=\sigma_{rf}=\sigma$,  and the patch size for the frequency domain term $d_f=5$ are used for all the tested noise level. For each local region, we use 20 iterations for the gradient descent algorithm of SFNLTV. The choice of other parameters are shown in Table \ref{paralsf} for general images except Barbara. 
 Since the Barbara image contains many textures, we use different choice for $\lambda$ and $\lambda_f$: $\lambda=1$, $\lambda_f=\sigma$.

 We compare L-SFNLTV with SFNLTV and other NLTV related methods, NLSTV\cite{lefkimmiatis2015nonlocal}, RNLTV \cite{li2017regularized}, BNLTV \cite{liu2017block} for different noise levels.
 For SFNLTV, $D=3$, $\sigma_r=\sigma$, $D_f=5$, $d_f=9$, $\sigma_{rf}=0.8\sigma$, $\lambda=0.55\sigma$, $\lambda_f=1.6+0.02\sigma$, $d$ is chosen as L-SFNLTV, and 50 iterations are used for the gradient descent algorithm.
 For NLSTV and BNLTV, the authors do not report concrete values for the regularization parameter $\tau$ or $\lambda$. 
 We choose the parameter which results in the best PSNR values in average for all our tested images,
 that is, $\tau=0.006,0.013,0.029,0.071$ or $\lambda=27, 11, 7.5, 4.3$ for $\sigma=10, 20, 30, 50$ respectively. For RNLTV, when $\sigma=10, 20$, we use the parameter for $\sigma=0.04, 0.08$ in the paper \cite{li2017regularized} (where the original images are normalized to the interval [0,1]); for $\sigma=30, 50$, we  also choose the parameter $\gamma$ which yields the best PSNR values in average for the  tested images Boats and Barbara from the authors' code, that is, $\gamma=0.6, 2.4$. Other parameters are chosen according to the corresponding papers.

\begin{figure} 
\center 
\renewcommand{\arraystretch}{0.5} \addtolength{\tabcolsep}{0pt} \vskip3mm %
\fontsize{8pt}{\baselineskip}\selectfont
\begin{tabular}{ccc}
\includegraphics[width=0.30\linewidth]{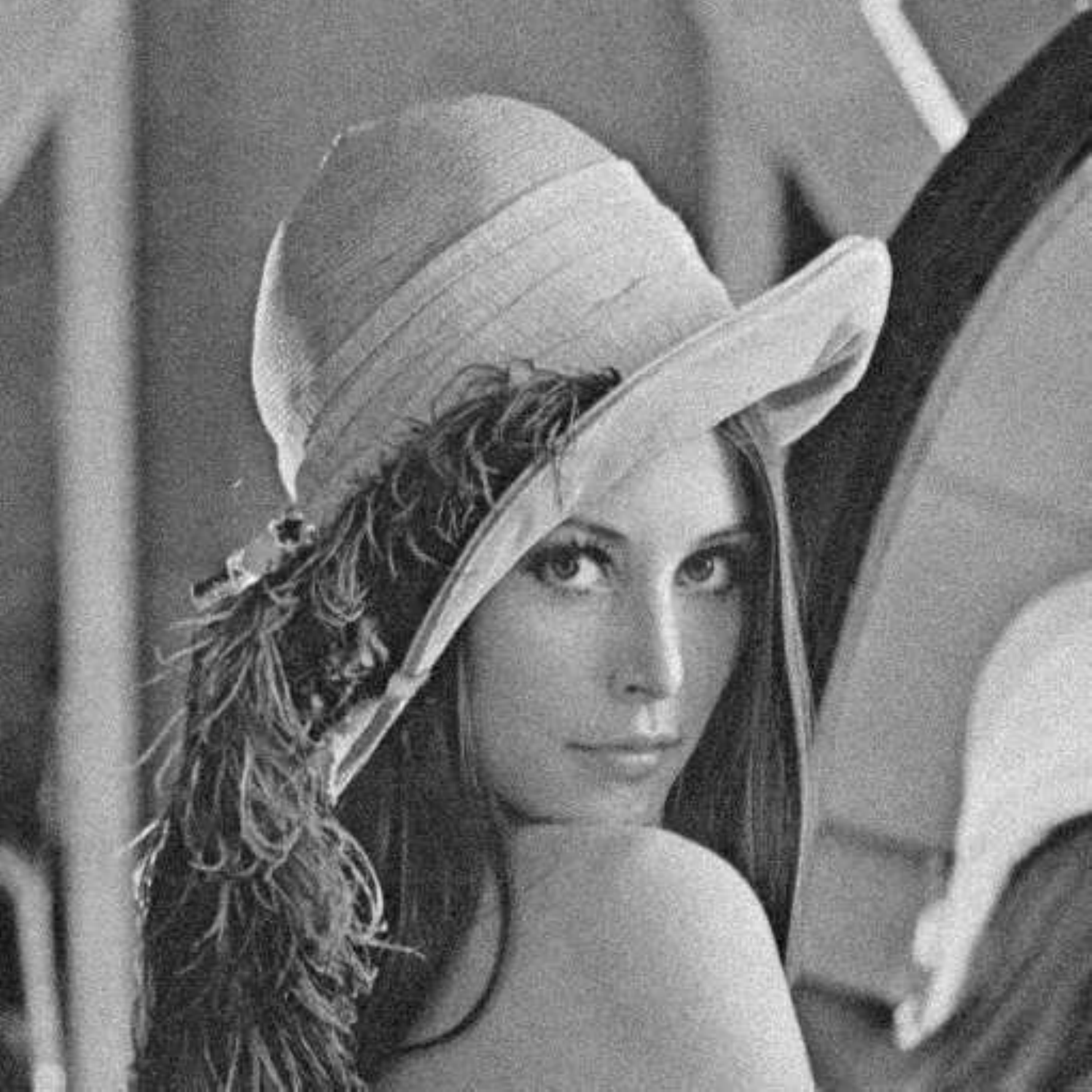}
&\includegraphics[width=0.30\linewidth]{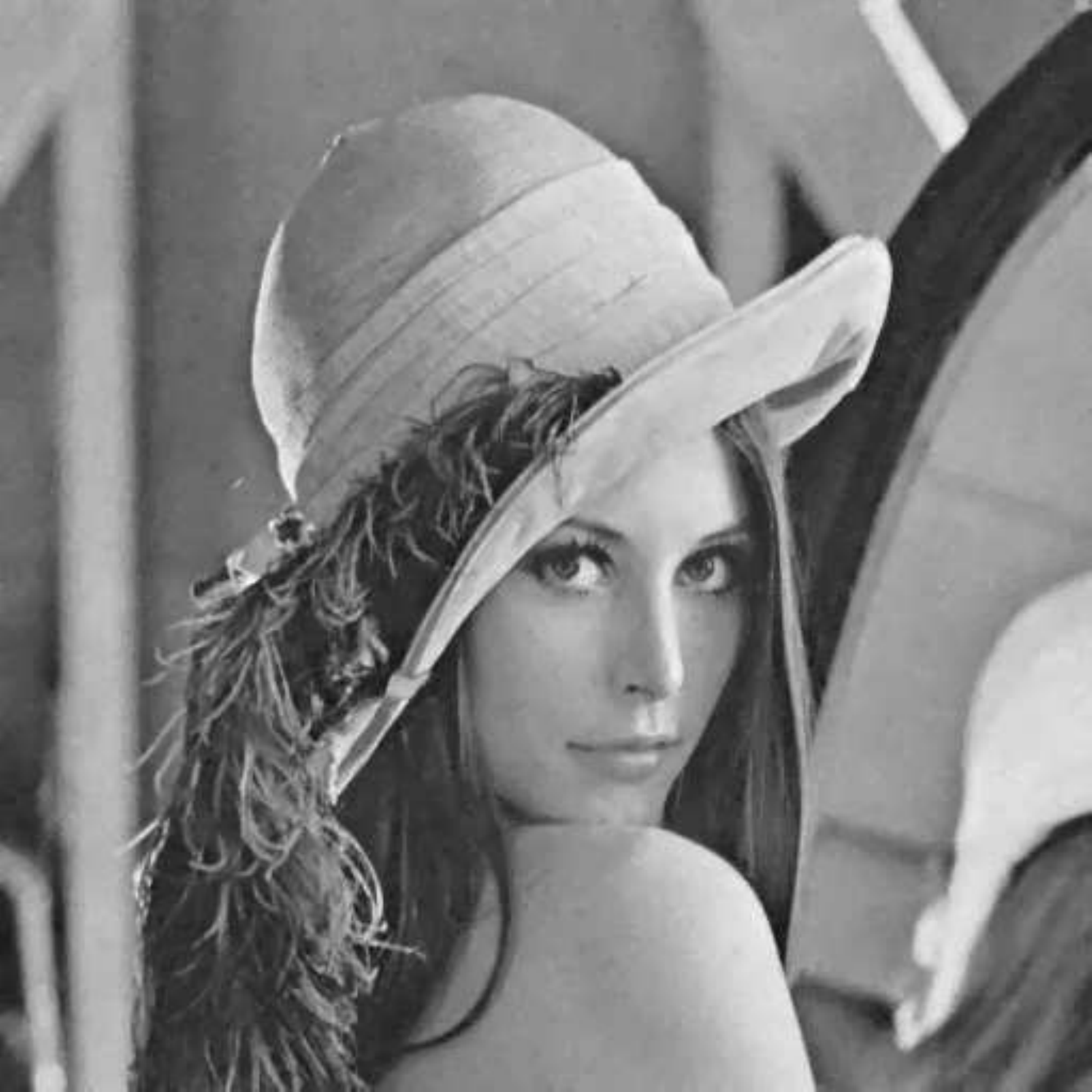}
&\includegraphics[width=0.30\linewidth]{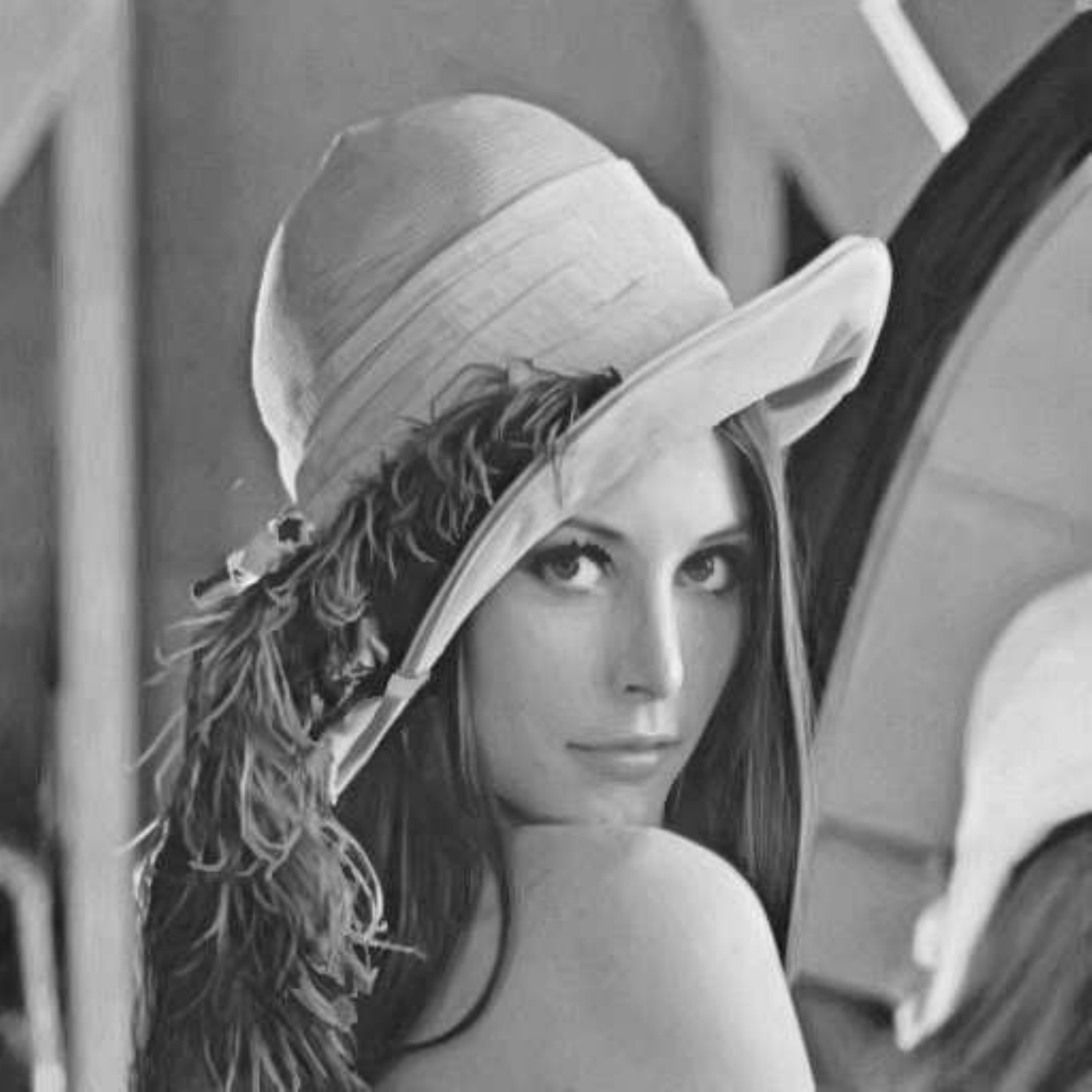}
\\
Noisy $\sigma=10$ & NLSTV PSNR=34.61 & RNLTV PSNR=34.17 \\
\includegraphics[width=0.30\linewidth]{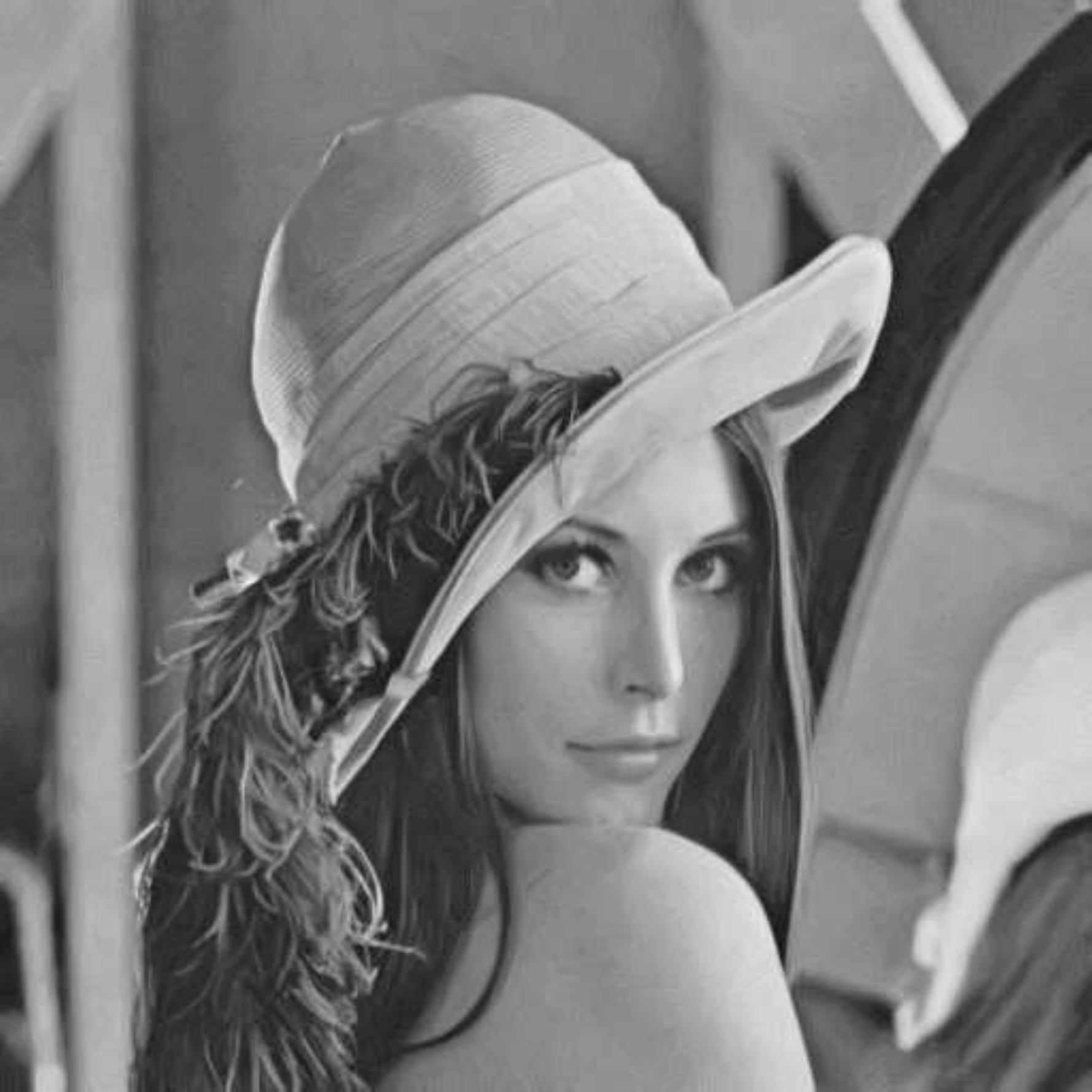}
&\includegraphics[width=0.30\linewidth]{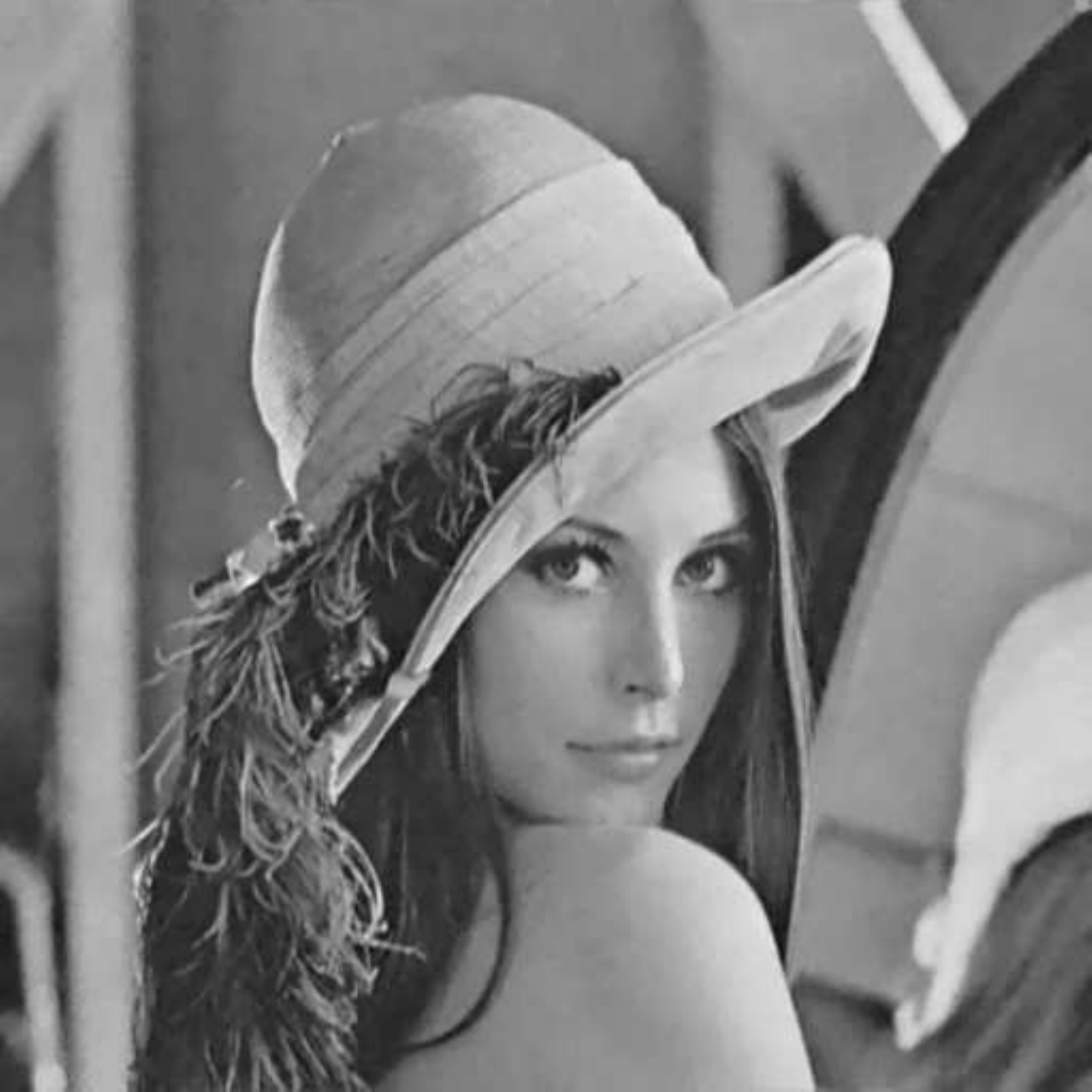}
&\includegraphics[width=0.30\linewidth]{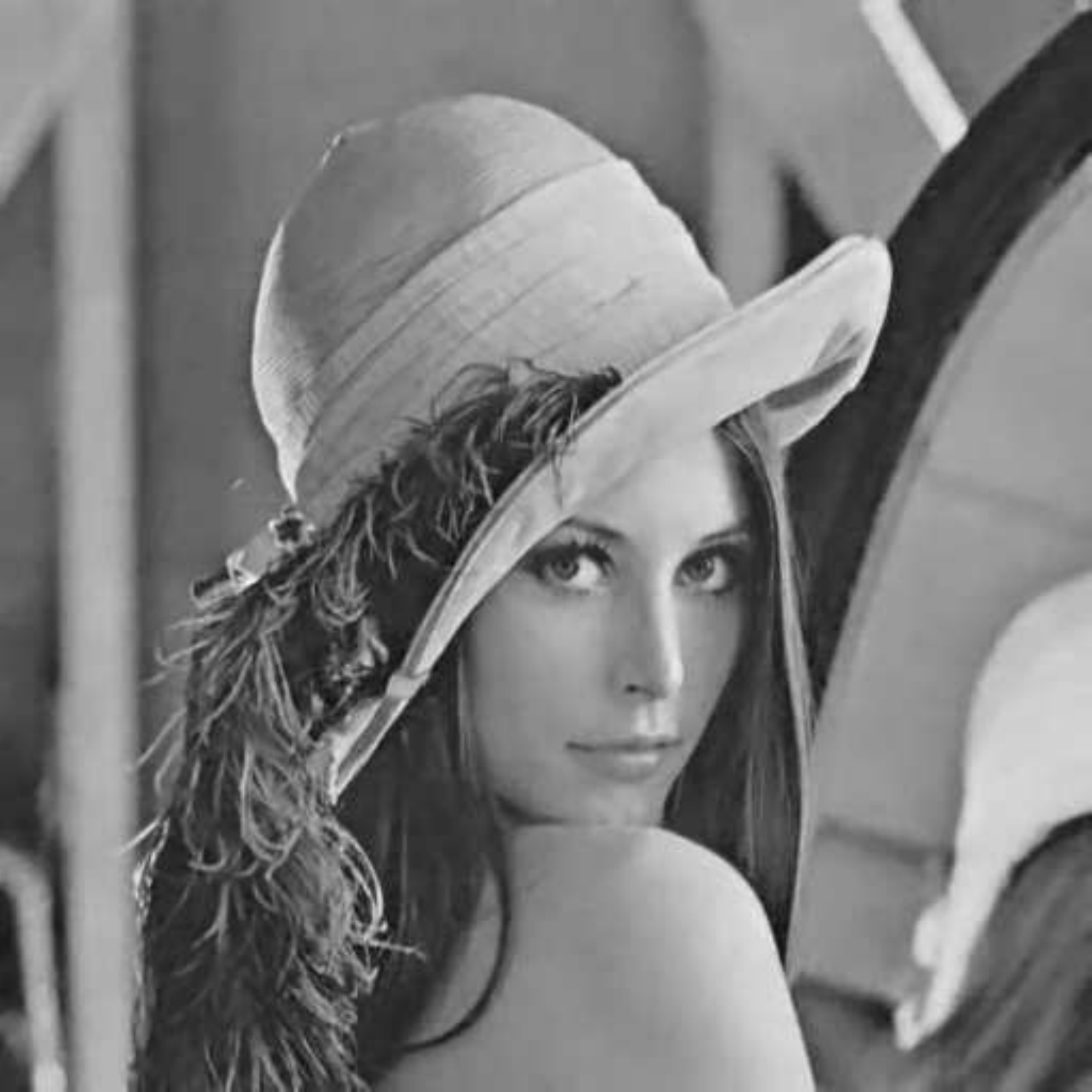}
\\
BNLTV PSNR=34.57 & SFNLTV PSNR=35.05 & L-SFNLTV PSNR=35.57   \\
\end{tabular}
\caption{Denoised Lena images  by L-SFNLTV,  NLSTV \cite{lefkimmiatis2015nonlocal}, RNLTV \cite{li2017regularized}, BNLTV \cite{liu2017block} and SFNLTV in the case $\sigma=10$.}
\label{lsfksvd10}
\end{figure}

\begin{figure} 
\center 
\renewcommand{\arraystretch}{0.5} \addtolength{\tabcolsep}{0pt} \vskip3mm %
\fontsize{8pt}{\baselineskip}\selectfont
\begin{tabular}{ccc}
\includegraphics[width=0.30\linewidth]{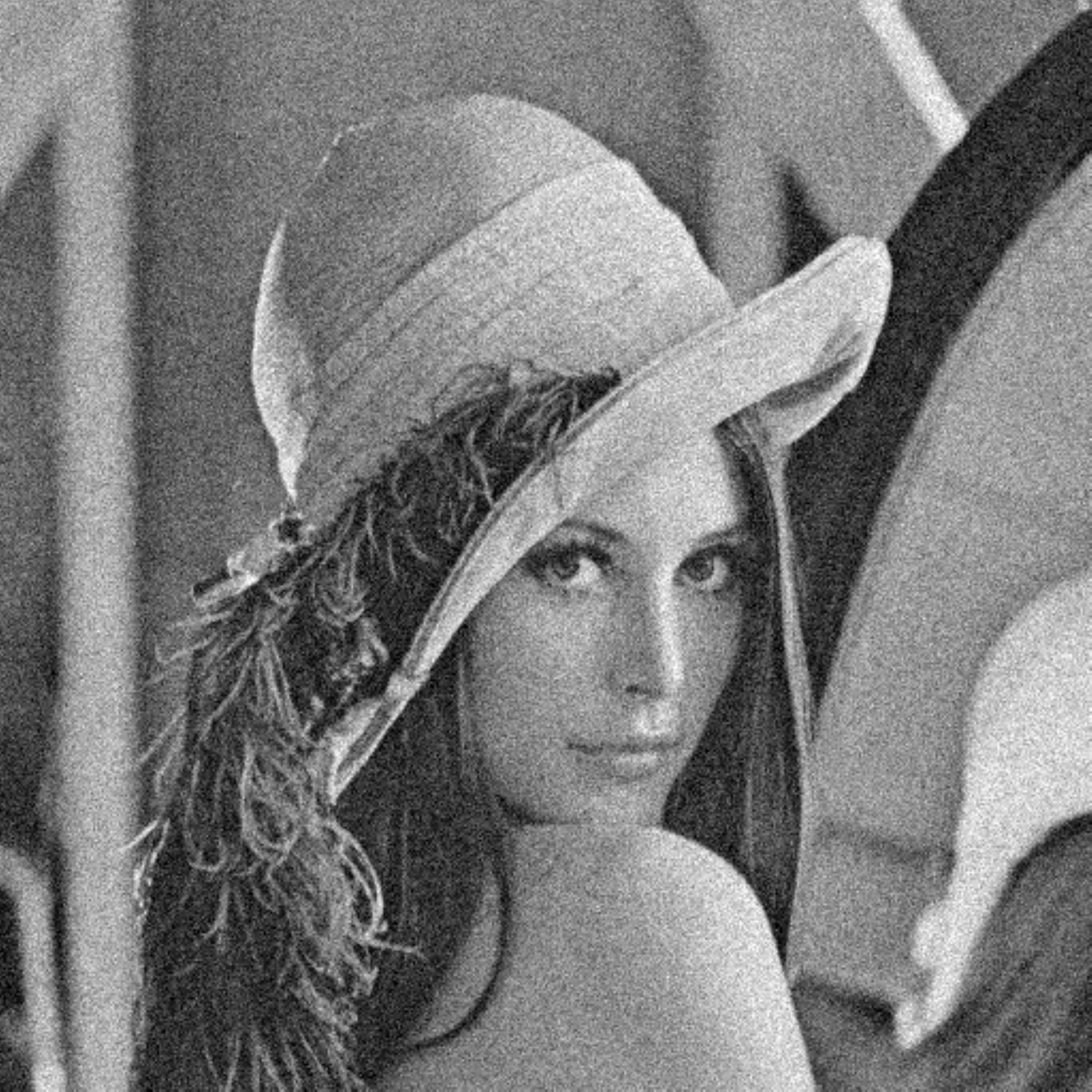}
&\includegraphics[width=0.30\linewidth]{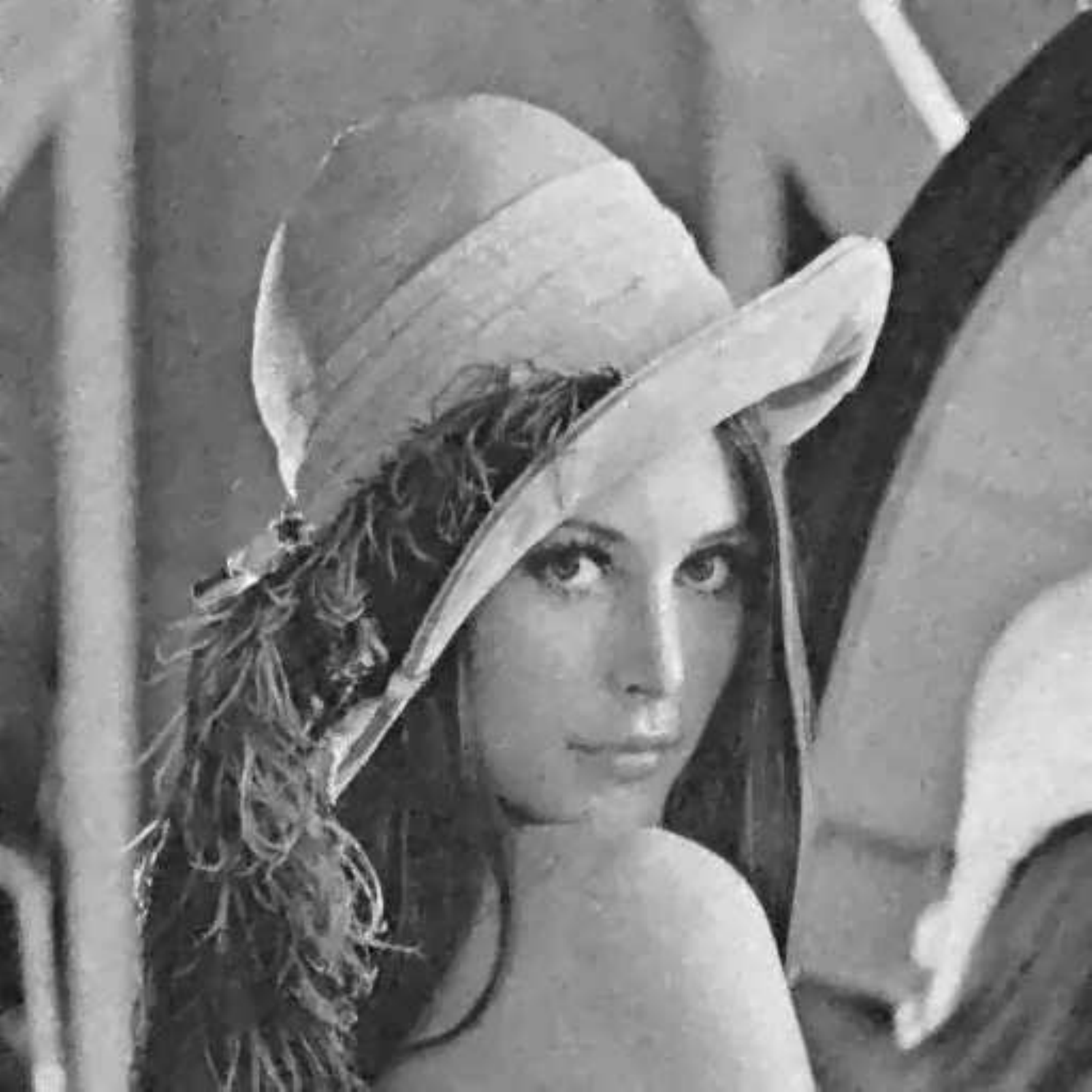}
&\includegraphics[width=0.30\linewidth]{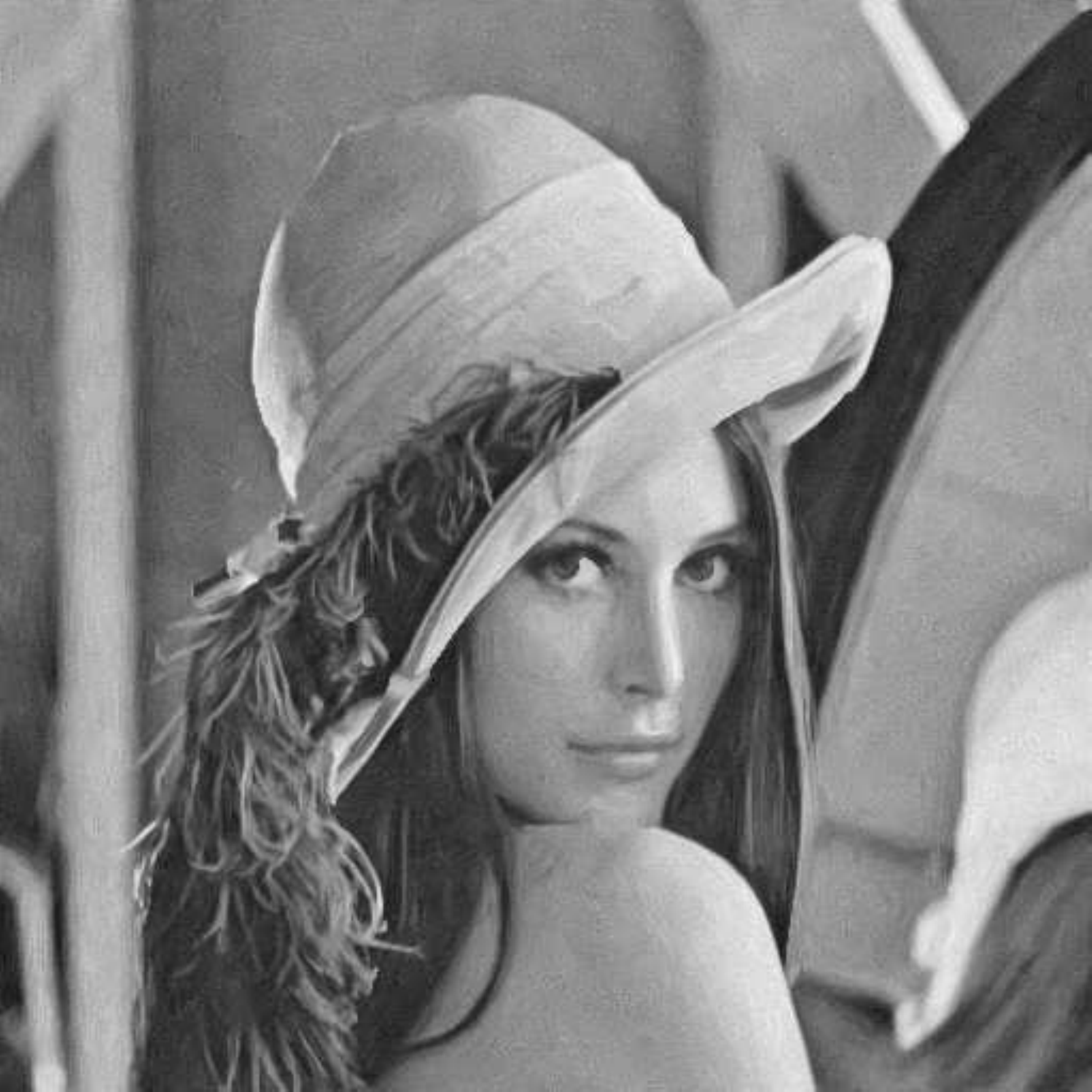}\\
Noisy $\sigma=20$& NLSTV PSNR=31.18& RNLTV PSNR=30.40\\
\includegraphics[width=0.30\linewidth]{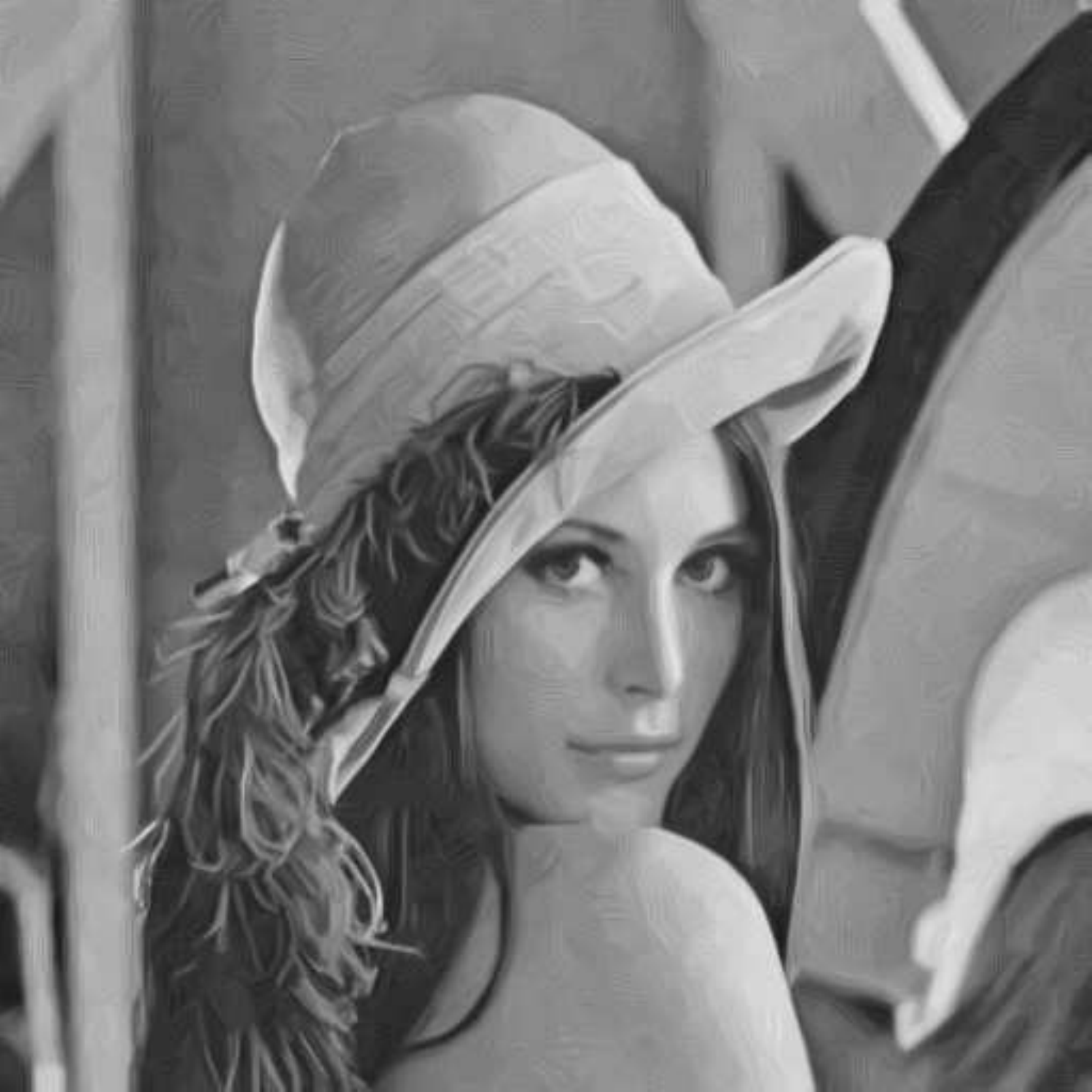}
&\includegraphics[width=0.30\linewidth]{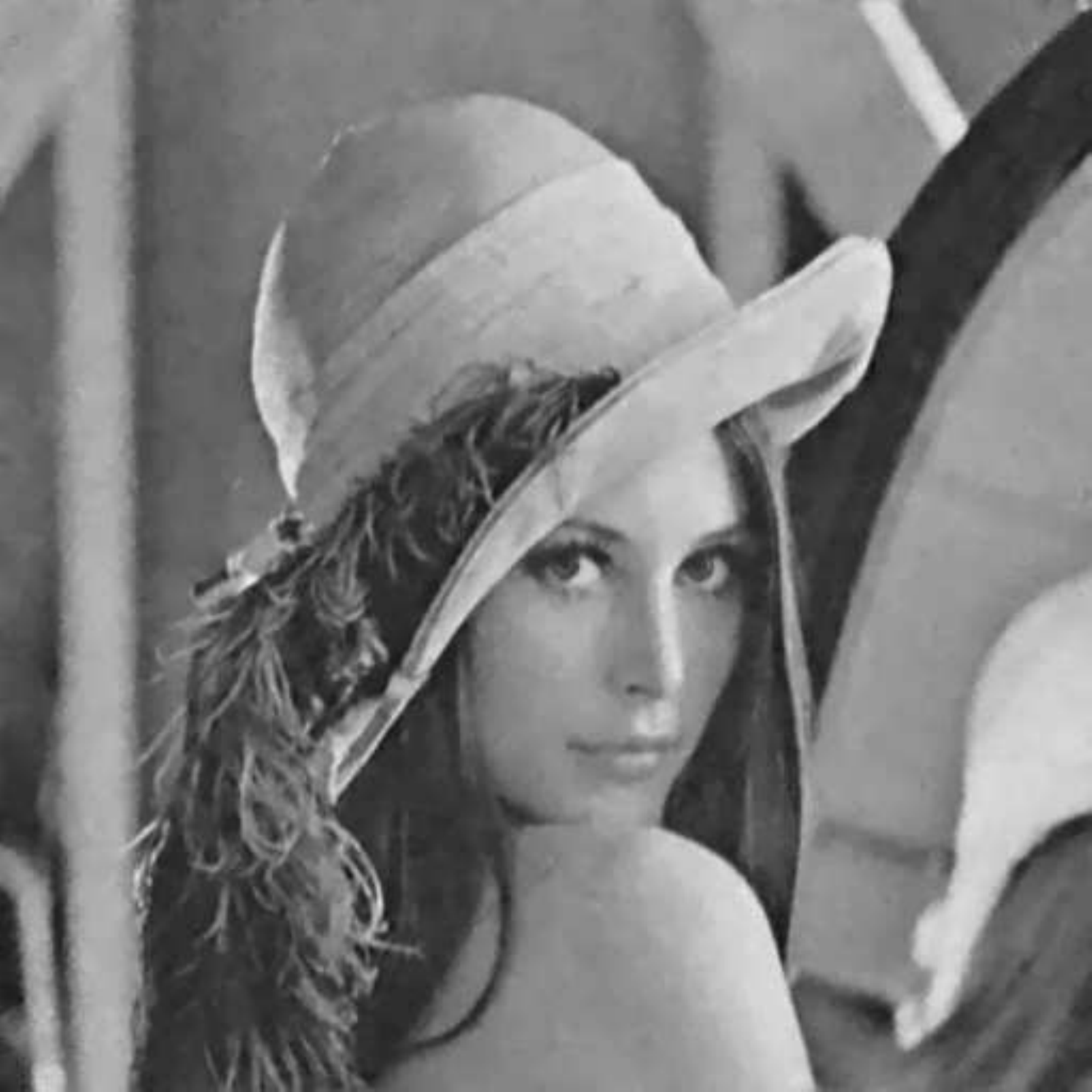}
&\includegraphics[width=0.30\linewidth]{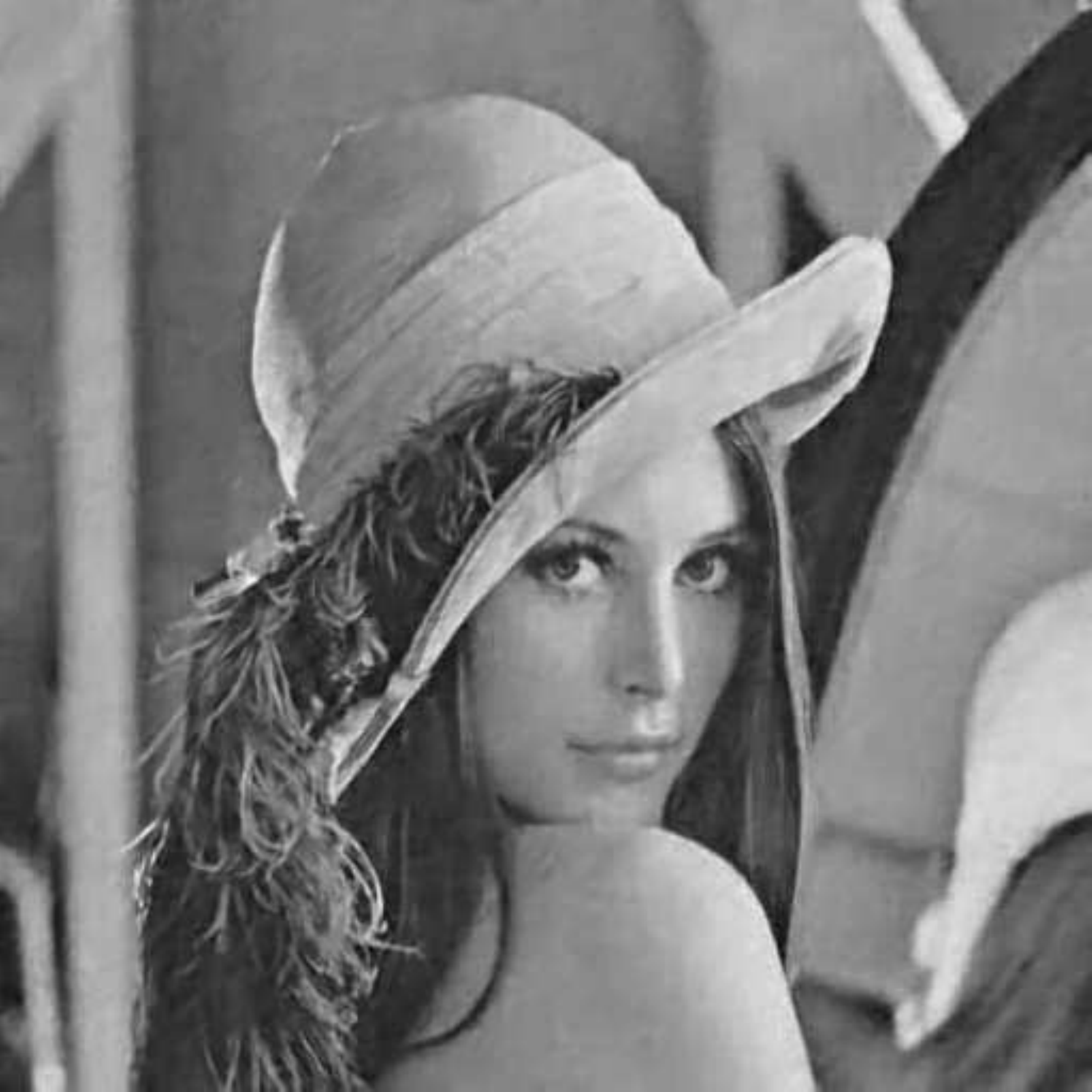}\\
 BNLTV PSNR=31.70 &SFNLTV PSNR=31.77 &L-SFNLTV PSNR=32.49\\
\end{tabular}
\caption{Denoised Lena images  by L-SFNLTV,  NLSTV \cite{lefkimmiatis2015nonlocal}, RNLTV \cite{li2017regularized}, BNLTV \cite{liu2017block} and SFNLTV in the case $\sigma= 20$.}
\label{lsfksvd20}
\end{figure}

\begin{figure} 
\center 
\renewcommand{\arraystretch}{0.5} \addtolength{\tabcolsep}{0pt} \vskip3mm %
\fontsize{8pt}{\baselineskip}\selectfont
\begin{tabular}{ccc}
\includegraphics[width=0.30\linewidth]{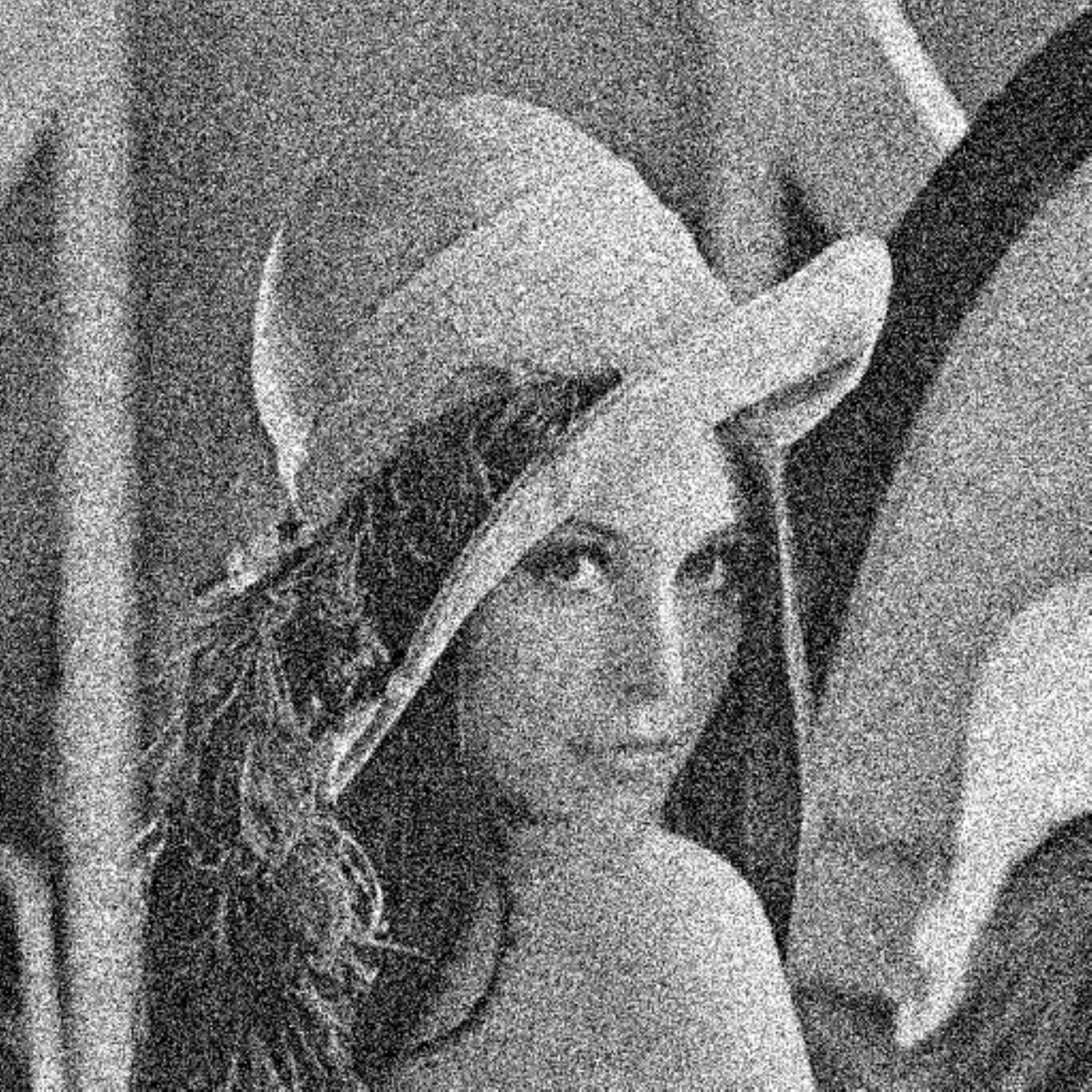}
&\includegraphics[width=0.30\linewidth]{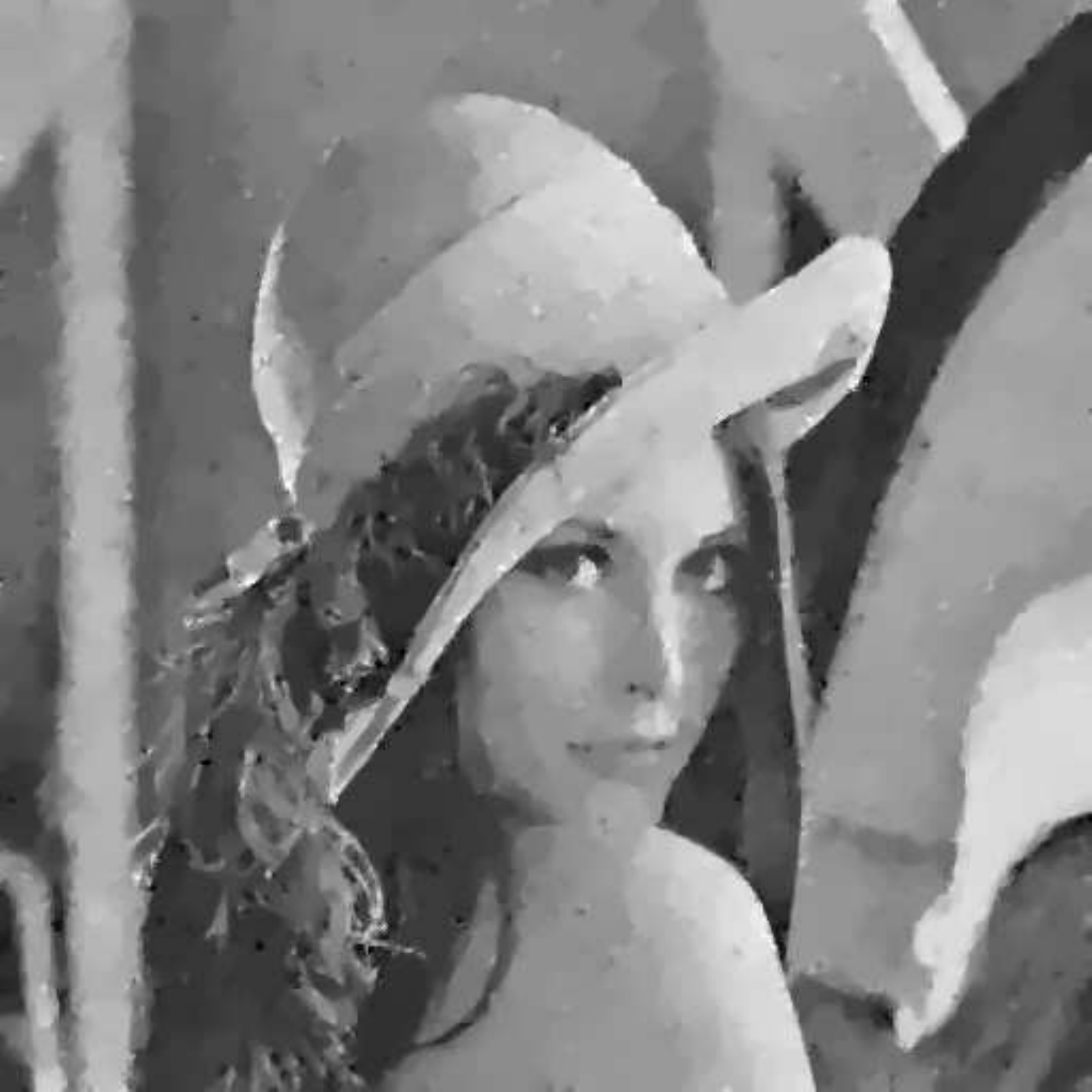}
&\includegraphics[width=0.30\linewidth]{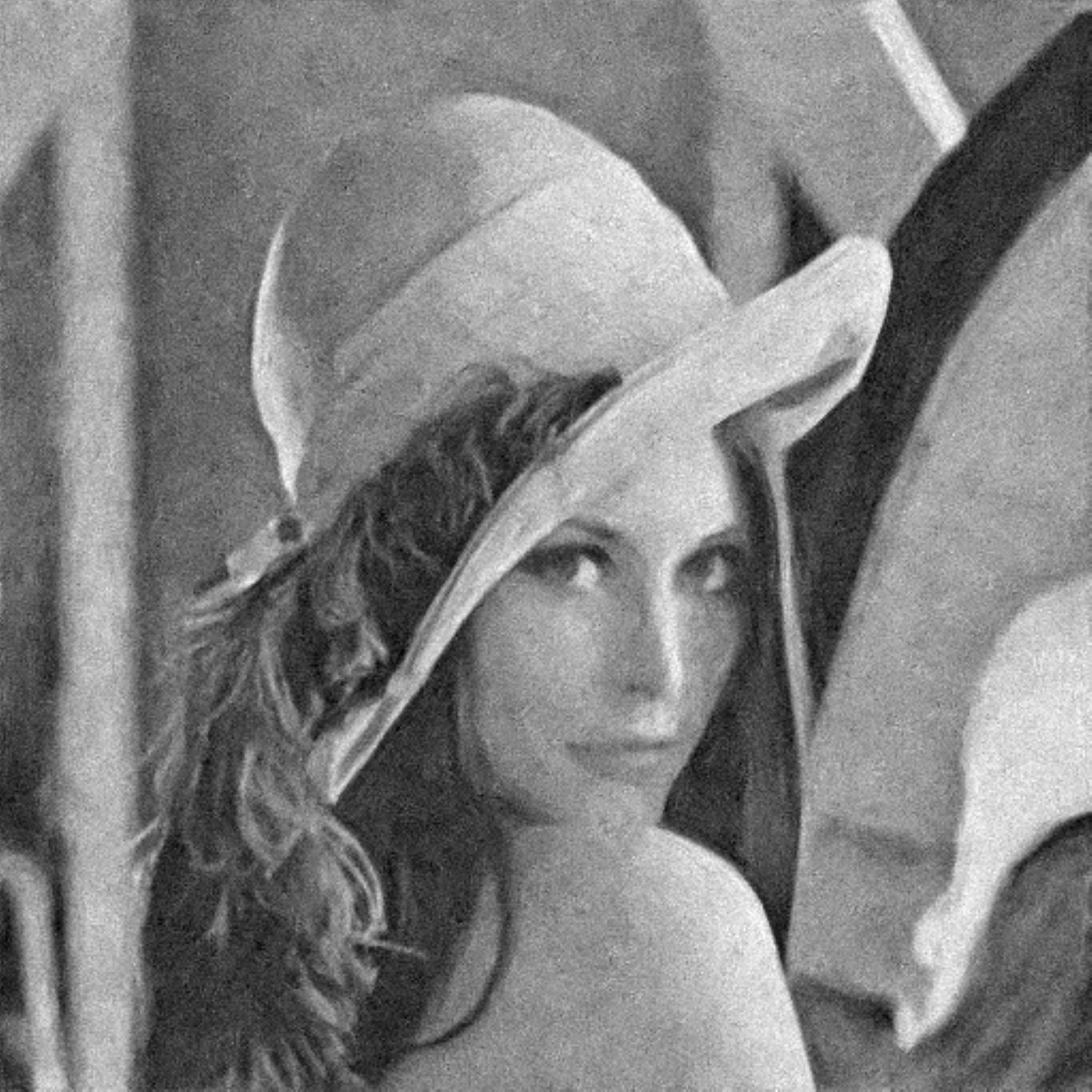}\\
Noisy $\sigma=50$&NLSTV PSNR=27.67 &RNLTV PSNR= 24.40\\
\includegraphics[width=0.30\linewidth]{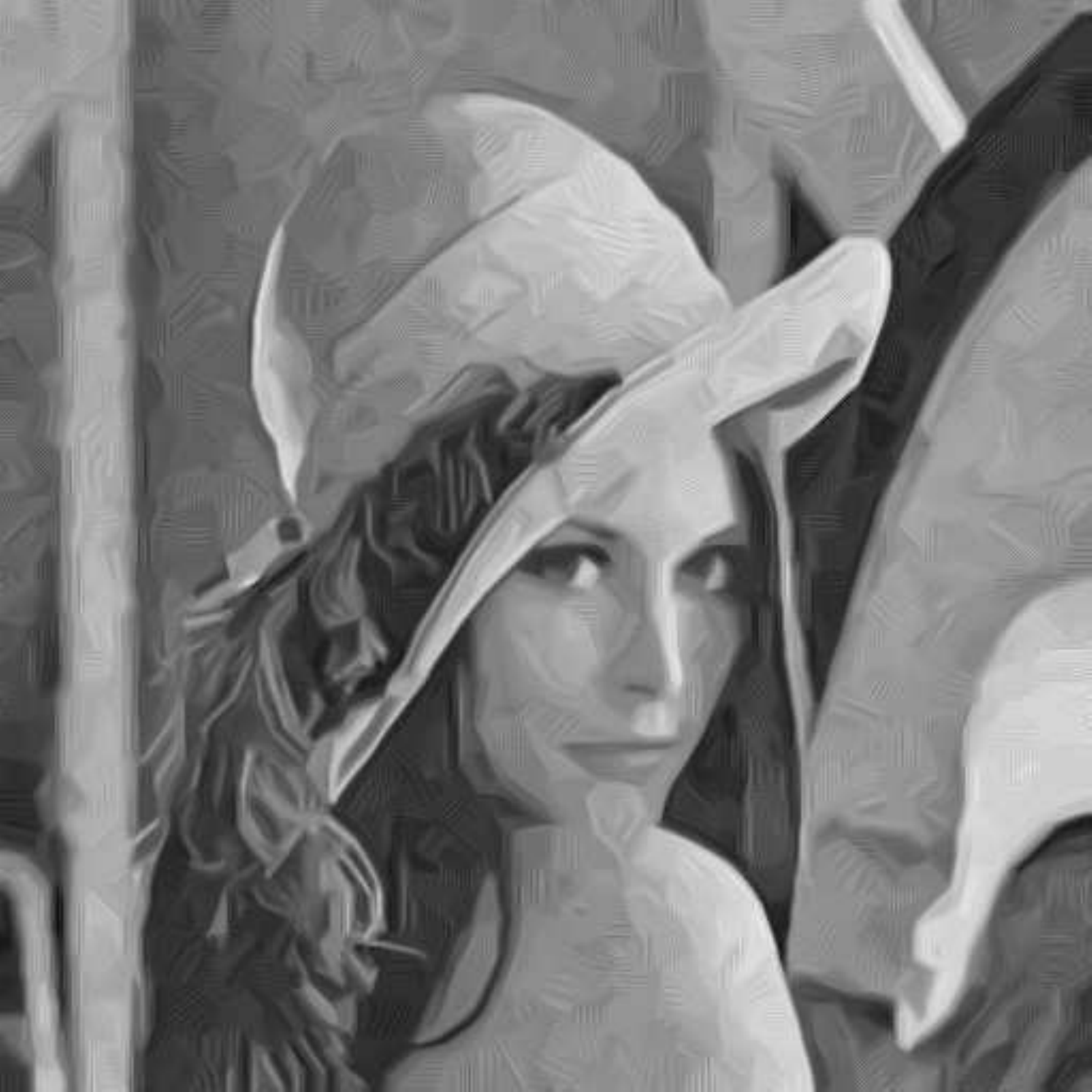}
&\includegraphics[width=0.30\linewidth]{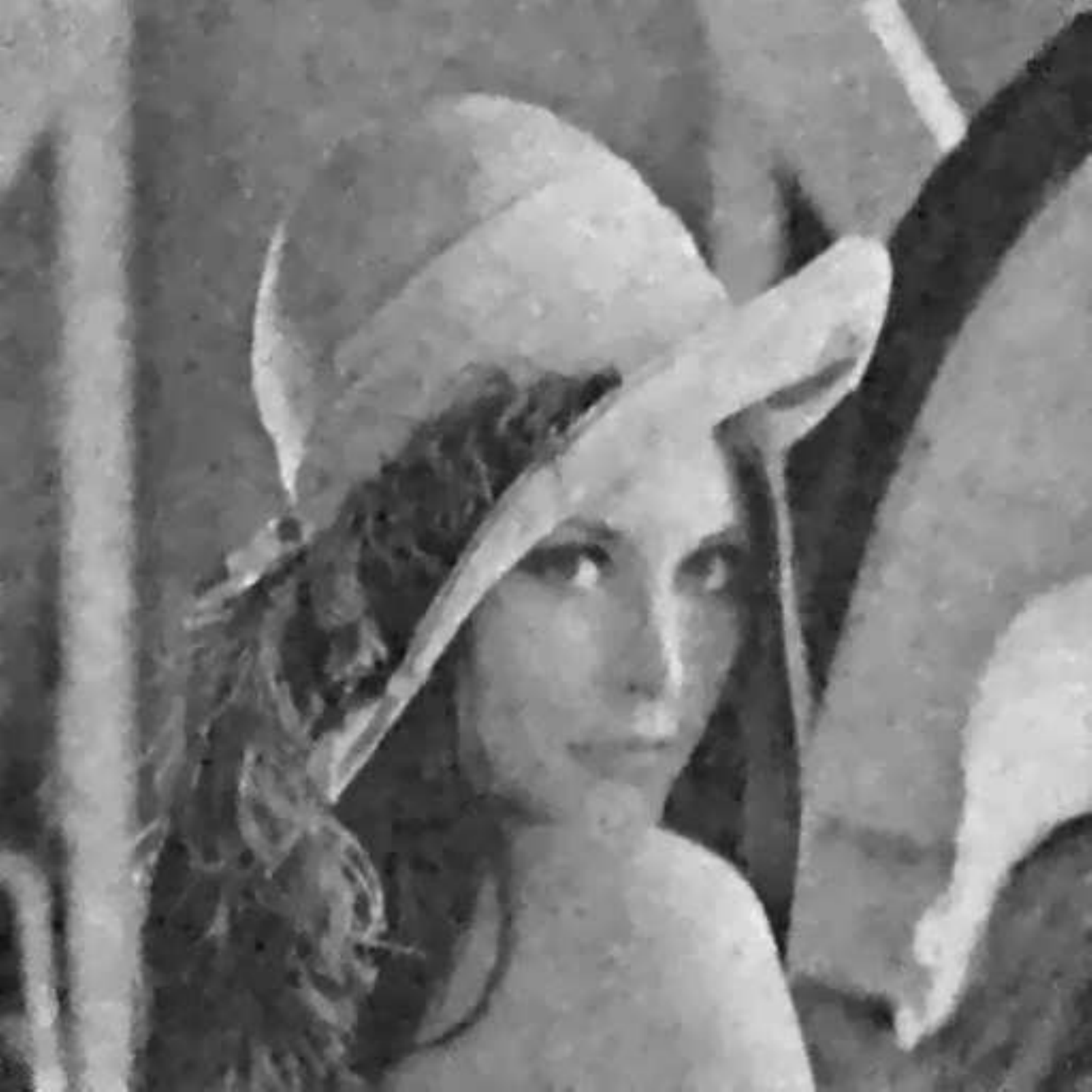}
&\includegraphics[width=0.30\linewidth]{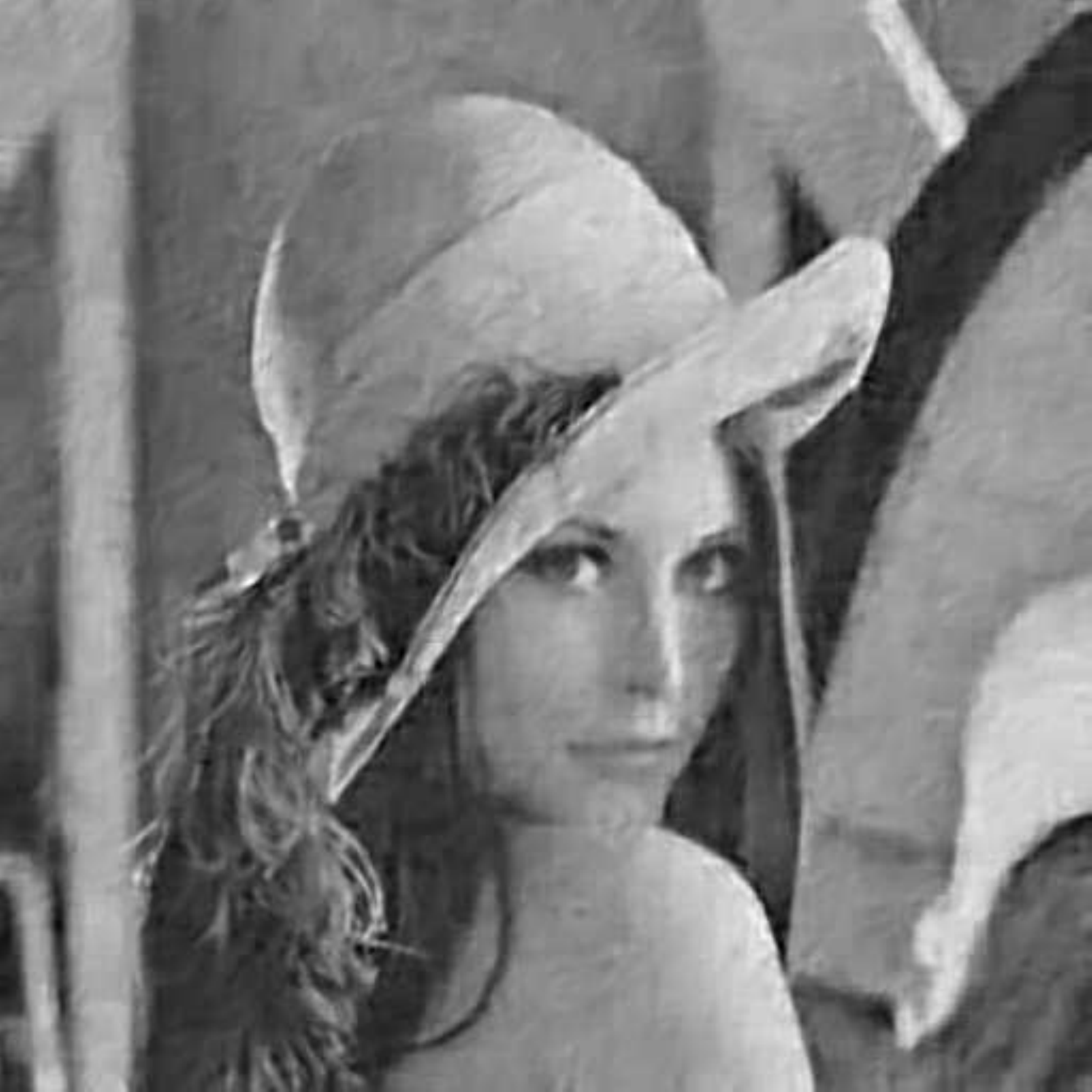}\\
BNLTV PSNR=27.92 & SFNLTV PSNR=27.61&L-SFNLTV PSNR=28.24 \\
\end{tabular}
\caption{Denoised Lena images  by L-SFNLTV,  NLSTV \cite{lefkimmiatis2015nonlocal}, RNLTV \cite{li2017regularized}, BNLTV \cite{liu2017block} and SFNLTV in the case $\sigma=50$.}
\label{lsfksvd50}
\end{figure}

 The comparison results of PSNR values  are shown in  Table \ref{localsig},  which shows that L-SFNLTV performs better than all the other NLTV related methods, NLSTV, RNLTV and BNLTV; and improves SFNLTV. The
 visual comparisons can be seen in Figures  \ref{lsfksvd10}, \ref{lsfksvd20}, and \ref{lsfksvd50}. NLSTV and SFNLTV models still contains stair-casing effects; RNLTV recovers thin structure very well, but homogenous regions are not smooth enough; BNLTV model recovers thin structures and homogenous regions well, but texture-like artifact exists. The artifacts of these methods are more evident when the noise level is high,e.g. $\sigma=50$.   Comparing with these methods, our method has less artifact, while recovers thin structure and homogenous regions well. 

\begin{figure} 
\center 
\renewcommand{\arraystretch}{0.5} \addtolength{\tabcolsep}{0pt} \vskip3mm %
\fontsize{8pt}{\baselineskip}\selectfont
\begin{tabular}{ccc}
\includegraphics[width=0.30\linewidth]{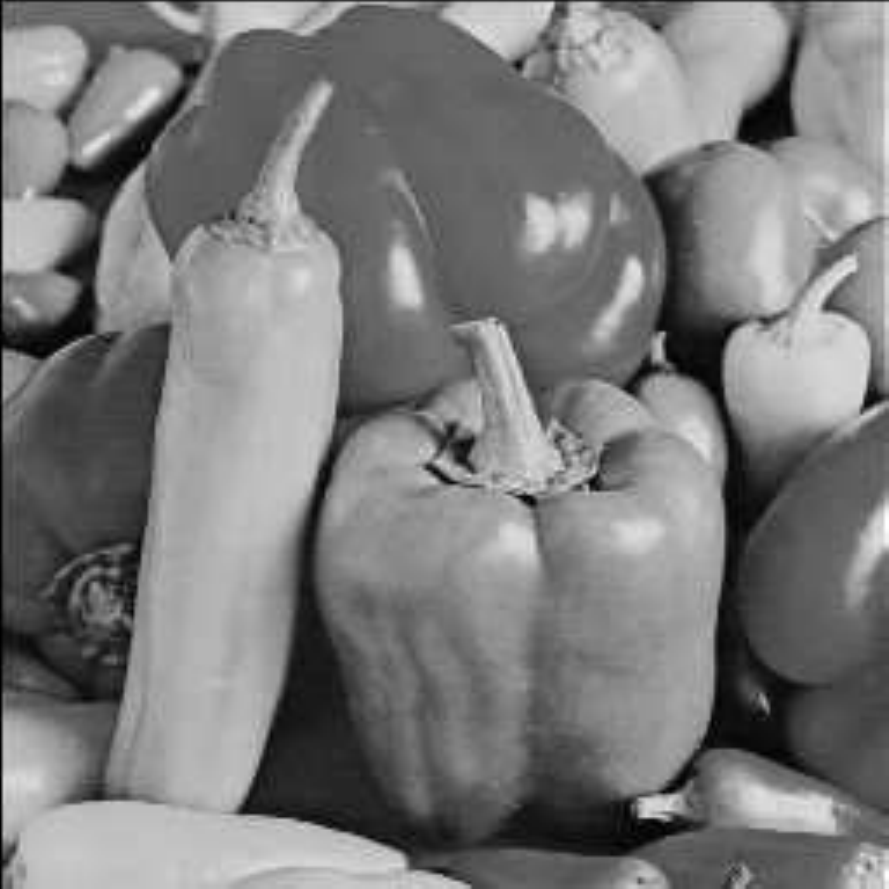}
& \includegraphics[width=0.30\linewidth]{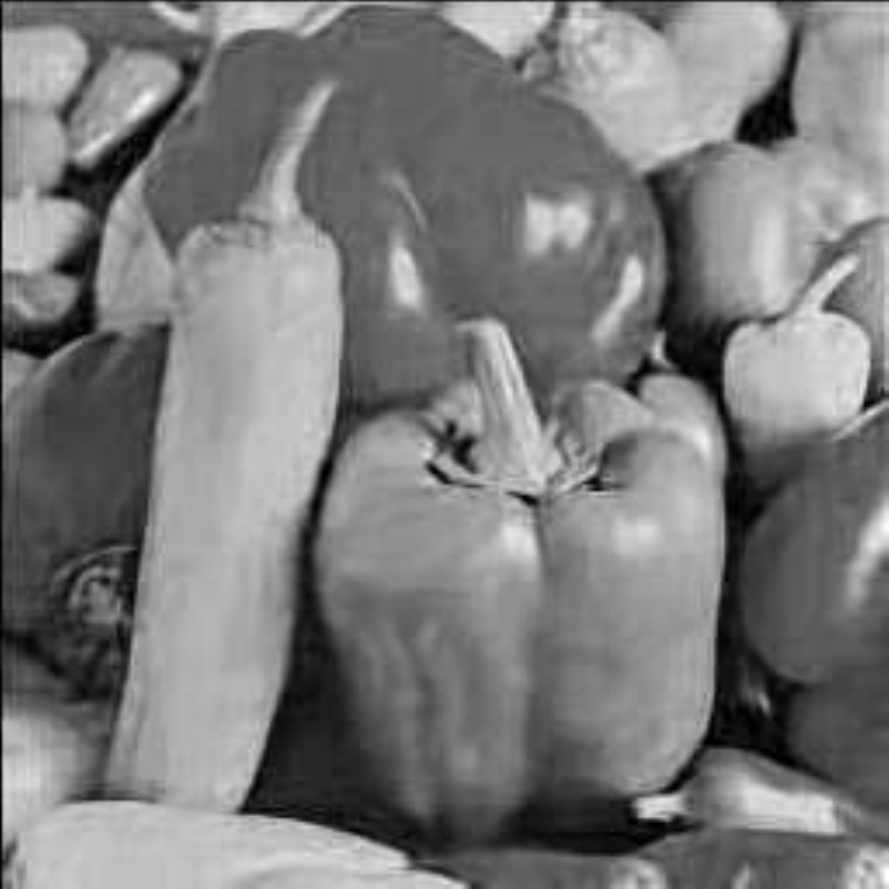}
&\includegraphics[width=0.30\linewidth]{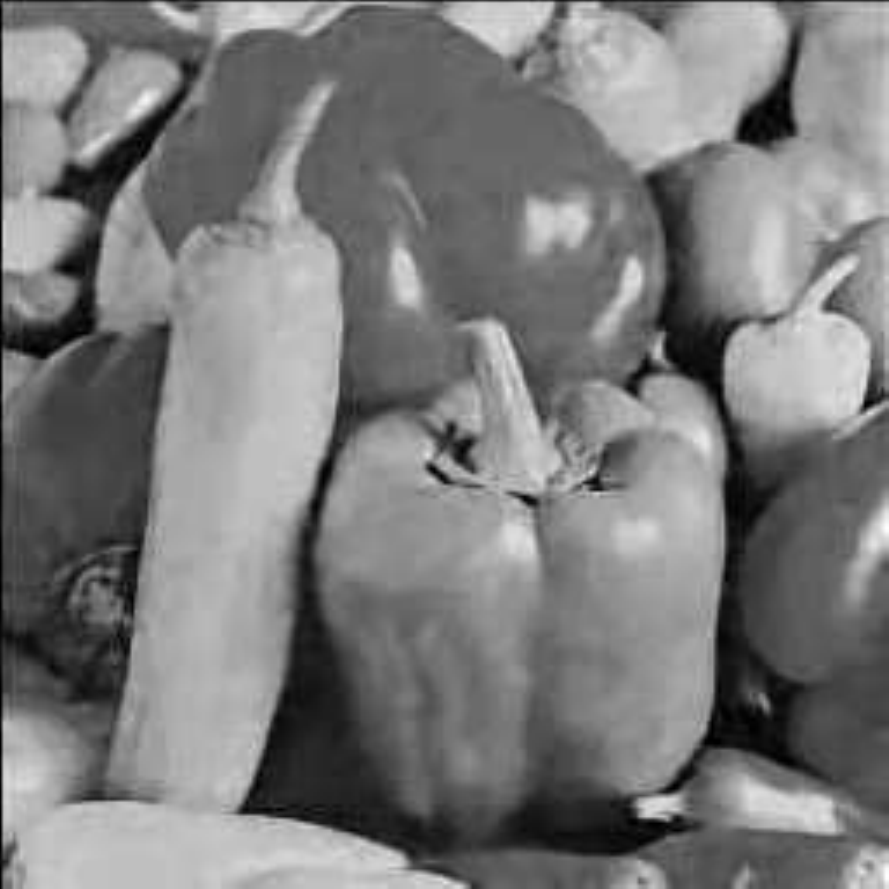}\\
Original (noise-free)&L-FNLTV PSNR=30.09 &L-SFNLTV PSNR=30.60 \\
\includegraphics[width=0.30\linewidth]{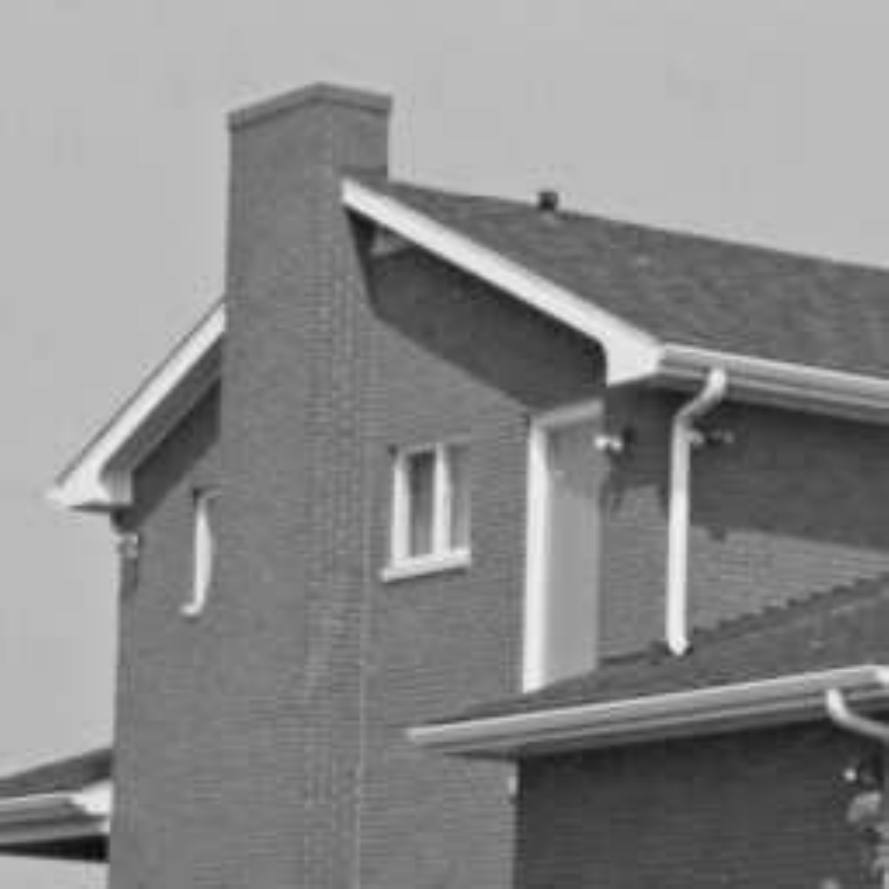}
&\includegraphics[width=0.30\linewidth]{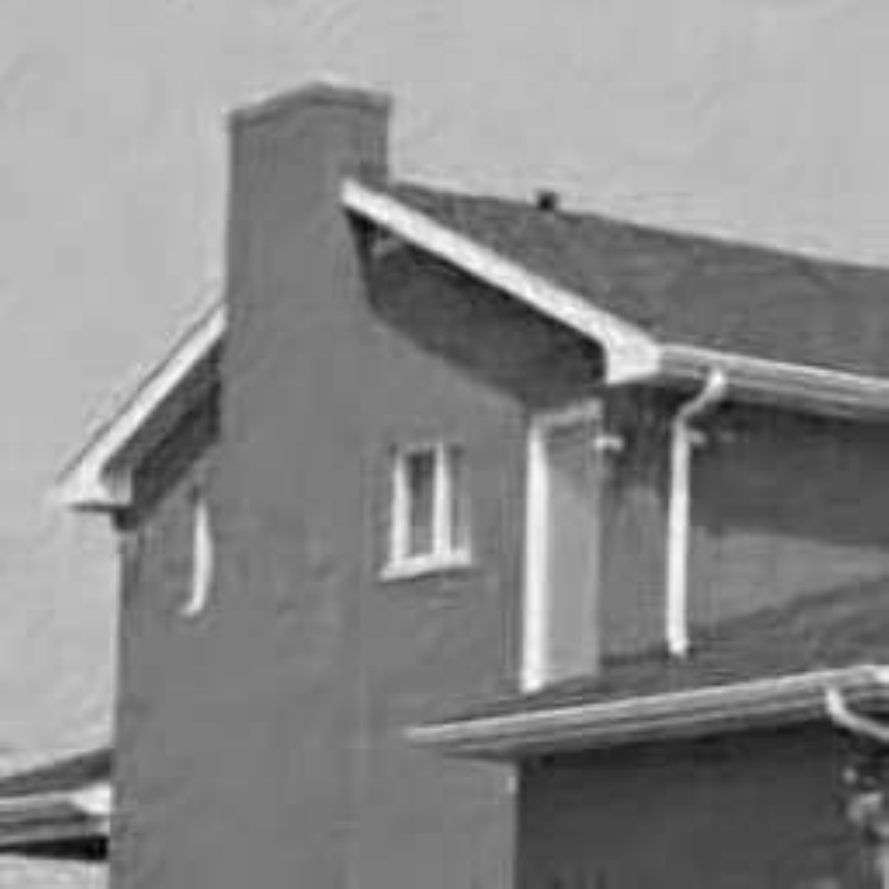}
&\includegraphics[width=0.30\linewidth]{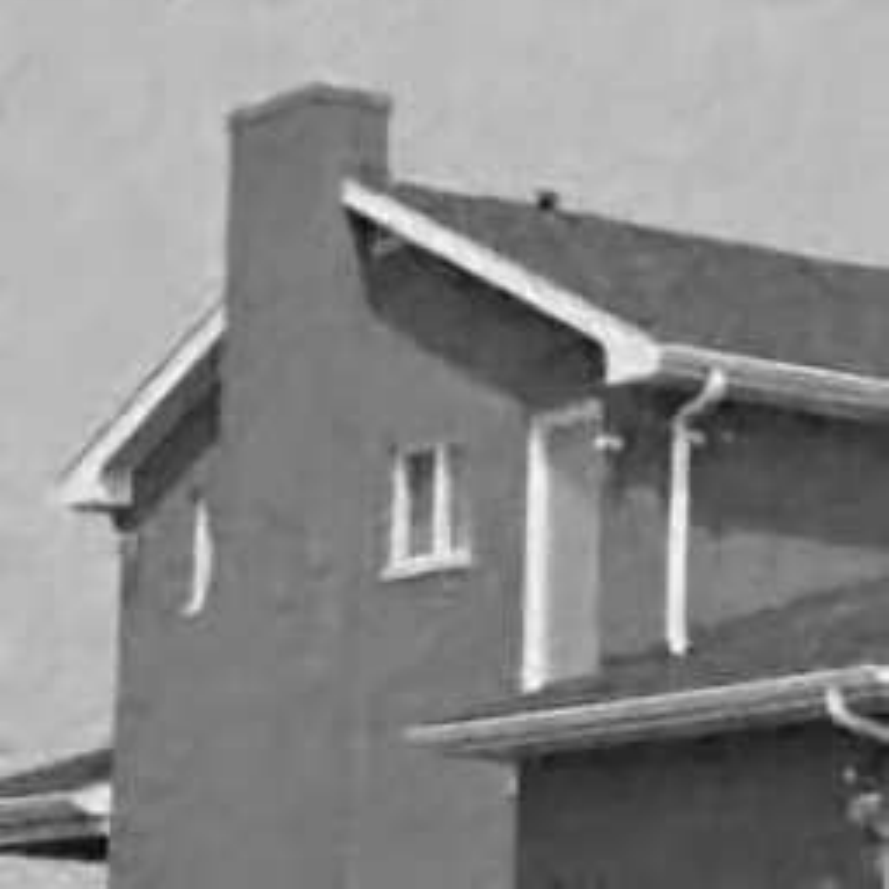}\\
Original (noise-free)&L-FNLTV PSNR=32.33& L-SFNLTV PSNR=32.51 \\
\end{tabular}
\caption{Denoised images Peppers and House by L-FNLTV and L-SFNLTV.}
\label{comft}
\end{figure}

In addition, note also that L-SFNLTV is better than L-FNLTV; see Figure \ref{comft}. In fact, the spatial domain nonlocal total variation helps attenuate the artifact created by the algorithms based on Fourier transform. For L-FNLTV, we use the regions of size $16\times 16$, the moving step $n_s=6$ and $\lambda_f=\sigma$.

Finally, we compare the running time of L-SFNLTV with other methods. All the  codes except NLSTV are run in the platform of MATLAB R2014a on a 2.40 GHz Intel Core i7 CPU processor under Windows system. NLSTV\footnote{The original code can not be implemented easily under Windows system, so we use a different computer installing Linux system.} is run in the platform of MATLAB R2011a on a 2.13 GHz Intel Core i3 CPU processor under Linux system. From Table 5,
we can see that L-SFNLTV also has advantage in running time comparing with the other NLTV related algorithms except SFNLTV.
\begin{table}
\begin{center}
\caption{Choice of parameters of L-SFNLTV for non-textured images.}
\begin{tabular}{ccccccccccc}
\noalign{\smallskip}\hline\noalign{\smallskip}
$\sigma$ & d & $\lambda$ & $\lambda_f$ \\
\noalign{\smallskip}\hline\noalign{\smallskip}
10 & 9& 4 & 6 \\
20 & 9 & 5 & 11 \\
30 & 11 & 7 & 21 \\
50 & 15 & 10 & 38\\ \noalign{\smallskip}\hline
\end{tabular}
\label{paralsf}
\end{center}
\end{table}

\begin{table}
\begin{center}
\caption{PSNR values for Lena, Barbara, Peppers, Boats, Bridge, House and Cameraman.}
\begin{tabular}{ccccccccccc} 
\hline\noalign{\smallskip}
Image & &Lena &Bar& Peppers &Boats &Bridge & House & Cam \\
\noalign{\smallskip}\hline \noalign{\smallskip}
$\sigma$=10&NLSTV &34.61&  31.29&  34.06&   33.15&   30.62&  34.52&   33.30\\
&RNLTV &34.17& 32.79&   33.11& 32.68&    29.14&   34.43&   31.97\\
&BNLTV &34.57&   33.77&   32.31&  32.83&   29.96&   34.97&   32.52\\
 &SFNLTV  & 35.05&   33.93&   33.82&   33.42&   30.86&  35.49&   33.45\\
 &L-SFNLTV & $\mathbf{35.57}$ & $\mathbf{34.60}$ &     $\mathbf{34.28}$ & $\mathbf{33.56}$ & $\mathbf{30.96}$ &$\mathbf{35.62}$ & $\mathbf{33.65 }$\\
\noalign{\smallskip}\hline\noalign{\smallskip}
$\sigma$=20&NLSTV &31.18&  27.23&   30.16&   29.80&  27.03&   30.93&   29.41\\
&RNLTV &30.40&   29.19&	29.64&   29.50&   26.63&   30.21&   28.54\\
&BNLTV &31.71&   30.40&   28.38&   29.55&   26.06&   32.17&   28.85\\
& SFNLTV & ${31.77}$ &  29.19 &   ${30.29}$ &   ${29.89}$ &  ${ 26.92}$ &  ${ 32.14}$ &   ${29.64}$ \\
&L-SFNLTV& $\mathbf{32.49}$ & $\mathbf{30.98}$ &$\mathbf{30.59}$&$\mathbf{30.40}$ &$\mathbf{27.15}$ & $\mathbf{32.50}$ & $\mathbf{29.69}$ \\
\noalign{\smallskip}\hline\noalign{\smallskip}
$\sigma$=30&NLSTV&29.86&   24.84&   28.51&   28.14&   25.04&   29.89&  $\mathbf{27.76}$\\
&RNLTV&27.89&   26.74&  27.26&   27.18&   24.93&   27.78&   26.55\\
&BNLTV &29.98&   28.59& 26.58&   27.94&   24.69&   30.36&   27.21\\
&SFNLTV  & 29.82&   26.55&   28.13&   27.93&   25.01&   29.97&   27.58\\
&L-SFNLTV&$\mathbf{30.64}$& $\mathbf{28.86}$ & $\mathbf{28.55}$ & $\mathbf{28.50}$ & $\mathbf{25.19}$ &$\mathbf{30.64}$ & 27.75 \\
\noalign{\smallskip}\hline\noalign{\smallskip}
$\sigma$=50 &NLSTV &27.67&   23.17&   ${26.00}$&   25.96&   23.12&   27.57&   $\mathbf{25.42}$\\
 &RNLTV& 24.40 & 23.30& 23.91&   23.94& 22.50& 24.12&    23.41&\\
&BNLTV & 27.92&  $\mathbf{26.21}$ &  24.39&   25.92&   23.03&   28.10&   25.08\\
&SFNLTV  & 27.61&   24.11&   25.48&  25.69&   23.18&   27.40&  24.83\\
&L-SFNLTV&$\mathbf{28.26}$ & ${26.20}$ & $\mathbf{ 26.08}$ &$\mathbf{26.25}$&$\mathbf{23.40}$& $\mathbf{28.16}$& 25.24\\
\noalign{\smallskip}\hline
\end{tabular}
\label{localsig}
\end{center}
\end{table}

\begin{table}
\begin{center}
\caption{Running time in second   with grayscale images of size $256\times 256$}
{\footnotesize\addtolength{\tabcolsep}{-3pt}
\begin{tabular}{|c|c|c|c|c|c|c|ccc} 
\hline
NLSTV & RNLTV & BNLTV & SFNLTV & L-SFNLTV \\ 
\hline
22 &3344  &11.7 & 2.4 &8.3\\
\hline
\end{tabular}
}
\end{center}
\label{runtime0}
\end{table}

\section{Conclusion} 
In this paper, we first studied the nonlocal total variation (NLTV) model for image denoising,
which was initially introduced for deconvolution in \cite{lou2010image}. We established
the relation between this model and neighborhood filters and derived the iterative formula
 of SURE (Stein’s Unbiased Risk
Estimation) for NLTV with gradient descent algorithm  to estimate the denoising performance without reference to original true
image. Then, we extended the NLTV model to a spatial-frequency domain nonlocal total variation model
(SFNLTV), which was showed to be better than NLTV for texture and fine details in images, and also
better than the NL-means algorithm in most cases. Finally, we proposed a local version of
SFNLTV, abbreviated as L-SFNLTV, to make full use of the advantage of Fourier transform and we
showed that L-SFNLTV improved SFNLTV greatly. Since L-SFNLTV can be implemented in parallel, it
leaded to a fast denoising algorithm. Experiments showed that L-SFNLTV had evident advantages both in
denoising performance and implementation speed, comparing with other recently proposed NLTV-related
methods.

\section*{References}





\end{document}